\documentclass[twocolumn]{aastex701}
\usepackage{graphicx,amsmath}

\newcommand{\Teff}{\ensuremath{T_{\rm eff}}}  
\newcommand{\logTeff}{\ensuremath{\log T_{\rm eff}}}  
   
\newcommand{\mesa}{\texttt{MESA}}
\newcommand{\MESA}{\mesa}
\newcommand{\rsp}{\texttt{RSP}}
\newcommand{\RSP}{\rsp}

\newcommand{\PL}{\ifmmode{P{\rm -}L}\else$P$--$L$\fi}
\newcommand{\PR}{\ifmmode{P{\rm -}R}\else$P$--$R$\fi}
\newcommand{\PA}{\ifmmode{P{\rm -}{\rm Age}}\else$P$--Age\fi}
\newcommand{\ML}{\ifmmode{M{\rm -}L}\else$M$--$L$\fi}

\newcommand{\MS}{\ifmmode{\,}\else\thinspace\fi{\rm M}\ifmmode_{\odot}\else$_{\odot}$\fi}
\newcommand{\LS}{\ifmmode{\,}\else\thinspace\fi{\rm L}\ifmmode_{\odot}\else$_{\odot}$\fi}
\newcommand{\RS}{\ifmmode{\,}\else\thinspace\fi{\rm R}\ifmmode_{\odot}\else$_{\odot}$\fi}
\newcommand{\feh}{\ifmmode{\rm [Fe/H]}\else[Fe/H]\fi}
\newcommand{\fcor}{\ifmmode{f_{\rm H}}\else$f_{\rm H}$\fi}
\newcommand{\fenv}{\ifmmode{f_{\rm env}}\else$f_{\rm env}$\fi}

\newcommand{\npg}{\ifmmode{^{14}\rm N (\rm p,\gamma) ^{15} \rm O}\else{$^{14}\rm N (\rm p,\gamma)^{15}\rm O$}\fi}
\newcommand{\cag}{\ifmmode{^{12}\rm C (\alpha,\gamma) ^{16} \rm O}\else{$^{12}\rm C (\alpha,\gamma)^{16}\rm O$ }\fi}

\defcitealias{Ziolkowska-2024}{Paper~I}

\begin{document}

\title{Toward a Comprehensive Grid of Cepheid Models with MESA. III. Evolutionary and Pulsation Relations for Models with Core and Envelope Overshooting}

\correspondingauthor{Radoslaw Smolec}
\author[orcid=0000-0001-7217-4884,sname='Smolec']{Radoslaw Smolec}
\affiliation{Nicolaus Copernicus Astronomical Centre, Polish Academy of Sciences, Bartycka 18, 00-716 Warszawa, Poland}
\email[show]{smolec@camk.edu.pl}  

\author[orcid=0000-0002-0696-2839,sname='Zi\'{o}\l{}kowska']{Oliwia Zi\'{o}\l{}kowska}
\email{oliwiakz@camk.edu.pl}
\affiliation{Nicolaus Copernicus Astronomical Centre, Polish Academy of Sciences, Bartycka 18, 00-716 Warszawa, Poland}

\author[orcid=0000-0002-7448-4285,sname='Rathour']{Rajeev Singh Rathour}
\email{rajeevsr@camk.edu.pl}
\affiliation{Université C\^{o}te d'Azur, Observatoire de la C\^{o}te d'Azur, CNRS, Laboratoire Lagrange, France}

\author[orcid=0000-0002-3643-0366,sname='Hocd\'{e}']{Vincent Hocd\'{e}}
\email{vhocde@camk.edu.pl}
\affiliation{Nicolaus Copernicus Astronomical Centre, Polish Academy of Sciences, Bartycka 18, 00-716 Warszawa, Poland}

\author[orcid=0000-0002-1662-5756,sname='Wielg\'orski']{Piotr Wielg\'orski}
\email{pwielgor@camk.edu.pl}
\affiliation{Nicolaus Copernicus Astronomical Centre, Polish Academy of Sciences, Bartycka 18, 00-716 Warszawa, Poland}

\begin{abstract}
Evolutionary tracks for 2--8\MS{} models, covering a \feh{}=$-1.0$ ($Z=0.0014$) to \feh{}=+0.2 ($Z=0.02$) metallicity range are computed with Modules for Experiments in Stellar Astrophysics, \MESA, to investigate evolutionary and pulsation properties of classical, fundamental mode Cepheids. We examine in detail the effects of convective overshooting from the Main Sequence core, as well as from the convective envelope on the Red Giant Branch. Mass loss is also included in a few model sets. Linear pulsation properties are derived consistently with a module of \MESA, Radial Stellar Pulsation, \rsp. We provide edges of the classical Instability Strip, as well as ages, crossing times through the Instability Strip and period change rates. Period--Luminosity, Mass--Luminosity, Period--Radius and Period--Age relations are provided, both in analytical and tabular form. Their dependence on metallicity, crossing number and overshooting parameters are investigated. Qualitative comparisons with classical Cepheids in the Milky Way and Magellanic Clouds as well as other theoretical relations are presented. We find satisfactory agreement for most of the observables and good match with other theoretical work, however reproducing short-period Cepheids in the Small Magellanic Cloud as well as Cepheid mass discrepancy pose a challenge for the presented models. Considering metallicity effect of the Period-Luminosity relation, we find $\gamma\approx-0.20$\,mag\,dex$^{-1}$, nearly independent on photometric pass band and in good agreement with recent observational studies. The magnitude of this effect depends on the underlying mass–luminosity relation, being stronger for relations that predict higher luminosities at a given mass.
\end{abstract}


\keywords{\uat{Cepheid variable stars}{218} --- \uat{Stellar evolution}{1599} --- \uat{Stellar evolutionary models}{2046} --- \uat{Stellar pulsations}{1625}}

\section{Introduction\label{sec:intro}} 

Classical Cepheids are one of the cornerstones of modern stellar astrophysics. They are bright, have characteristic light curves with high amplitude variability, and are commonly found in young stellar systems. Most of classical Cepheids are core-helium burning stars, performing a nearly horizontal blue loop in the Hertzsprung-Russell diagram (HRD), gradually depleting helium in their cores. During this evolution, they develop large amplitude radial pulsations as they cross the Instability Strip (IS) -- a region in the HRD in which stars are unstable to radial pulsation. Pulsations in single radial fundamental (F) or radial first overtone (1O) modes are most common. In general two IS crossings are possible at core helium burning, second (2nd) during blue-ward evolution and third (3rd) during red-ward evolution. The first (1st) crossing of the IS occurs earlier, during post Main Sequence (MS) evolution, as star evolves towards Red Giant Branch (RGB). This phase is much faster than the blue loop phase as there are no core nuclear energy sources. Consequently, most of the observed Cepheids are expected to follow blue loop evolution. For general overview of Cepheids, see eg., \cite{Catelan-book, Bono-2024Review}. 

Owing to the Period-Luminosity (\PL) relation, the Leavitt law \citep{Leavitt-1912}, which makes them excellent standard candles, classical Cepheids constitute a crucial step in the local cosmic distance ladder \citep[eg.,][]{Freedman-2001,Riess-2022}. Since these are relatively simple pulsators, they also provide an excellent testing ground for theories of stellar evolution and pulsation \citep[eg.,][]{Bono-2000,MoskalikDziembowski-2005,Pietrzynski-2010,DeSomma-2022,Rathour-2025,Deka-2025}. As a result of their well constrained location in the HRD, they also follow relations such as Period-Radius (\PR) or Period-Age (\PA), which, along with their distances, can be used in galactic structure and galactic archaeology studies \citep[eg.,][]{Jacyszyn-2016,DeSomma-2025}.

Although relatively well understood, Cepheids still pose several unsolved problems. The oldest is the Cepheid mass discrepancy. The masses of Cepheids predicted on the basis of evolutionary calculations are higher compared to the masses determined on the basis of pulsation theory \citep[][]{Stobie-1969,Cox-1980}. The discrepancy has been partially removed by the revision of the opacity tables \citep[eg.,][]{IglesiasRogers-1991, Moskalik-1992}, and its current level is $\sim20$\,\% \citep[][]{Keller-2008}. Precise determinations of Cepheid masses in eclipsing binary systems \citep{Pietrzynski-2010, Pilecki-2018} indicate that the solution to this problem lies in evolution theory. 

Another current problem is the dependence of the \PL{} relation on metallicity, \feh, quantified with the $\gamma$ (in mag\,dex$^{-1}$) parameter, see eg., \cite{Storm-2011,Gieren-2018,Breuval-2021,Breuval-2022,Bhardwaj-2023,Trentin-2024,Ripepi-2025}. The sign and strength of the metallicity effect directly affects the distance scale, so it is an important factor discussed in connection with the Hubble tension \citep[][]{Breuval-2025}. While most of the studies seem to converge on a negative value of $\gamma\approx-0.2$\,mag\,dex$^{-1}$ \citep[][and references therein]{Breuval-2025}, meaning that metal rich Cepheids are intrinsically brighter than the metal poor ones, other work indicate $\sim$twice as strong metallicity effect \citep{Ripepi-2025} or negligible metallicity effect \citep{MadoreFreedman-2025}.

Theoretical evolutionary and pulsation studies are essential to address the above problems and they have been undertaken with a variety of evolution and pulsation codes \citep[eg.,][]{Bono-2000b,Bressan-2012,Georgy-2013,Hidalgo-2018}. The blue loop phase is challenging to model, as whether the loop develops, or not, is a subject to adopted input physics and numerical setup (eg., resolution) of the models \citep[eg.,][]{Lauterborn-1971,Walmswell-2015,XuLi-2004a,XuLi-2004b}. In some parameter regimes, the behavior of the loops seem to be erratic (eg., non-monotonous behavior with increasing mass) or challenging to reconcile with observations (eg., no loops or too short loops at certain metallicities).

Solutions to the mass discrepancy problem \citep[see][for different possibilities]{Bono-2006} were investigated in terms of overshooting \citep[eg.,][]{CassisiSalaris-2011, PradaMoroni-2012}, rotation \citep[eg.,][]{Anderson-2014, Anderson-2016, Miller-2020}, or pulsation induced mass loss \citep[eg.,][]{Neilson-2011}. Some of the above studies addressed several possible factors simultaneously. Cepheids in double-lined (SB2) eclipsing binary systems, and stringent constraints they impose on the models, were also used in addressing the above problem \citep[eg.,][]{CassisiSalaris-2011, PradaMoroni-2012, NeilsonLanger-2012, Marconi-2013, Neilson-2015, Deka-2025}. While some of the above studies indicate the mass discrepancy is no longer an issue, all observational constraints, including eg., period change rates, in particular the ratio of the blue- and red-ward evolving Cepheids, are not always met or taken into account. It is also important to stress that 1D codes come with their limitations: there are degeneracies between different phenomena, as their effects, possible to test with observation, may be the same. This is largely the case for overshooting and rotation which, due to additional mixing, increase the core size leading to increased luminosity at a given mass. There is however no unique way to implement rotation \citep[eg.,][]{MaederMeynet-2000,Paxton-2013}. Comprehensive studies, addressing all observational constraint and including the effects of overshooting, rotation and pulsation induced mass loss are needed.

Concerning metallicity dependence of the $\PL$ relation, theoretical work seem to converge on the negative value of $\gamma$ \citep[eg.,][]{Anderson-2016,DeSomma-2022,Khan-2025}, in agreement with most of the recent observational analyses, see \cite{Breuval-2025}. 

This paper is third in a series in which we address evolutionary Cepheid models (more generally, models of intermediate mass stars, till end of core helium burning) with Modules for Experiments in Stellar Astrophysics, \MESA{} \citep{Paxton-2011,Paxton-2013,Paxton-2015,Paxton-2018,Paxton-2019,Jermyn-2023}. While several studies regarding classical Cepheids were published with \MESA{} and \MESA-\RSP{} \citep[eg.,][]{EspinozaArancibia-2022,Kurbah-2023,Hocde-2024,Deka-2024}, comprehensive analysis of evolutionary and pulsation properties of Cepheid models was not conducted yet. In the first paper of the series, \cite{Ziolkowska-2024} (hereafter Paper~I), we defined numerical and micro-physical setup for the reference model which we adopt for the present investigation. We also conducted a convergence study showing that the results are little sensitive to parameters controlling the numerical resolution of the model. The most important result was the assessment of uncertainty of the evolutionary tracks, at different evolutionary phases, arising from the freedom in choosing various options regarding eg., Mixing Length Theory (MLT), atmosphere boundary conditions, nuclear reactions, reference solar mixture, opacity interpolation method, or the scheme convective boundaries are determined. These uncertainties are taken into account while deriving evolutionary relations in the present study. In the second paper, \cite{Ziolkowska-2026}, we investigated the corresponding uncertainties for surface abundances. In the present work, we computed a large grid of evolutionary models covering 2--8\MS{} mass range and metallicities from $\feh=-1.0$ to $+0.2$, with dense sampling both in metallicity and mass. We explore the effects of MS core overshooting and envelope overshooting. For some of the model sequences we include standard mass loss (Reimers formula). The effects of \npg{} nuclear reaction rate as well as different reference solar mixture on the morphology of blue loops are also investigated. Based on evolutionary and pulsation calculations we derive $\PL$, $\ML$, $\PR$ and $\PA$ relations for various scenarios and compare these with other theoretical work and observations. We also investigate the metallicity dependence of the $\PL$ relation. We focus on F-mode relations only, deferring the 1O study to future work. The present study is primarily theoretical. The comparisons with observations included here are limited and qualitative; more detailed analyses will be presented in follow-up studies.

Structure of the paper is the following. In Sect.~\ref{sec:methods} we present our tools and methods. Results are given in Sect.~\ref{sec:results}, topics of which include morphology of the blue loops and its dependence on overshooting (Sect.~\ref{subsec:ov}, \ref{subsec:loops}), ages and crossing times (Sect.~\ref{subsec:pcr}), \PL{} relation and its metallicity dependence (Sect.~\ref{subsec:pl}), \ML{} relation (Sect.~\ref{subsec:ml}), \PR{} relation (Sect.~\ref{subsec:pr}) and \PA{} relation (Sect.~\ref{subsec:pa}), including their metallicity dependence. The paper is concluded with summary and conclusions in Sect.~\ref{sec:discussion}.

\section{Methods\label{sec:methods}}

\subsection{Modeling tools -- stellar evolution\label{subsec:tools}}

In all our evolutionary calculations we use \MESA, version r-21.12.1. We build on our earlier work, in particular micro-physical and numerical setup of the models is the same as in the reference model elaborated in \citetalias{Ziolkowska-2024}. Here we provide a brief summary. 

We use OPAL opacity tables \citep{Iglesias-1993, Iglesias-1996} supplemented with \cite{Ferguson-2005} tables at lower temperatures. Type 2 tables are used during and after core helium burning. Cubic interpolation in tables of different composition, $X$/$Z$, is used. The relative distributions of metals follows scaled solar composition as given by \cite{Asplund-2009} (A09 in the following). A linear helium enrichment law is adopted with $\Delta Y/\Delta Z=1.5$ and protosolar helium abundance, $Y_{\rm p}=0.2485$ \citep{Komatsu-2011}.

 We use one of the built in \mesa{} nuclear reaction networks, \texttt{pp\_and\_cno\_extras.net}, which follows evolution of 25 isotopes. Nuclear reaction rates come from NACRE \citep{nacre} except for \cag{}, that comes from \cite{Kunz-2002}, and the slowest reaction in the CNO cycle, \npg{}, that comes from JINA REACLIB \citep{jina}.

For the outer layers, \MESA{} built-in PHOENIX atmosphere tables are used \citep{Hauschildt-1999a, Hauschildt-1999b} along with \cite{Castelli-2003} models. We use standard \mesa{} equation of state, which is a blend of the OPAL \citep{Rogers2002}, SCVH \citep{Saumon1995}, FreeEOS \citep{Irwin2004}, HELM \citep{Timmes2000}, PC \citep{Potekhin2010}, and Skye \citep{Jermyn2021} EOSes -- see \cite{Paxton-2019}.
 
Convective layers are modeled with MLT following \cite{Henyey1965}. Calibration of the mixing length parameter was conducted in \citetalias{Ziolkowska-2024} resulting in $\alpha_{\rm MLT}=1.77$. 

For convective overshooting we use exponential prescription, see \cite{Herwig-2000}, and \cite{Paxton-2011} for \mesa{} implementation. We consider MS convective core overshooting of different extent, characterized by $\fcor$ parameter and convective envelope overshooting (during the RGB evolution), characterized by $\fenv$ parameter. The explored ranges are $0.0-0.03$ for $\fcor$ and $0.0-0.06$ for $\fenv$. Overshooting may also be described with step function, in which its extent is characterized with $\beta$ parameter, a fraction of the local pressure scale height, $\beta H_p$. In Sect.~\ref{secapp:stepvsexp} in the Appendix, to facilitate comparison with other studies, we derive relation between $f$ and $\beta$. The relation slightly depends on metallicity; for rough estimates, a factor of $11.5$ may be used to translate exponential to step overshooting parameter.

\mesa{} offers several parameters to control numerical convergence of the models. We adopt the same set of controls as in the reference model described in \citetalias{Ziolkowska-2024}, which assures numerical convergence in the explored mass and metallicity range.

For mass loss on RGB we use Reimers formula \citep{Reimers-1975MSRSL}, with mass loss rate scaled with $\eta$ parameter, for which we explore $0.0-0.6$ range. In this study we neglect atomic diffusion and rotation.

\MESA{} is a rapidly developing software. In this work we use its older version, r21.12.1, for a few reasons. First, we want to stay consistent with our earlier work (\citetalias{Ziolkowska-2024}). Second, computing and analyzing a large grid of stellar evolutionary models is a time consuming task; the project time is longer than the interval between consecutive releases. The version we are using however, is a fully mature version of \mesa; relatively few bug fixes were reported since its release and none is critical for the scope of our work\footnote{\url{https://docs.mesastar.org/en/latest/changelog.html}}. In Sect.~\ref{secapp:oldvsnew} in the Appendix, we demonstrate that our results fully hold with the more recent release version of \MESA, 24.08.1.

\subsection{2\MS{} models\label{subsec:twoms}}

2\MS{} models go through helium flash in electron degenerate core, which is numerically difficult to handle. With the adopted numerical solver tolerances, some of the models cannot converge during helium flash and relaxation of convergence criteria is needed. For that purpose we use a control implemented in \mesa, which relaxes the otherwise stringent convergence criteria for the luminosity equation (\texttt{convergence\_ignore\_equL\_residuals=.true.}). We note that for models that do converge with and without this control enabled, the tracks are qualitatively the same; only small shifts (of the order of 0.001\,dex in $\log(L/\LS)$/$\log\Teff$) may be present.

\subsection{Modeling tools -- stellar pulsation}\label{subsec:pulsation}

To compute linear pulsation properties of low order radial modes, their periods and growth rates, we use Radial Stellar Pulsation (\RSP) tool available as part of \MESA{} \citep{Paxton-2019,sm08a}. Model calculations use exactly the same micro-physical data as evolutionary calculations, in particular the equation of state and opacities are the same. Since radial pulsations are driven and achieve largest amplitudes at surface layers, it is not necessary to consider full evolutionary models. Instead, \RSP{} constructs chemically homogeneous envelope models, with physical parameters (mass, effective temperature, $\Teff$, luminosity, $L$, hydrogen and metal content, $X$ and $Z$) that originate from evolutionary tracks. The envelope extends till the temperature of $2\times 10^6$\,K and is constructed of 200 Lagrangian mass cells. The top 60 cells have constant mass, down till the anchor zone, in which temperature is fixed to $11\times 10^3$\,K (which assures smooth variation of model properties along a sequence of models). Below the anchor, cell mass increases geometrically inward. 

To include time-dependent convection-pulsation coupling, \RSP{} adopts one-equation \cite{Kuhfuss-1986} model which contains several free parameters. Their detailed description can be found in \cite{Paxton-2019}. Here we use four sets of convective parameters, A, B, C and D, exactly the same as in \cite{Paxton-2019}, see their tab.~4. Set A is the most basic set of convective parameters. In sets B and D the effects of radiative losses are included, while sets C and D include the effects of turbulent pressure and kinetic turbulent energy. While these four sets of convective parameters are by no means universal, nor calibrated to observation, as demonstrated by \cite{Paxton-2019}, they lead to reasonable IS in the HRD and reproduce the shape of light and radial velocity curves of classical Cepheids reasonably well \citep[see also][]{Kurbah-2023}. 

The properties that we record along the tracks are linear radial F- and 1O-mode periods ($P_0$ and $P_1$) and the corresponding growth rates ($\gamma_0$ and $\gamma_1$). The latter are used to determine the edges of the IS that depend not only on metallicity and adopted convective parameter set, but also on physical properties of the underlying evolutionary models, see Sect.~\ref{subsec:is}. 

Periods are little sensitive to convective parameter sets and for all evolutionary-pulsation relations discussed in the following Sections, set A is adopted to derive $P_0$. We use linear pulsation periods; while observed periods correspond to nonlinear period, the expected nonlinear period correction is small, below 1\% for the considered mass range \citep[see eg.,][]{BMS-1999}. In our coming paper, Ziolkowska et al., in prep., we explicitly give nonlinear period corrections computed with \RSP{} for a few Cepheids in eclipsing binary systems from \cite{Pilecki-2018}, all below 0.3\%.

We stress that all the relations we derive in this paper are based on linear pulsation calculations. We do not study mode selection, a process that determines the final full-amplitude pulsation state in cases where two (or more) modes are linearly unstable, see eg., \cite{Smolec-2014} for a review. This effect is not important for the construction of \PL, \PR, or \PA{} relations, which depend primarily on the stellar structure. Nonlinear calculations are essential to study eg., light curve morphology and are important for population synthesis studies, which should take into account relative distribution of F and 1O pulsators within the IS.

\subsection{Evolutionary model grids\label{subsec:grid}}

Our attention is focused on $2-8\MS$ models. We do not compute higher masses due to difficulties associated with development of thin convective shells described in detail in Sect.~3.3 of \citetalias{Ziolkowska-2024}. Metallicities range from $Z=0.0014$ to $Z=0.02$ which corresponds to $\feh=-1.0$ to $\feh=+0.2$ on the A09 scale. Altogether we use eleven $Z$ values; the corresponding helium and hydrogen mass content, and \feh{} are given in Tab.~\ref{tab:fehz}. 

\begin{deluxetable}{rrrr}
\tablecaption{Metal, $Z$, helium, $Y$, and hydrogen, $X$, mass fractions and corresponding metallicity, \feh{} (A09 scale), for models presented in the paper.\label{tab:fehz}}
\tablehead{
\colhead{$Z$}    &  \colhead{$Y$}   & \colhead{$X$}  & \colhead{\feh{}}
}
\startdata
0.0200 & 0.2785 & 0.7015 &   $0.197$  \\
0.0160 & 0.2725 & 0.7115 &   $0.094$  \\
0.0140 & 0.2695 & 0.7165 &   $0.033$  \\
0.0120 & 0.2665 & 0.7215 &  $-0.037$  \\
0.0100 & 0.2635 & 0.7265 &  $-0.119$  \\
0.0080 & 0.2605 & 0.7315 &  $-0.219$  \\
0.0060 & 0.2575 & 0.7365 &  $-0.347$  \\
0.0040 & 0.2545 & 0.7415 &  $-0.526$  \\
0.0030 & 0.2530 & 0.7440 &  $-0.652$  \\
0.0020 & 0.2515 & 0.7465 &  $-0.830$  \\
0.0014 & 0.2506 & 0.7480 &  $-0.985$  \\
\enddata
\end{deluxetable}

Depending on physical parameter settings of the models (eg., overshoot/mass-loss parameters), two types of grids are computed, as summarized in Tab.~\ref{tab:grids}. In GC grid we use a larger step in mass, $0.5\MS$. In GF grid, starting from $3\MS$, we use an $0.1\MS$ step. In both grids, for each mass we consider 11 $Z$ values which yields a total of 143 models for GC grid and a total of 583 models for GF grid.

To study loop properties and determine various evolutionary and pulsation relations in this work, we use the GC grids. Their computation is relatively fast and allows us to explore a wide range of parameters, eg., to probe the overshooting parameter space in detail in Sect.~\ref{subsec:loops}. The main purpose of computing the GF grids, which have a finer spacing in mass, is to facilitate future studies, such as modeling of individual Cepheids or more detailed comparisons with observations, as discussed further in the concluding section of the paper.

\begin{deluxetable}{lll}
\tablecaption{Range of physical parameters, masses, $M$, and metallicities, $Z$, for two types of model grids, GC and GF, considered in this paper.
\label{tab:grids}}
\tablehead{\colhead{id} & \colhead{masses} & \colhead{metallicities}
}
\startdata
GC & 2.0, 2.5, 3.0, \ldots, 8.0     & Tab.~\ref{tab:fehz} (11 values) \\
   & (step 0.5\MS; 13 values)       &  \\
GF & 2.0, 2.5, 3.0, 3.1\ldots, 8.0  & Tab.~\ref{tab:fehz} (11 values) \\
   & (step 0.1\MS; 53 values)       &  \\
\enddata
\end{deluxetable}

\section{Results} \label{sec:results}

\subsection{Overview of evolutionary tracks and online resources\label{subsec:online}}

In Tab.~\ref{tab:models} we summarize all model sets computed and analyzed in this study. Model set designation is given in the first column. Model sets differ primarily in MS core overshooting ($\fcor$, second column), envelope overshooting ($\fenv$, third column) and mass loss ($\eta$, fourth column). All model sets were computed with GC grid. For model sets in which finer grid (GF) was used a note is included in the last column.

\begin{deluxetable}{lcccc}
\tablewidth{0pt}
\tablecaption{Description of model sets considered in this study.\label{tab:models}}
\tablehead{
  \colhead{id} & \colhead{$\fcor$} & \colhead{$\fenv$} & \colhead{$\eta$} & \colhead{remarks}
}
\startdata
\multicolumn{5}{l}{{\it Core and envelope overshooting, no mass loss}}\\
 O00    & 0.00  &  0.00  & 0.0 & GF \\ 
 O02    & 0.00  &  0.02  & 0.0 &    \\ 
 O04    & 0.00  &  0.04  & 0.0 &    \\ 
 O06    & 0.00  &  0.06  & 0.0 &    \\ 
 O10    & 0.01  &  0.00  & 0.0 &    \\ 
 O12    & 0.01  &  0.02  & 0.0 &    \\ 
 O14    & 0.01  &  0.04  & 0.0 &    \\ 
 O16    & 0.01  &  0.06  & 0.0 &    \\ 
 O20    & 0.02  &  0.00  & 0.0 & GF \\ 
 O22    & 0.02  &  0.02  & 0.0 &    \\ 
 O24    & 0.02  &  0.04  & 0.0 & GF \\ 
 O26    & 0.02  &  0.06  & 0.0 &    \\ 
 O30    & 0.03  &  0.00  & 0.0 &    \\ 
 O32    & 0.03  &  0.02  & 0.0 &    \\ 
 O34    & 0.03  &  0.04  & 0.0 &    \\ 
 O36    & 0.03  &  0.06  & 0.0 &    \\ 
\hline
\multicolumn{5}{l}{{\it tracks with Reimers mass loss}}\\
O24\_ML2 & 0.02  &  0.04  & 0.2 & GF \\ 
O24\_ML4 & 0.02  &  0.04  & 0.4 & GF \\ 
O24\_ML6 & 0.02  &  0.04  & 0.6 &    \\ 
\hline
\multicolumn{5}{l}{{\it nuclear reactions -- \npg{} from NACRE}}\\
 O00\_AB & 0.00  &  0.00  & 0.0 &   \\ 
 O24\_AB & 0.02  &  0.04  & 0.0 &   \\ 
\hline
\multicolumn{5}{l}{{\it reference solar composition -- GS98}}\\
 O00\_AC & 0.00  &  0.00  & 0.0 &   \\ 
 O24\_AC & 0.02  &  0.04  & 0.0 &   \\ 
\hline
\multicolumn{5}{l}{{\it increased helium abundance, $\Delta Y/\Delta Z=2.0$}}\\
 O00\_AE & 0.00  &  0.00  & 0.0 &   \\ 
 O24\_AE & 0.02  &  0.04  & 0.0 &   \\ 
 \enddata
 \end{deluxetable}

All computed evolutionary tracks and exemplary \texttt{inlist} are available via Zenodo (\url{https://doi.org/10.5281/zenodo.17987357}). Individual track name includes model set id, mass, metal and helium content values, eg., \texttt{history.dat\_O12\_3.0\_0.0014\_0.2506}. Content (columns) of the evolutionary track files is described in Tab.~\ref{tab:history}.

Computed tracks include those with different nuclear reaction rate for \npg{} \citep[NACRE;][]{nacre} (*\_AB sets) and different reference solar composition \citep[GS98;][]{GS-98} (*\_AC sets). These models either do not include overshooting (O00 sets) or include both core and envelope overshooting (O24 sets). In both cases, noticeable differences are recorded only for models without overshooting and with solar metallicity. These differences are visualized and briefly discussed in Sect.~\ref{secapp:net} (\npg{} reaction rate) and in Sect.~\ref{secapp:gs98} (reference solar mixture) in the Appendix.  These models are not discussed further, however the computed tracks are available in the online repository.

Computed tracks also include those with slightly increased helium abundance (*\_AE sets). For these models we used $\Delta Y/\Delta Z=2.0$ (compared to 1.5 used in all other tracks), and recomputed $X$, $Y$ and $Z$ to match the same \feh{} values as given in the last column of Tab.~\ref{tab:fehz}. This is a slight increase in helium abundance resulting in tracks that are barely different from those in which we assumed $\Delta Y/\Delta Z=1.5$. The morphology of all tracks, including the presence and extent of the blue loops, is the same in both model sets. The largest differences are recorded at solar metallicity for which the tracks with increased helium abundance are slightly ($\simeq0.02$\,dex) brighter. We do not discuss these models further, however the computed tracks are available in the online repository.

\begin{deluxetable}{rll}
\tablewidth{0pt}
\tablecaption{Description of the columns of the evolutionary track files stored in online repository.\label{tab:history}}
\tablehead{
\colhead{No} & \colhead{Label} & \colhead{Explanation}
}
\startdata
 1 & model\_number & model number \\
 2 & star\_age     & model age in yrs\\
 3 & star\_mass    & model mass (solar units) \\
 4 & log\_Teff     & log effective temperature  \\
 5 & log\_L        & log absolute luminosity (solar units) \\
 6 & log\_R        & log radius (solar units) \\
 7 & log\_g        & log surface gravity (cgs units) \\
 8 & log\_cntr\_P  & log central pressure \\
 9 & log\_cntr\_Rho& log central density \\
10 & log\_cntr\_T  & log central temperature \\
11 & center\_mu    & central mean molecular weight\\ 
12 & center\_h1    & central $^{1}$H  mass fraction \\ 
13 & center\_he4   & central $^{4}$He mass fraction \\ 
14 & center\_c12   & central $^{12}$C mass fraction \\
15 & center\_n14   & central $^{14}$N mass fraction \\
16 & center\_o16   & central $^{16}$O mass fraction \\
17 & surface\_h1   & surface $^{1}$H  mass fraction  \\
18 & surface\_he4  & surface $^{4}$He mass fraction \\
19 & surface\_c12  & surface $^{12}$C mass fraction \\
20 & surface\_n14  & surface $^{14}$N mass fraction \\
21 & surface\_o16  & surface $^{16}$O mass fraction \\
22 & abs\_mag\_V   & absolute magnitude in V band\\
23 & abs\_mag\_I   & absolute magnitude in I band\\
24 & abs\_mag\_J   & absolute magnitude in J band\\
25 & abs\_mag\_H   & absolute magnitude in H band\\
26 & abs\_mag\_K   & absolute magnitude in K band\\
\enddata
\end{deluxetable}

\subsection{Edges of the classical IS\label{subsec:is}}

To compute the edges of the IS, pulsation properties were computed along evolutionary tracks with a $50$\,K step in \Teff{}. Then, we interpolated for the zero growth rate of the F mode.

\begin{figure*}[ht!]
\includegraphics[width=\linewidth]{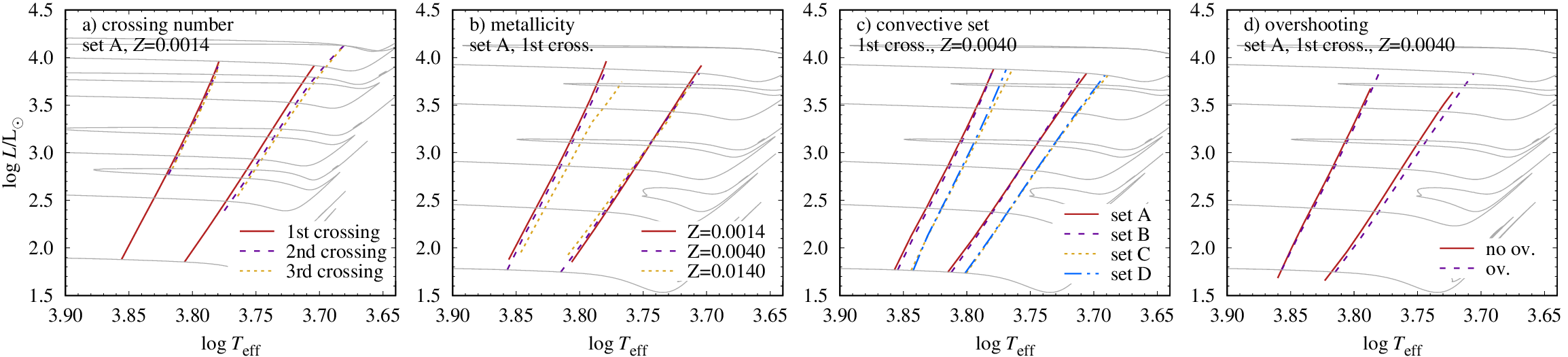}
\caption{Location of the F-mode IS determined for: panel a) different crossing numbers, panel b) different metallicities, panel c) different parameter sets describing convection-pulsation coupling and, panel d), models with and without convective core overshoot. In all panels evolutionary tracks of O24 set and 2, 3, 4, 6 and 8\MS{} are plotted for a reference.
\label{fig:is}}
\end{figure*}

In Fig.~\ref{fig:is} we explore how the location of the IS depends on model properties: crossing number (panel a), metallicity (panel b), parameters describing pulsation-convection coupling (panel c) and MS convective core overshooting (panel d).

The differences in IS locations we see in panels a) b) and d) come from differences in evolutionary tracks and physical parameters of the models. In panel c) we observe the differences arising from different parametrization of the convection-pulsation coupling implemented in \RSP{} (Sect.~\ref{subsec:pulsation}).

In panel a) of Fig.~\ref{fig:is} we show the dependence on crossing number. Convective parameters are those of set A, $Z=0.0014$ and models include both MS core and envelope overshooting (O24). Location of the blue edge is very little sensitive to crossing number. A very slight shift towards lower \Teff{} is present with increasing crossing number. This is a consequence of increased luminosity at the 2nd and 3rd crossing as compared to 1st crossing. Similar for the red edge, although the shift towards lower \Teff{} for 2nd/3rd crossing is more noticeable. As compared to 1st crossing, the red edge is shifted on average by $0.005$\,dex in $\logTeff$ for the 2nd crossing and by $0.008$\,dex for the 3rd crossing. Results are qualitatively the same for other convective parameter sets and metallicities as well as for models without overshooting. Consequently, for the following discussion we focus on the 1st crossing models (as they allow to investigate IS for the largest luminosity extent; 1st crossing is present for each mass and metallicity).

In panel b) of Fig.~\ref{fig:is} we show the dependence on metallicity. Convective parameters are those of set A and models include both MS core and envelope overshooting (O24). At the blue edge we observe a shift towards lower \Teff{} with increasing metallicity (on average by $0.002$\,dex for $Z=0.004$ and by $0.009$\,dex for $Z=0.014$). At the red edge we rather observe a change in the slope. The overall effect is a slightly narrower IS with increasing metallicity.

In panel c) of Fig.~\ref{fig:is} we show the dependence on convective parameter set. Models include both MS core and envelope overshooting (O24) and $Z=0.004$. Models split into two groups: those that do not include turbulent pressure and turbulent flux (sets A and B) and those that include these effects (sets C and D). IS for the latter models is significantly shifted towards lower \Teff{}. Comparing sets A and C the shift at the blue edge is on average $0.013$\,dex and at the red edge $0.015$\,dex. The inclusion of radiative losses (compare results for sets A and B, and C and D) has negligible effect on the IS (one however has to keep in mind that the inclusion of radiative losses requires an adjustment of eddy viscosity to assure similar pulsation amplitudes and width of the IS).

In panel d) of Fig.~\ref{fig:is} we compare IS computed based on models without (O00) and with overshooting (O24). Otherwise the figure shows models of set A, 1st crossing and $Z=0.004$. We observe a slight shift of the IS towards lower \Teff{} for overshooting models. Just as in panel a) this is the effect of increased luminosity of these models. The shift is more pronounced at the red edge and on average amounts to $0.005$\,dex.

The mass loss on RGB, even at the highest considered rate ($\eta=0.6$) has negligible effect on the IS. As discussed in Sect.~\ref{subsec:massloss} (see also Fig.~\ref{fig:HRMassLoss}), mass loss has a very small effect on evolutionary tracks. All crossings occur at the same luminosity; just the mass is slightly lower, by up to 2\%. This leads to a negligible effect on the IS. In the scale such as that displayed in Fig.~\ref{fig:is}, IS determined based on O24 and O24\_ML4 ($\eta=0.4)$ models, nearly overlap. There is a systematic shift towards lower \Teff{} at the red edge, the largest for 3rd crossing (the largest loss of mass), but even for the highest mass loss rate the maximum difference in the location of IS is well below $0.001$\,dex in $\logTeff$. Consequently, the IS determined based on evolutionary sets without mass loss may be safely applied for models with mass loss included.

\begin{figure}
\includegraphics[width=\linewidth]{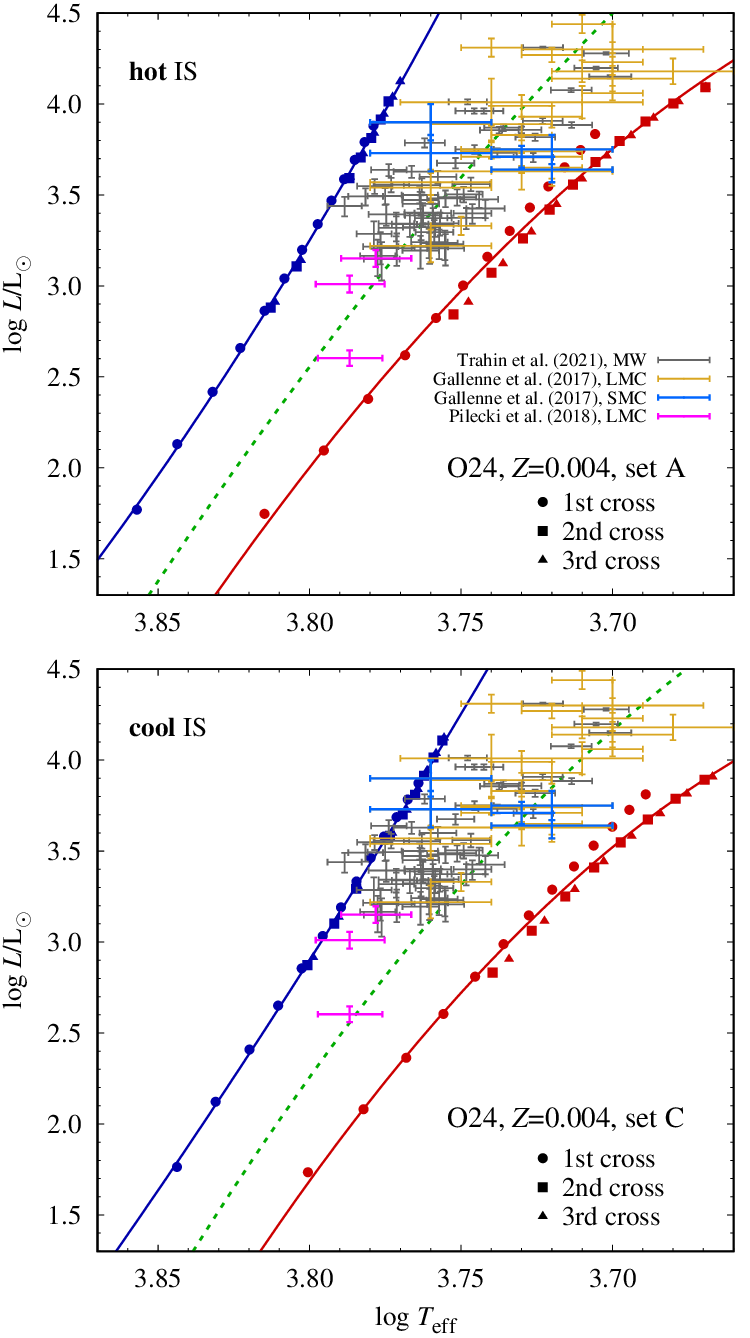}
\caption{Fiducial hot (upper panel) and cool (lower panel) IS edges on the HRD along with the midline (green curve). Data used to determine the edges (O24, $Z=0.004$) are plotted with filled symbols of different shapes corresponding to crossing number. Over-plotted are data for F-mode classical Cepheids from sources indicated in the top panel.
\label{fig:fis}}
\end{figure}

For further derivation of relevant evolutionary relations, it is convenient to define fiducial instability strips and their corresponding midlines, ie., curves located {\it between} the blue and red edges of a given IS. Two such IS are considered, the hot IS and the cool IS. This distinction is motivated by the strong dependence of the IS location on the adopted convective parameter set -- see Fig.~\ref{fig:is}, panel c).     To determine the corresponding edges we use data points with zero growth rate along evolutionary tracks of the O24 set with $Z=0.004$ computed with convective parameter sets A ({\it hot IS}) and C ({\it cool IS}) -- see Fig.~\ref{fig:fis}.  For each IS we also define a midline located midway between its blue and red edges as a function of effective temperature. Data from all crossings are used in the fit of the following 2nd order polynomial relation:
\begin{equation}
\log L/\LS=a(\logTeff-x_0)^2+b(\logTeff-x_0)+c\label{eq:poly2}
\end{equation}
Coefficients for the cool and hot IS, as well as for corresponding midlines are collected in Tab.~\ref{tab:ais}. Since the crossings occur at different luminosities, the derived edges closely follow the 1st crossing data at lower luminosities and then gradually bend towards 2nd/3rd crossing data. In Fig.~\ref{fig:fis} we plot these IS in the HRD, along with data points used in their fit and over-plot data for F-mode classical Cepheids from \cite{Gallenne-2017} (obtained with Baade-Wesselink, BW, method), \cite{Pilecki-2018} (Cepheids in eclipsing binaries) and \cite{Trahin-2021} (BW method). For cool IS, some of the observed Cepheids are placed beyond the blue edge of the IS and the stars are, in general, shifted towards the blue part of the IS. The hot IS encompasses data for all Cepheids and they occupy a central part of the IS. There are no more direct determinations of \Teff{} and $L$ for less luminous Cepheids (and determinations based on color-magnitude data are less certain due to reddening and models involved). We conclude that our fiducial IS are of reasonable width. Unless otherwise stated, relations in the following are based on the hot IS.

\begin{deluxetable*}{llrrrr}
\tablewidth{0pt}
\tablecaption{Analytic expressions, see eq.~\eqref{eq:poly2}, for edges and midlines of the IS specified in the first column. Coefficients are given to the last significant digit. \label{tab:ais}}
\tablehead{
\colhead{specification}  & \colhead{edge}   &  \colhead{$a$} & \colhead{$b$} & \colhead{$c$} & \colhead{$x_0$}
}
\startdata
fiducial, hot IS    & blue & $36.86$  & $-26.64$ & $2.87$ & $3.814$ \\
                    & red  & $-38.11$ & $-16.91$ & $3.11$ &  $3.742$ \\
                    & midline  & $-27.60$ & $-20.98$ & $3.03$ & $3.778$ \\
fiducial, cool IS   & blue &  $17.72$ & $-26.18$ & $2.90$ & $3.800$ \\
                    & red  & $-46.64$  & $-16.01$ & $3.15$ & $3.725$ \\
                    & midline  & $-41.99$ & $-20.07$ & $3.08$ & $3.762$ \\
\multicolumn{6}{l}{\it{Recommended range of applicability:}}\\
\multicolumn{6}{l}{$\log L/\LS\in(1.5,\,4.3)$}\\
\enddata  
\end{deluxetable*}

\subsection{The effects of MS core and envelope overshooting\label{subsec:ov}}

\begin{figure*}
\includegraphics[width=\linewidth]{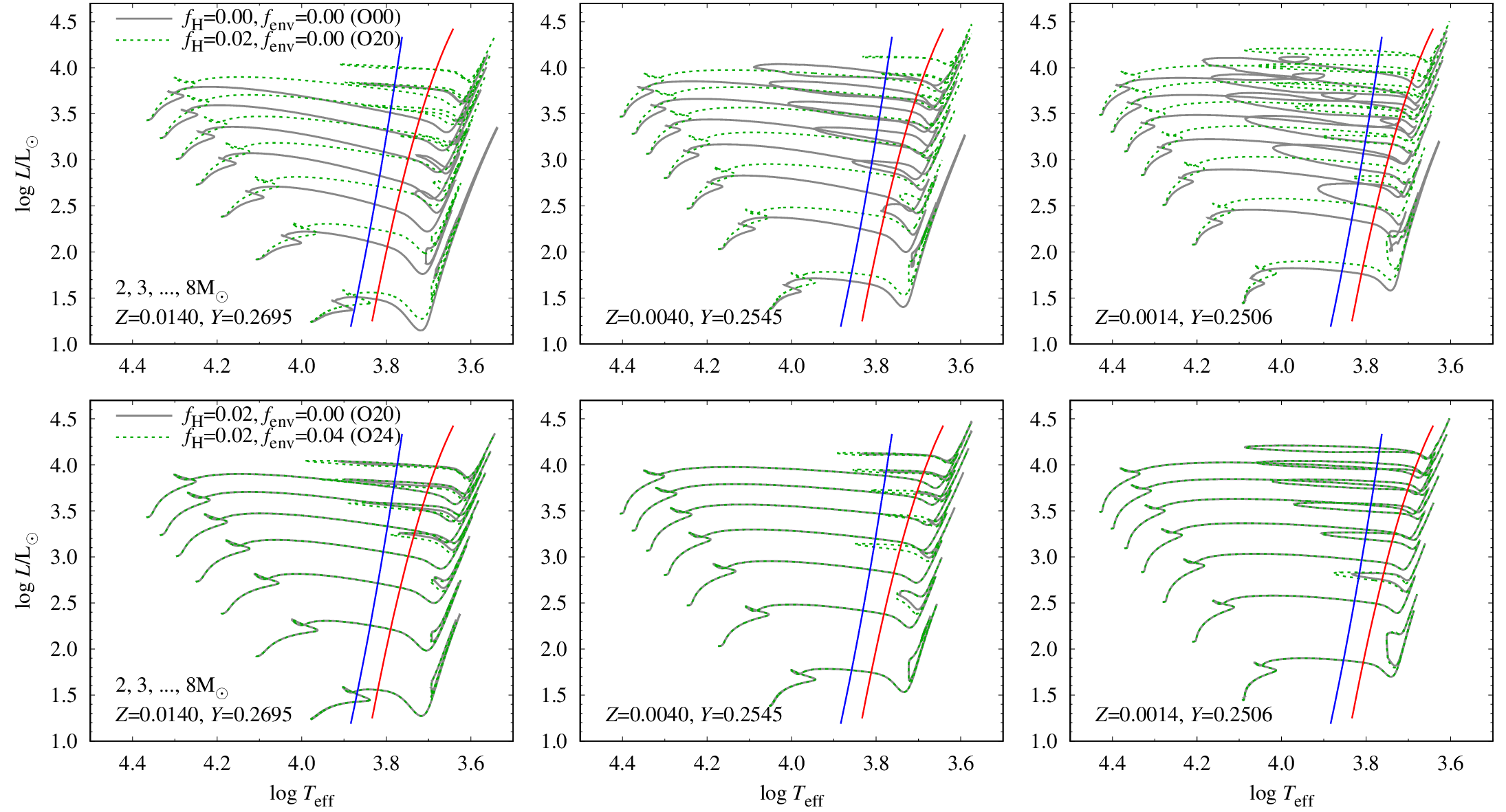}
\caption{The effects of core (top row) and envelope (bottom row) overshooting on evolutionary tracks of $2-8\MS$ models of high ($Z=0.014$, left column), intermediate ($Z=0.004$, middle column) and low ($Z=0.0014$, right column) metallicity. Fiducial hot IS strip is over-plotted for a reference.
\label{fig:HRov}}
\end{figure*}

The effects of MS core and envelope overshooting on evolutionary tracks are explored in Fig.~\ref{fig:HRov}. For a reference, we over-plot the edges of the fiducial hot IS. The consecutive columns show results for high ($Z=0.014$), intermediate ($Z=0.004$) and low ($Z=0.0014$) metallicity.

In the top row of Fig.~\ref{fig:HRov} we compare models without convective overshooting (solid gray lines, O00 models) with models in which convective core overshooting is included ($\fcor=0.02$; no envelope overshooting, O20 models). The most notable difference is an overall increase in luminosity through the whole evolution when MS core overshooting is included. A significant effect on the temperature extent of the blue loops is also observed and depends on metallicity. For high metallicity (top left panel), the inclusion of overshooting leads to shorter loops for $3$ and $4\MS$ models (these loops do not enter IS in both scenarios). For higher mass overshooting models, the loops do develop and are long enough to reach the IS, in contrast to models without overshooting, which, at this $Z$, and except $8\MS$, do not develop loops at all. This behavior is characteristic for solar metallicity only. For lower metallicities (top middle and top right panels) inclusion of core overshooting leads to significantly shorter loops that may not enter the IS ($Z=0.004$, $M<7\MS$), while corresponding models without overshooting easily reach the IS at this $Z$. For the lowest $Z$, when core overshooting is included, secondary loops ($M\geq6\MS$) do not develop during red-ward evolution. 

In the bottom row of Fig.~\ref{fig:HRov} we explore the effects of envelope overshooting in models in which convective core overshooting is included ($\fcor=0.02$). Models have either disabled envelope overshooting ($\fenv=0.0$; solid gray lines, O20) or enabled envelope overshooting ($\fenv=0.04$; dashed green lines, O24). The envelope overshooting operates during RGB evolution and hence it does not affect the MS phase and the luminosity level of the tracks. It modifies the loop extent, leading in general to longer loops. The effect is best visible for high and intermediate metallicities. In particular, for $Z=0.004$, inclusion of envelope overshooting is essential for the loops to enter the IS for $4-6\MS$ models. For the lowest metallicity, inclusion of envelope overshooting has essentially no effect on the tracks.

While inclusion of core overshooting increases the luminosity of the blue loops, inclusion of the envelope overshooting affects the extent of the loops for high and intermediate metallicity models. The underlying parameters, $\fcor$ and $\fenv$, may be used to adjust the luminosity and extent of the blue loops; both factors are essential eg., for the $\ML$ relation. This is just a qualitative description of the overall effects of MS core and envelope overshooting on the evolutionary tracks, including metallicity dependence. The detailed investigation on how $\fcor$, $\fenv$ and $Z$ affect the blue loops and their properties will be presented in the following Section and in Sect.~\ref{subsec:pcr}.

\subsection{Blue loop morphology\label{subsec:loops}}

\begin{figure*}[ht!]
\includegraphics[width=\linewidth]{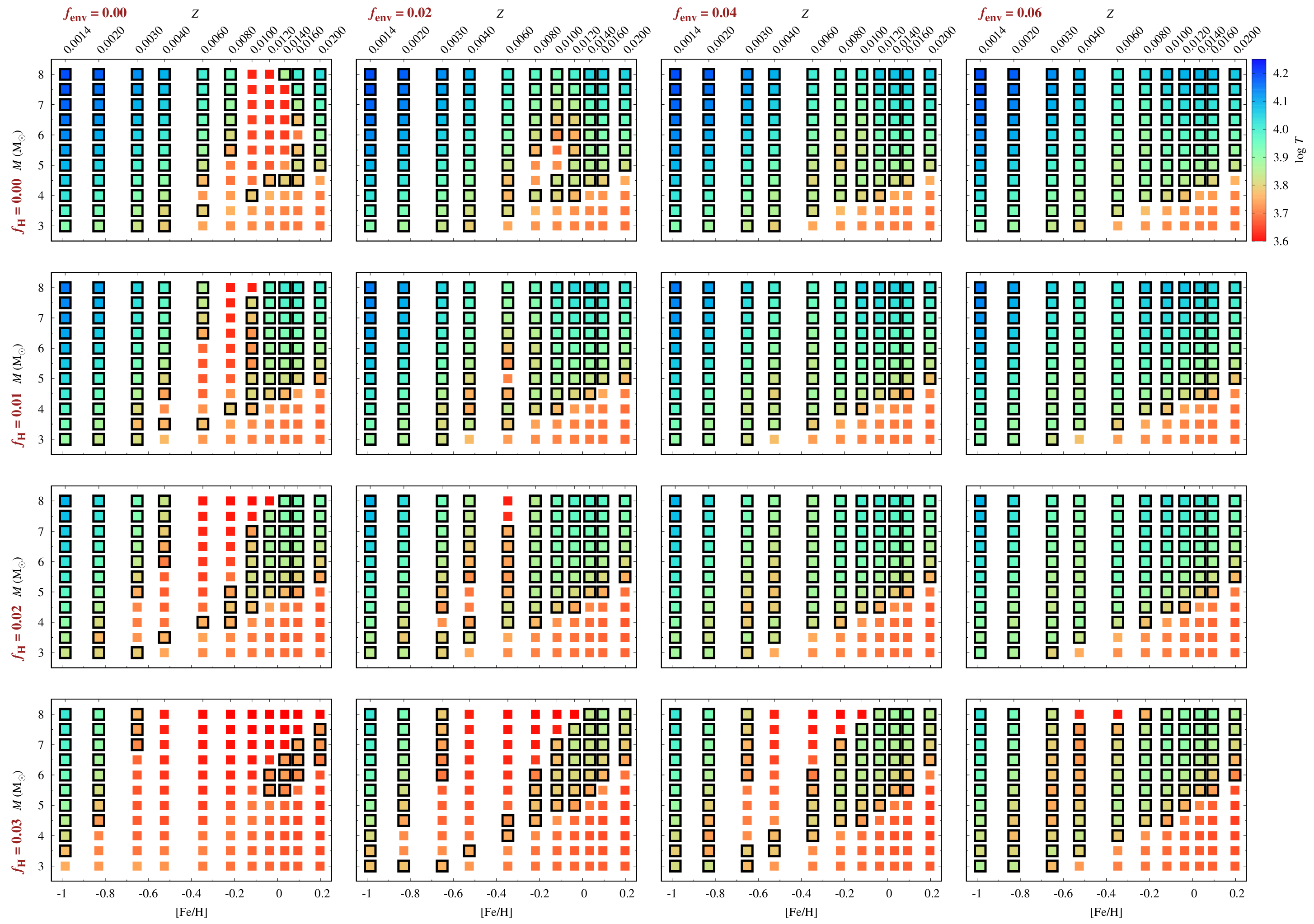}
\caption{Temperature extent of the blue loop across mass (vertical axis in each panel), metallicity (horizontal axis in each panel) and overshooting scenarios (rows and columns). The extent of MS core overshooting increases in rows, while the extent of envelope overshooting increases in columns. In a single panel, overshooting parameters are fixed and color of each square codes the effective temperature of the tip of the blue loop for models of different $M$/$Z$. In case the loop enters the hot IS, a black frame is drawn around square.
\label{fig:loopext}}
\end{figure*}

In Fig.~\ref{fig:loopext} we investigate the blue loop morphology across mass, metallicity and overshooting scenarios. For every single panel in this figure, the overshooting parameters are fixed. Focusing on a single panel, it records the maximum temperature extent of the blue loop for a model of given metallicity (along horizontal axis) and mass (along vertical axis). The effective temperature of the hottest point on the loop (its tip) is color coded according to scale displayed on the right side of the figure. Blue tones correspond to long loops (hot tip), while red tones to short loops (cool tip). In addition, if the loop enters the IS (fiducial, hot IS), a black frame is drawn around the corresponding point. In the figure, the extent of MS core overshooting increases in rows, from no core overshooting in the top row ($\fcor=0.0$, O0* models) to $\fcor=0.01$ (O1*) in the second row, $\fcor=0.02$ (O2*) in the third row and finally $\fcor=0.03$ (O3*) in the fourth row. Envelope overshooting increases in columns, from no envelope overshooting in the left-most column ($\fenv=0.0$, O*0 models) to $\fenv=0.02$ (O*2) in the second column, $\fenv=0.04$ (O*4) in the third column and finally $\fenv=0.06$ (O*6) in the fourth column.

Two general tendencies may be inferred from Fig.~\ref{fig:loopext}. First, as MS core overshooting increases (moving from top to bottom rows) the loops get shorter (cooler). At the same time, the luminosity of the loops increases, see top panel in Fig.~\ref{fig:HRov}. As envelope overshooting increases (moving from left to right columns) the loops get systematically longer (hotter). At the same time, the luminosity level of the loops remains constant (see bottom panel in Fig.~\ref{fig:HRov}).

Focusing on the no overshooting case (top left panel in Fig.~\ref{fig:loopext}), we observe that the extent of the loop and consequently whether it may enter the IS, may vary in a strongly non-monotonic way, both at fixed mass and at fixed metallicity. For example, for $4.5\MS$ model, it does not enter the IS for the highest metallicity. Then, for metallicities in a range 0.016--0.012 the loop gets slightly longer and enters the IS, so its tip is inside. For lower metallicities the loop gets shorter and does not enter IS, down to $Z\leq 0.006$, at which the loop gets systematically longer end crosses the IS. Focusing on $Z=0.016$, we observe that the loop is too cool to enter the IS for lower mass models ($M\leq 4.0\MS$) but also for $6\MS$ model. For $4.5-5.5\MS$ and $\geq 6.5\MS$, the loop is long enough to enter the IS. The behavior we see is non-monotonic but the smooth variation of loop extent on the $M$/$Z$ plane indicates it is not erratic. Similar non-monotonic behavior is well visible in all models without envelope overshooting (left column in Fig.~\ref{fig:loopext}) and with weak envelope overshooting (second column in the figure). An increase in envelope overshooting systematically extends the loops and leads to more monotonic behavior. For $\fcor=0.0$, $\fenv=0.06$ case (top right panel), we observe a nearly monotonic behavior: at fixed metallicity the loop gets longer with increasing mass and at fixed mass the loop gets longer with decreasing metallicity. Increase in MS core overshooting tends to introduce non-monotonic behavior.

Based on the above analysis, when deriving various evolutionary and pulsation relations in the following sections, we focus on two sets of models. The first without overshooting (O00 model set) and the second with $\fcor=0.02$ and $\fenv=0.04$ (O24 model set). For this set, the loop behavior in the $M-Z$ plane is largely monotonic, and, except in the low-mass, high-metallicity regime, the loops are sufficiently extended to enter the IS. The MS core overshooting with $\fcor=0.02$ corresponds to 0.23$H_p$ in the step overshooting formalism (see Sect.~\ref{secapp:stepvsexp}), which is a value close to adopted in the literature with other codes and models without rotation, eg., in \cite{Pietrinferni-2004} ($0.2H_p$) or \cite{Bressan-2012} ($\sim 0.25H_p$).

We note that the extent of the loop is not its only characteristics. Another interesting property is the loop's thickness -- the separation in luminosity between the 2nd and 3rd crossing, which may vary considerably with effective temperature. In general, the lower the metallicity, the larger the loop's thickness, as is visible eg., in Fig.~\ref{fig:HRov}. In the context of Cepheid mass discrepancy, \cite{Anderson-2014, Anderson-2016} noted, that inclusion of rotation not only increases the loop luminosity, but also affects the luminosity separation between the 2nd and 3rd crossings, which largely resolves the mass discrepancy. A more detailed discussion will be presented in our next paper, Smolec et al., in prep., in which we include rotation in \MESA{} models.

\subsection{The effects of mass loss\label{subsec:massloss}}

We use \cite{Reimers-1975MSRSL} formula for mass loss on the RGB followed by \cite{Bloecker-1995} formula on the AGB. For the context of Cepheid evolution, only the first matters. This is not the pulsation driven mass loss postulated in the literature \citep[eg.,][]{Neilson-2011}; the effects of pulsation on mass loss are neglected in this study. We explore low ($\eta=0.2$), intermediate ($\eta=0.4$) and high mass loss rates ($\eta=0.6)$.

\begin{figure*}[ht!]
\includegraphics[width=\linewidth]{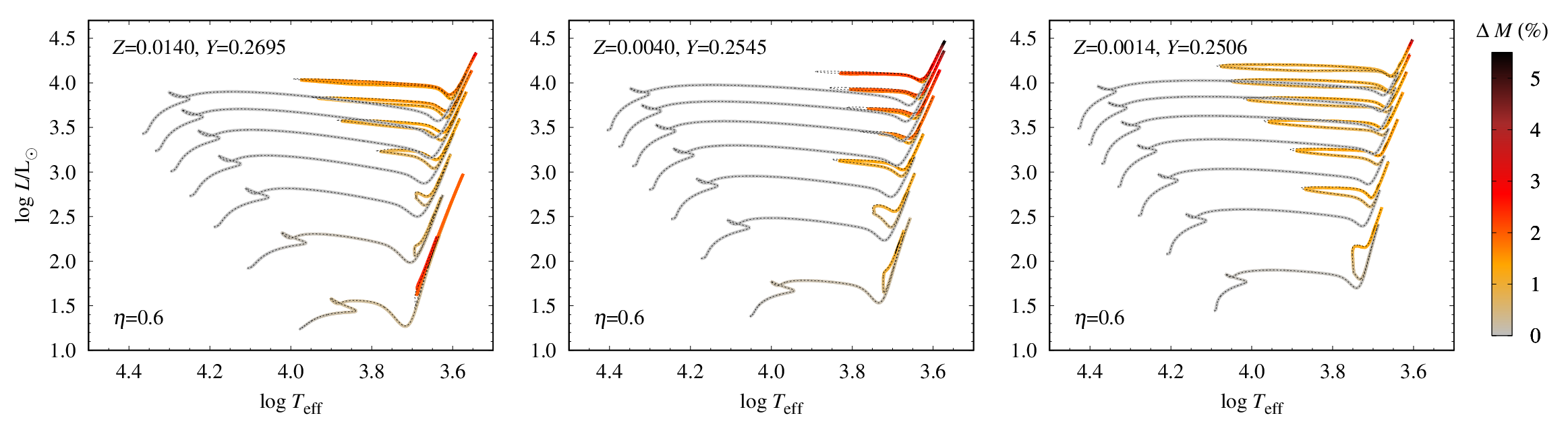}
\caption{The effects of mass-loss ($\eta=0.6$) on evolutionary tracks of high ($Z=0.014$, left panel) intermediate ($Z=0.004$, middle panel) and low metallicity ($Z=0.0014$, right panel). The models include overshooting both from the MS core and envelope (O24\_ML6 models). Tracks are color codded according to a fraction of mass lost, $\Delta M$. Tracks without mass loss (O24) are plotted with dashed black lines for a reference. 
\label{fig:HRMassLoss}}
\end{figure*}

The effect of mass loss on evolutionary tracks is illustrated in Fig.~\ref{fig:HRMassLoss}. The models plotted include convective overshooting (O24) and explore three metallicities $Z=0.014$ (left column) $Z=0.004$ (middle column) and $Z=0.0014$ (right column). We just show the results for the highest mass loss rate, $\eta=0.6$ (O24\_ML6). Tracks including mass loss are color codded according to fraction of mass lost. For a reference, tracks without mass loss are plotted with black dashed lines. The effects of mass loss become apparent only once RGB evolution commences. Even for the highest considered mass loss rate, only up to 2\% of mass is lost at the end of 3rd crossing (and up to 5\%, when model is stopped on AGB). The effects on evolutionary tracks are very small, but noticeable, in particular for the highest computed mass loss rate and $Z=0.004$ (middle column in Fig.~\ref{fig:HRMassLoss}). Qualitatively, the mass loss leads to a slight shortening of the blue loop and decrease of its luminosity. Both effects are very small. Considering most extreme differences, for 8\MS{}, $Z=0.004$ and $\eta=0.6$, the extent of the blue loop is shortened from $3.892$ to $3.830$ in \logTeff{}. The luminosity at the 2nd and at the 3rd crossing is lower by $0.01$ and $0.015$\,dex, respectively, for the track with mass loss. As the model leaves the IS, its mass is $7.82$\MS{}, so 2.3\% of mass was lost up to that point.

\subsection{Crossing times and period change rates\label{subsec:pcr}}

\begin{figure*}
\includegraphics[width=\linewidth]{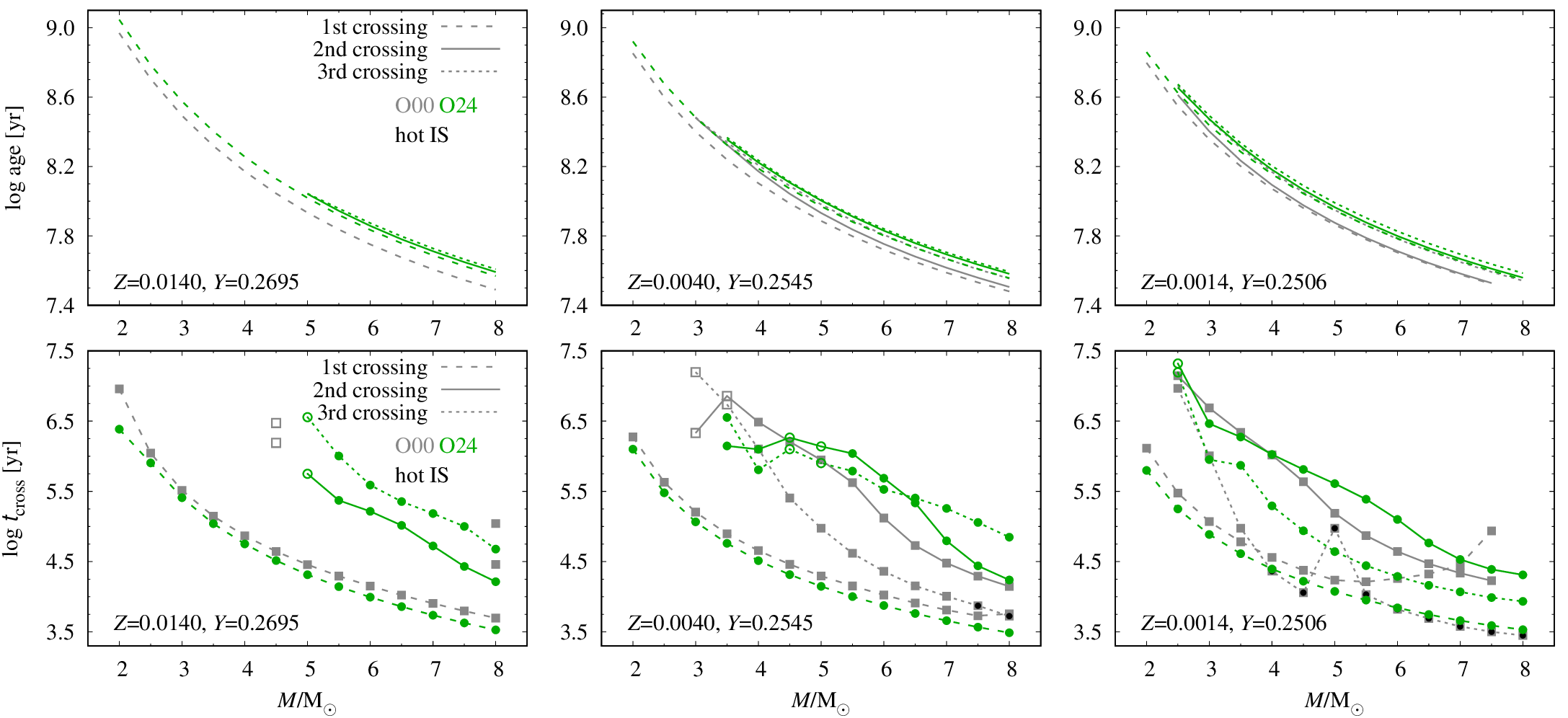}
\caption{The age at the entry of the IS (top) and crossing time through the IS (bottom) for models of different mass (horizontal axis) and metallicities ($Z=0.014$, $Z=0.004$ and $Z=0.0014$ in the left, middle and right columns, respectively). Data for O00 models (no overshooting) are plotted with gray, while data for O24 models (including core and envelope overshooting) are plotted in green. Different line styles correspond to different crossing numbers, as indicated in the legend. Determinations are based on the boundaries of hot IS.
\label{fig:agextime}}
\end{figure*}

In the top panels of Fig.~\ref{fig:agextime} we show age of the model at the entry of the IS (at the blue edge for 1st and 3rd crossing and at the red edge for 2nd crossing) as a function of model mass and in the bottom panels we show crossing time through the IS, $t_{\rm cross}$. The columns correspond to solar ($Z=0.014$), intermediate ($Z=0.004$) and low metallicity ($Z=0.0014$), from left to right, respectively. Data for all crossings and for O00, and O24 models are presented. When the tip of the loop is inside the IS, the values for the 2nd and 3rd crossings are calculated using the age of the model at the tip. For this case, open symbols are used in the bottom panels of Fig.~\ref{fig:agextime}.

The $\PA$ relation will be discussed later in Sect.~\ref{subsec:pa}. Here we just note general trends related to Cepheid's ages. For a given mass, due to significantly longer MS phase, overshooting models are older. The higher the metallicity, the older the model for a given mass. It reflects the MS evolution, during which metal-rich models are slightly cooler and fainter than their metal poor counterparts.

The crossing time, $t_{\rm cross}$ (bottom panel in Fig.~\ref{fig:agextime}), in general decreases with mass and is at least order of magnitude longer for the 2nd and 3rd crossings as compared to 1st crossing. There are several notable exceptions however. The primary factor influencing the crossing time is nuclear age during the IS crossing, characterized with central helium content. When helium gets depleted in the core, evolution speeds up. This is well visible for low metallicity 3rd crossing models without overshooting. The models in which central helium is depleted are indicated with a black dot. For the lowest metallicity and $M>4.5\MS$ helium is depleted in the core already before 3rd crossing starts. The crossing time is comparable or even shorter than for the 1st crossing. A spike at 5\MS{} is a consequence of a {\it breathing pulse}, injection of helium into the core resulting in a small secondary loop inside the IS -- see the corresponding evolutionary track in Fig.~\ref{fig:HRov}, top right panel. Crossing time is also relatively longer when tip of the loop is inside the IS (corresponding models are marked with open symbols in Fig.~\ref{fig:agextime}) or beyond the IS, but close to its blue edge. Crossing times for 2nd and 3rd crossing are typically longer for models including overshooting.

\begin{figure*}
\includegraphics[width=\linewidth]{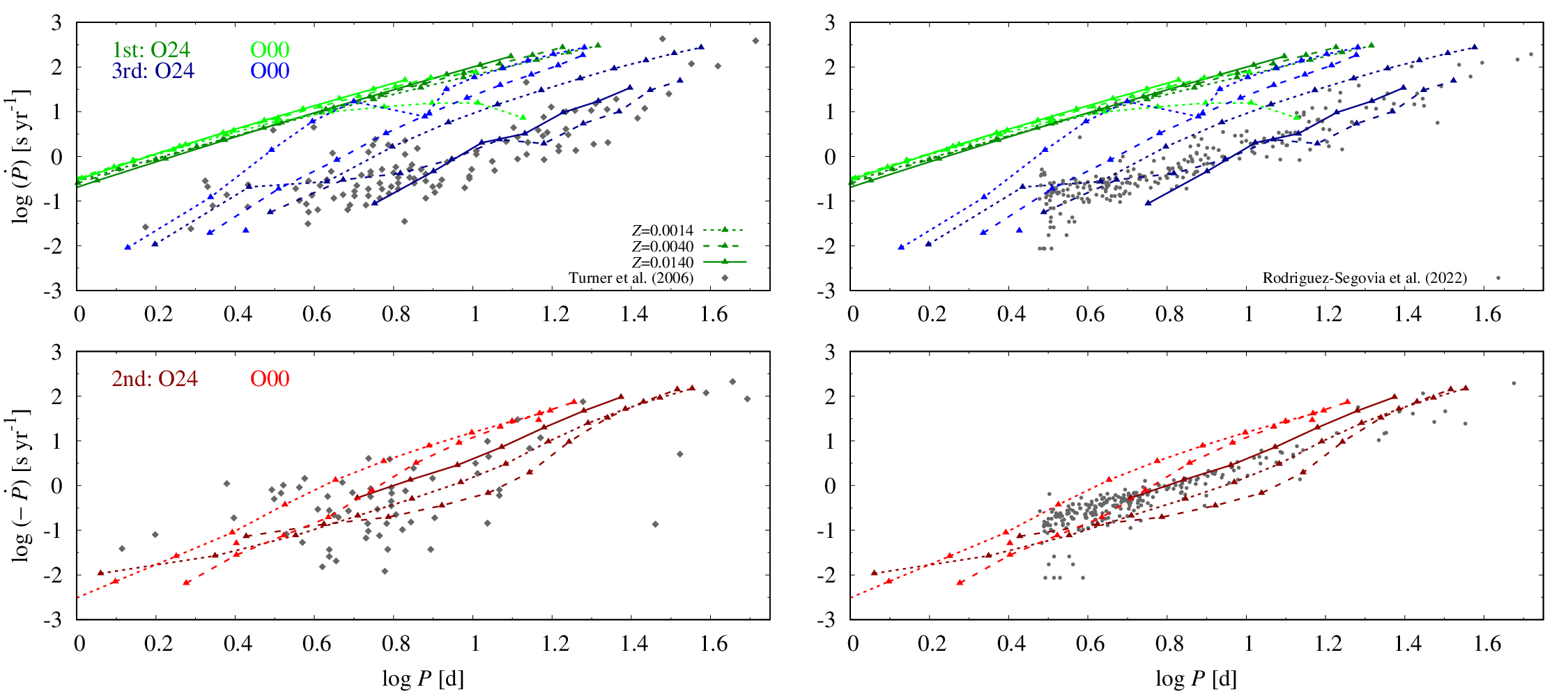}
\caption{Positive (top) and negative (bottom) period change rates predicted form evolutionary models (lines) confronted with determinations for Galactic Cepheids from \cite{Turner-2006} and LMC Cepheids from \cite{Rodriguez-Segovia-2022}.
\label{fig:pcrobs}}
\end{figure*}

Crossing times are related to period change rates (PCR) which may be directly compared to observations. In Fig.~\ref{fig:pcrobs} we plot $\dot{P}$ vs.\ $P$, separately for positive PCR (1st and 3rd crossing, top panels) and negative PCR (2nd crossing; $\log (-\dot{P})$ is plotted). Values were determined for the midline of the IS. Data for three metallicities and OO, and O24 models are plotted. Models are compared with data for Galactic Cepheids from a compilation of \cite{Turner-2006} (left panels) and PCR determinations for LMC Cepheids of \cite{Rodriguez-Segovia-2022} (right panels). We observe a qualitative agreement between the observed and model PCR. Both for MW and LMC Cepheids, comparison indicates that nearly all stars are during blue loop phase. Also, models with overshooting (O24) reproduce the observed PCR best, both for Galactic and LMC Cepheids. Models without overshooting predict too high PCRs as compared with observations. For Galactic Cepheids, those with lowest PCRs are not matched with the models. For the LMC, the overshooting models alone, fail to predict PCR at $\log P\approx 0.6$ for 3rd crossing. This indicates a possible problem in reproducing the ratio of the Cepheids with positive and negative period change rates, an important constraint in the context of Cepheid mass discrepancy problem, see eg., \cite{Neilson-2012}. However, we refrain from a more detailed comparison with observations that would require population synthesis, which is beyond the scope of this paper.

Age at the entry of the IS, crossing times and the values of $\dot{P}/P$ (for the corresponding midpoints of the IS crossing) are collected in Tab.~\ref{tab:agextime}, available in its full extent in the electronic version of the Journal and on Zenodo. Two versions are available, for the hot and  cool IS. Data for all model sets listed in Tab.~\ref{tab:models}, except the last six, are included.

\begin{deluxetable*}{lcclllllllll}
\tablecaption{Logarithm of age (yrs) at the entry of the IS, logarithm of crossing time (yrs), $\log t_{\rm cross}$, and period change rate, $\dot{P}/P$ (yr$^{-1}$; at the midpoint of the IS) for the three IS crossings. The first three columns identify model set, metallicity and model mass (initial mass for models with mass loss), respectively. For the 2nd and 3rd crossing, asterisk at crossing time indicates the tip of the loop falls inside the IS (and the crossing time is computed based on the model's age at the tip of the loop and respective edge of the IS). \label{tab:agextime}}
\tablehead{
    &     &     & \multicolumn{3}{c}{ 1st crossing} & \multicolumn{3}{c}{2nd crossing} & \multicolumn{3}{c}{3rd crossing}\\
\colhead{set} & \colhead{$Z$} & \colhead{$M/\MS$} & \colhead{log age} & \colhead{$\log t_{\rm cross}$} & \colhead{$\dot{P}/P$} & \colhead{log age} & \colhead{$\log t_{\rm cross}$} & \colhead{$\dot{P}/P$}  & \colhead{log age} & \colhead{$\log t_{\rm cross}$}  & \colhead{$\dot{P}/P$}
}
\startdata
O00 & 0.0200 & 2.00000 &    9.0015  & 7.3404  & 1.9355e-08 &          -  &      -  &           - &          -  &      -  &          - \\
O00 & 0.0200 & 2.50000 &    8.7381  & 6.2085  & 1.7132e-07 &          -  &      -  &           - &          -  &      -  &          - \\
O00 & 0.0200 & 3.00000 &    8.5221  & 5.6257  & 5.9411e-07 &          -  &      -  &           - &          -  &      -  &          - \\
O00 & 0.0200 & 3.50000 &    8.3430  & 5.2301  & 1.4459e-06 &          -  &      -  &           - &          -  &      -  &          - \\
O00 & 0.0200 & 4.00000 &    8.1915  & 4.9366  & 2.9096e-06 &          -  &      -  &           - &          -  &      -  &          - \\
O00 & 0.0200 & 4.50000 &    8.0606  & 4.7023  & 5.2465e-06 &          -  &      -  &           - &          -  &      -  &          - \\
O00 & 0.0200 & 5.00000 &    7.9469  & 4.5070  & 8.7743e-06 &     8.0136  & 6.0555* & -4.2279e-07 &     8.0184  & 6.5786* & 8.7244e-08 \\
O00 & 0.0200 & 5.50000 &    7.8470  & 4.3393  & 1.3830e-05 &     7.9135  & 5.8184  & -5.6853e-07 &     7.9314  & 6.0285  & 4.1984e-07 \\
O00 & 0.0200 & 6.00000 &    7.7586  & 4.1915  & 2.0803e-05 &     7.8233  & 5.7198  & -7.8430e-07 &     7.8451  & 5.6765  & 9.2996e-07 \\
O00 & 0.0200 & 6.50000 &    7.6797  & 4.0588  & 3.0033e-05 &     7.7404  & 5.5216  & -1.3222e-06 &     7.7660  & 5.3650  & 1.9501e-06 \\
O00 & 0.0200 & 7.00000 &    7.6094  & 3.9373  & 4.2147e-05 &     7.6652  & 5.1467  & -3.8587e-06 &     7.6940  & 4.9934  & 5.3948e-06 \\
O00 & 0.0200 & 7.50000 &    7.5459  & 3.8260  & 5.7408e-05 &     7.5983  & 4.7167  & -1.0780e-05 &     7.6281  & 4.7878  & 9.0592e-06 \\
O00 & 0.0200 & 8.00000 &    7.4890  & 3.7226  & 7.6491e-05 &     7.5376  & 4.4236  & -2.1727e-05 &     7.5681  & 4.5983  & 1.4396e-05 \\
\multicolumn{12}{l}{\ldots}\\
O24 & 0.0014 & 2.00000 &    8.8590  & 5.7984  & 5.2429e-07 &          -  &      -  &           - &          -  &      -  &          - \\
O24 & 0.0014 & 2.50000 &    8.6234  & 5.2495  & 1.8819e-06 &     8.6542  & 7.3207* & -8.8501e-08 &     8.6739  & 7.1946* & 5.1655e-08 \\
O24 & 0.0014 & 3.00000 &    8.4377  & 4.8865  & 4.6622e-06 &     8.4732  & 6.4644  & -1.3276e-07 &     8.4934  & 5.9524  & 1.6547e-07 \\
O24 & 0.0014 & 3.50000 &    8.2851  & 4.6127  & 9.4535e-06 &     8.3163  & 6.2742  & -2.2225e-07 &     8.3347  & 5.8707  & 1.0740e-06 \\
O24 & 0.0014 & 4.00000 &    8.1558  & 4.3946  & 1.7061e-05 &     8.1815  & 6.0261  & -4.3707e-07 &     8.2029  & 5.2932  & 2.7011e-06 \\
O24 & 0.0014 & 4.50000 &    8.0440  & 4.2218  & 2.7680e-05 &     8.0657  & 5.8119  & -8.1812e-07 &     8.0894  & 4.9391  & 6.7426e-06 \\
O24 & 0.0014 & 5.00000 &    7.9467  & 4.0765  & 4.1716e-05 &     7.9653  & 5.6103  & -1.3626e-06 &     7.9906  & 4.6423  & 1.3835e-05 \\
O24 & 0.0014 & 5.50000 &    7.8611  & 3.9520  & 5.9925e-05 &     7.8773  & 5.3878  & -2.4558e-06 &     7.9039  & 4.4423  & 2.3241e-05 \\
O24 & 0.0014 & 6.00000 &    7.7850  & 3.8421  & 8.2838e-05 &     7.7996  & 5.1001  & -6.1403e-06 &     7.8266  & 4.2885  & 3.5167e-05 \\
O24 & 0.0014 & 6.50000 &    7.7167  & 3.7464  & 1.0987e-04 &     7.7300  & 4.7652  & -1.4775e-05 &     7.7568  & 4.1627  & 4.9774e-05 \\
O24 & 0.0014 & 7.00000 &    7.6551  & 3.6606  & 1.4185e-04 &     7.6675  & 4.5309  & -2.7328e-05 &     7.6943  & 4.0699  & 6.5568e-05 \\
O24 & 0.0014 & 7.50000 &    7.5991  & 3.5894  & 1.7787e-04 &     7.6108  & 4.3898  & -4.2943e-05 &     7.6373  & 3.9881  & 8.3841e-05 \\
O24 & 0.0014 & 8.00000 &    7.5481  & 3.5311  & 2.1826e-04 &     7.5592  & 4.3117  & -6.0671e-05 &     7.5854  & 3.9337  & 1.0082e-04 \\
\multicolumn{12}{l}{\ldots}\\
\enddata
\tablecomments{This table is published in its entirety in the electronic 
edition of the {\it Astrophysical Journal Supplement Series} and on Zenodo. Results for hot and cool IS are available in separate tables online. A portion is shown here for guidance regarding its form and content.}
\end{deluxetable*}

\subsection{Data for evolutionary and pulsation relations\label{ssec:datarel}}

In the following sections we will analyze Period-Luminosity ($\PL$), Mass-Luminosity ($\ML$), Period-Radius ($\PR$) and Period-Age ($\PA$) relations. Relations will be defined either along the edges of the IS, or along its midline. Relations may be determined separately for each considered model set, crossing number and metallicity. Data for the relations are collected in Tab.~\ref{tab:data_all}, sample of which is available in the Appendix, and in full extent in the electronic version of the Journal and on Zenodo. Content of Tab.~\ref{tab:data_all} is described in Tab.~\ref{tab:datatab_content}. Data for all model sets in Tab.~\ref{tab:models}, except the last six, with 0.5\MS{} resolution in mass, are included in the table. Data were computed from evolutionary tracks by determining their intersection with edges and midline of the IS, both for its fiducial hot and cool versions (see Tab.~\ref{tab:ais}). For absolute magnitudes in $V$, $I$, $J$, $H$ and $K$ bands, bolometric corrections from \cite{Lejeune-1998} were used. These adopt the \cite{Buser-1978} ($V$), \cite{Bessel-1979} (Cousins $I$) and \cite{BesselBrett-1988} ($JHK$) filter transmission functions.  Linear pulsation periods were computed with \RSP{}.  These data may be used to directly plot any of the discussed relations for a given metallicity, crossing number, model set and reference line (IS edge/midline).

\begin{deluxetable}{rll}
\tablewidth{0pt}
\tablecaption{Content of Tab.~\ref{tab:data_all} with data for evolutionary and pulsation relations.\label{tab:datatab_content}}
\tablehead{
\colhead{Col.} & \colhead{Label} & \colhead{Explanation}}
\startdata
1 & edge & IS identifier (b/r/m -- blue/red/midline) \\
2 & set & Model set identifier, see Tab.~\ref{tab:models} \\
3 & cross. & crossing number \\
4 & $M/\MS$ & model mass \\
5 & $Z$ & metal mass fraction \\
6 & $X$ & hydrogen mass fraction \\
7 & log age & logarithm of age \\
8 & $\log\Teff$  & logarithm of effective temperature \\
9 & $\log L/\LS$ & logarithm of luminosity \\
10 & $\log R/\RS$ & logarithm of radius \\
11 & $Y_c$ & central helium mass fraction\\
12 & $P_0$ & F mode pulsation period \\
   &       & (`x' if \RSP{} model didn't converge) \\
13 & $P_1$ & 1O pulsation period \\
   &       & (`x' if \RSP{} model didn't converge) \\
14 & $V$ & absolute magnitude in $V$ band \\
15 & $I$ & absolute magnitude in $I$ band \\
16 & $J$ & absolute magnitude in $J$ band \\
17 & $H$ & absolute magnitude in $H$ band \\
18 & $K$ & absolute magnitude in $K$ band \\
\enddata
\end{deluxetable}

In the following sections, we also provide analytical fits based on the data listed in Tab.~\ref{tab:data_all}. In deriving these relations, we account for uncertainties in the evolutionary tracks, as determined in \citetalias{Ziolkowska-2024}\footnote{The numbers given slightly differ from those in tab.~6 of \citetalias{Ziolkowska-2024}. When computing uncertainties for the 1st and 3rd crossings (not included in \citetalias{Ziolkowska-2024}) we realized that some tracks were, by mistake, omitted from original calculations.}. For models including MS core overshooting at 1st/2nd/3rd crossing, uncertainties are 1.3/0.7/0.8\% for $\log L/\LS$, accordingly 0.65/0.35/0.4\% for $\log R/\RS$, and  0.2/0.1/0.1\% for $\log{\rm Age}$. For O00 models the corresponding numbers are: 0.5/0.5/1.5\% for $\log L/\LS$, accordingly 0.25/0.25/0.75\%,  for $\log R/\RS$, and 0.1/0.1/0.1\% for $\log{\rm Age}$. 

The relations are modeled using linear fits in log–log space, which provide an excellent representation of the data but remain approximations. For some relations, most notably $\PA$, the scatter about the fit is dominated by intrinsic dispersion. This intrinsic scatter, estimated from an initial fit, is then added in quadrature to the formal uncertainties to obtain reliable estimates of the uncertainties of the fitted parameters.

This paper is devoted to the presentation of theoretical relations. Comparisons with observations, if included, are qualitative, while detailed quantitative analyses are deferred to dedicated follow-up papers.

\subsection{Period-Luminosity relation\label{subsec:pl}}

We first focus on $\PL$ relation using Wesenheit $W_{VI}$ index as reddening-free luminosity indicator \citep{Madore-1982}, where $W_{VI}=I-1.55(V-I)$. 

\begin{figure*}[ht!]
\includegraphics[width=\linewidth]{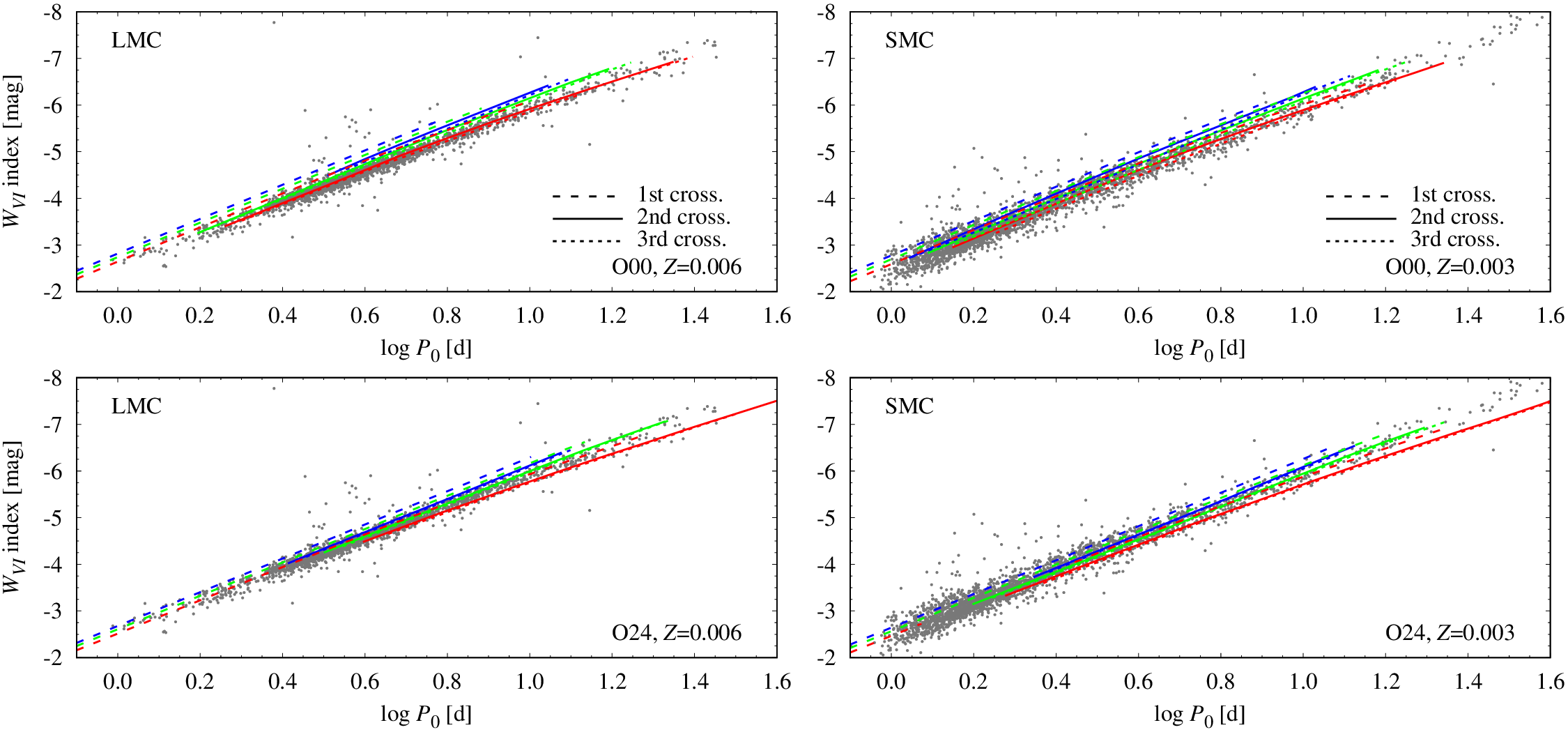}
\caption{Comparison of the observed and model \PL{} relations expressed in terms of Wesenheit index, $W_{VI}$, for the LMC (left) and SMC (right). OGLE data are compared to $Z=0.006$ models (LMC) and $Z=0.003$ (SMC) without overshooting (O00, top) and including overshooting (O24, bottom). Model relations are shown for the blue and red edges, and midline of the IS, with blue, red and green colors, respectively. Relations for different crossings are plotted with different line styles, as indicated in the legend (relations for the 2nd and 3rd crossings largely overlap).
\label{fig:pwi_obs}}
\end{figure*}

In Fig.~\ref{fig:pwi_obs} we compare the model $\PL$ relations, plotted separately for all crossings and the blue, and red edge, and midline of the IS, with OGLE data for the Magellanic Cloud F-mode Cepheids \citep{Soszynski-2015, Soszynski-2017, Soszynski-2019}. Adopted distance moduli are from \cite{Pietrzynski-2019} for the LMC and from \cite{Graczyk-2020} for the SMC. In the top panels, models do not include overshooting (O00), while in the bottom panels overshooting is included (O24). We use $Z=0.006$ and $Z=0.003$ models to compare with the LMC and SMC data, respectively. 

For the LMC the agreement is satisfactory. Majority of the observed stars lay in between model relations corresponding to the blue loop phase, both for models without and including overshooting. For the latter, the agreement is better in terms of luminosity: models without overshooting are on average a bit more luminous at a given period as compared to observations. The short period tail is well encompassed with 1st crossing relation. Relations for the loop start at $\log P\approx 0.4$; above this period a bulk of LMC Cepheids are located. The comparison is no longer as favorable for the SMC. Here, we notice a problem at short period end. The loop relations start at $\log P \approx 0.2$, however period distribution of the SMC Cepheids is shifted towards significantly shorter pulsation periods. The models would indicate that a large fraction of the SMC Cepheids should be 1st crossing stars, which is unlikely. 

In Fig.~\ref{fig:pwi_obs}, we also observe that \PL{} relations for different crossings of the IS are nearly parallel to each other. For the blue loop relations (2nd and 3rd crossing) the separation is very small, so these relations nearly overlap in Fig.~\ref{fig:pwi_obs}. For the midline, the separation it is typically 0.02--0.03\,mag. The separation is larger between the 1st crossing and blue loop relations and is typically around 0.2\,mag. While comparable to the observed scatter in the relation, we note that 1st crossing Cepheids are necessarily rare, hence contribution of Cepheids on different crossings to the observed scatter of the \PL{} relation must be very small.

\begin{figure}
\includegraphics[width=\linewidth]{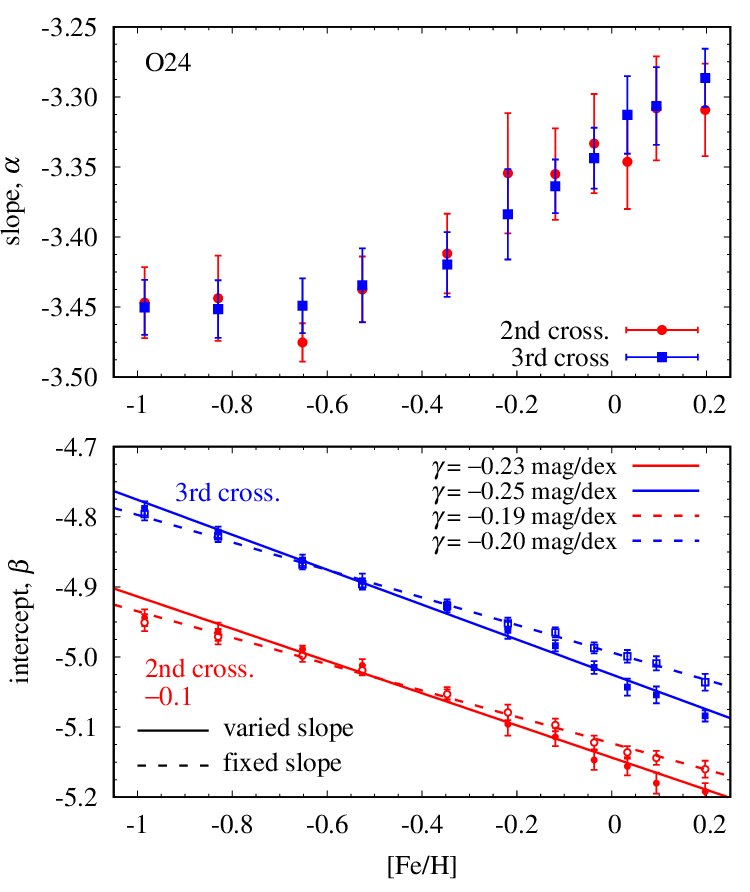}
\caption{Slope, $\alpha$, and intercept, $\beta$ (see eq.~\eqref{eq:pl}), as a function of metallicity for $\PL$ relation for O24 models on the 2nd and 3rd crossing of the IS. In the bottom panel, linear fits determine the metallicity effect, $\gamma$. Open symbols (and fits plotted with dashed line) correspond to $\PL$ relation of fixed slope. In the bottom panel, relations for the 2nd crossing were shifted by 0.1 for better visibility.
\label{fig:pwi_feh}}
\end{figure}

Thanks to the large number of metallicities included in our model grid, we can examine the metallicity dependence of the $\PL$ relation (see Sect.~\ref{sec:intro}). We can express the $\PL$ relation as ($m$ represents luminosity):
\begin{align}
    m &= \alpha (\log P - \log P_0) +\beta\,,\label{eq:pl}\\
      &= \alpha (\log P - \log P_0) +\delta + \gamma {\rm [Fe/H]}\,. \label{eq:plfeh}
\end{align}
We focus our attention on models including overshooting (O24). For each of the 11 metallicities, we independently fit eq.~\eqref{eq:pl} to the $W_{VI}$ data corresponding to the midline of the hot IS and the 2nd, and 3rd crossings. Adopting a cool version of IS leads to similar results. Fig.~\ref{fig:pwi_feh} shows shows how the slope, $\alpha$, and intercept, $\beta$, of the relation depend on metallicity. The slope remains constant, up to $\feh=-0.4$ and then increases linearly with \feh{}, while the zero point decreases with increasing \feh{} in a linear manner. By fitting the linear relationship to the intercept data (solid line and filled symbols in Fig.~\ref{fig:pwi_feh}), we obtain a metallicity effect of $\gamma=-0.23$\, mag\,dex$^{-1}$ and $\gamma=-0.25$\,mag\,dex$^{-1}$ for the 2nd and 3rd crossing, respectively. Following the observational work \citep[eg.,][]{Gieren-2018,Breuval-2021,Breuval-2022}, we can fix the slope of the relation to one that corresponds to the LMC metallicity ($Z=0.006$). The metallicity effect is slightly smaller then, $\gamma=-0.19$ and $\gamma=-0.20$\,mag\,dex$^{-1}$, for the 2nd and 3rd crossing, respectively (dashed lines and open symbols in Fig.~\ref{fig:pwi_feh}).

Considering models without overshooting (O00), and repeating the same procedure, we get significantly smaller effect, $\gamma=-0.11$ and $\gamma=-0.19$\,mag\,dex$^{-1}$, for the 2nd and 3rd crossing and with fixed slope of the relation. Since we observe that the metallicity effect is stronger for the 3rd crossing compared to the 2nd crossing, and in models including overshooting (O24 vs.\ O00), we conclude that the strength of the metallicity effect depends on the underlying $\ML$ relation of the models. The larger the luminosity for a given mass, the stronger the metallicity effect.

\begin{figure}
\includegraphics[width=\linewidth]{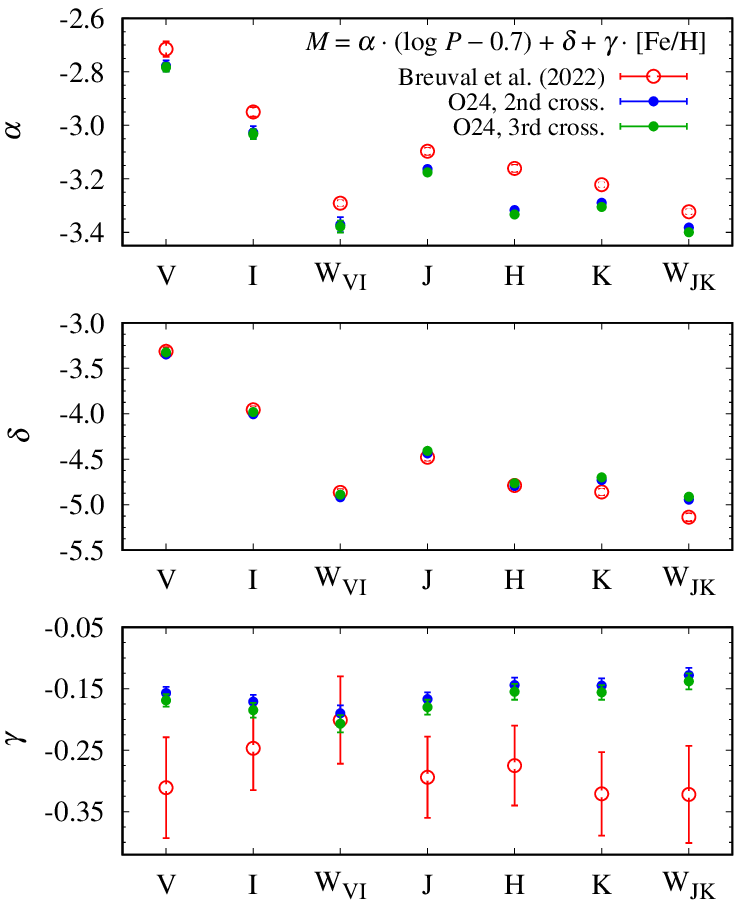}
\caption{Coefficients of the $\PL$ relation, $\alpha$, $\delta$ and $\gamma$ (top, middle and bottom, respectively, see eq.~\eqref{eq:plfeh}) as a function of pass band. Data for O24 models on the 2nd/3rd crossings are compared with results from \cite{Breuval-2022}.
\label{fig:breuval22}}
\end{figure}

We can also investigate the metallicity effect in different pass bands. In order to compare with the observational results from \cite{Breuval-2022}, we adopt a similar methodology. In addition to $V$, $I$, $J$, $H$ and $K$ bands\footnote{We note that \cite{Breuval-2022} uses $K_s$ band. Since our comparison is qualitative, we ignore the difference between $K$ and $K_s$ bands, which according to transformation equations given in \cite{Koen-2007} is very small.}, we include two Wesenheit indices, $W_{VI}$ and $W_{JK}$ \citep[defined as in tab.~3 in][]{Breuval-2022}, in the analysis. First, for each band, we fit eq.~\eqref{eq:pl} to $Z=0.006$, 2nd crossing data. The resulting slope is then fixed for the corresponding band \citep[][fixed the slope to that corresponding to LMC, hence we use $Z=0.006$]{Breuval-2022}, and relation \eqref{eq:plfeh} is fitted to the data to determine $\gamma$. For this comparison, which we present in Fig.~\ref{fig:breuval22}, we use O24 tracks and data for the 2nd and 3rd crossings. The three panels show the slope, $\alpha$, intercept, $\delta$, and metallicity term, $\gamma$, as a function of band (organized so that the effective wavelength increases) for observations (open circles) and 2nd/3rd crossing models (filled symbols). Concerning the slope and intercept, we observe a good qualitative agreement -- while there are systematic differences (eg., model slopes are systematically lower) the models follow exactly same trends with pass band. Concerning the metallicity effect, $\gamma$, both models and observations show essentially no dependence on pass band and negative sign of $\gamma$. Except for $W_{VI}$, for which models and observations match, the models predict systematically weaker metallicity effect. Averaging across pass bands and the two crossings, models yield $\gamma=-0.164$\,mag\,dex$^{-1}$ (standard deviation, $\sigma=0.022$), while average from observations is $-0.282$\,mag\,dex$^{-1}$ ($\sigma=0.045$). 

Repeating the above procedure for models without overshooting (O00), average across pass bands is $\gamma=-0.117$\,mag\,dex$^{-1}$, so significantly smaller than for models including overshooting and in agreement with our previous observation that metallicity effect depends on underlying $\ML$ relation.

Based on the above observations, in particular relatively weak dependence of slope of the \PL{} relation on metallicity, we derive the \PL{} relations for the two Wesenheit indices, $W_{VI}$ and $W_{VK}$, and absolute $K$-band luminosity, in the functional form of eq.~\eqref{eq:plfeh} for the three crossings of the IS. Coefficients are collected in Tab.~\ref{tab:plrelation}. The fits are based on data for all 11 metallicities in our grid and relations are given both for overshooting (O24) and no overshooting (O00) models. In addition to relations computed along the midline of the IS, for $W_{VI}$ and O24 set, we provide relations computed along the blue and red edge of the IS. In all relations hot IS is assumed. Recommended range in which the relations apply is given in the last column of Tab.~\ref{tab:plrelation}. The range corresponds to the 3rd and 97th percentiles of the distribution of periods used in the fit. It is approximate and depends on the metallicity. For the lowest metallicities, the applicability range can be shifted by about 0.1\,dex toward shorter periods.

It is evident from Tab.~\ref{tab:plrelation} that metallicity term is always negative. Considering the color, ie., comparing the relations at the blue, midline and red edge of the IS (for $W_{VI}$ and O24), the metallicity term is steeper for relations computed along red edge as compared to the midline and then blue edge.

Relations for the cool IS are collected in Tab.~\ref{tabapp:plrelation} in Sect.~\ref{secapp:coolIS} of the Appendix. Comparing the corresponding coefficients for the blue loop relations (2nd and 3rd crossing) we note that the slopes are systematically less steep and the zero points are dimmer for the cool IS relations. The metallicity term has similar values. It results in lower brightness at a given period for cool IS relations. As an example, for O24, 2nd crossing relations in $W_{VI}$, the cool relation is dimmer by $0.02$\,mag and $0.014$\,mag at the short ($\log P=0.4$) and long period ($\log P=1.4$) ends, respectively, assuming LMC metallicity. For the $K$-band relations the numbers are $0.05$\,mag and $0.012$\,mag at the short and long period ends.

\begin{table*}
\caption{Coefficients of the $\PL$ relations as given by eq.~\eqref{eq:plfeh}. First two columns identify data set, crossing number and edge along which relation is computed (`b/m/r' are for the blue, midline and red edge). In the last column, we provide the recommended parameter range in which the relation is applicable. All relations adopt hot IS. \label{tab:plrelation}}
\begin{tabular}{lcrrrrcl}
\tablecolumns{8}
data & crossing & $\alpha$  & $\delta$ & $\gamma$ & $\log P_0$ & rms & remarks\\
\hline
\multicolumn{8}{l}{$W_{VI}=I-1.55(V-I)$}\\
\hline
O00 & 1c/m & $-3.658\pm0.007$ & $-3.918\pm0.004$ & $-0.178\pm0.009$ & $0.308$ & $0.024$ & $\log P_0\in(-0.72,0.94)$ \\
O00 & 2c/m & $-3.566\pm0.016$ & $-5.450\pm0.008$ & $-0.097\pm0.014$ & $0.789$ & $0.031$ & $\log P_0\in(0.08,1.18)$ \\
O00 & 3c/m & $-3.595\pm0.030$ & $-5.420\pm0.017$ & $-0.180\pm0.027$ & $0.789$ & $0.031$ & $\log P_0\in(0.20,1.23)$ \\
O24 & 1c/b & $-3.618\pm0.010$ & $-4.739\pm0.009$ & $-0.142\pm0.015$ & $0.556$ & $0.014$ & $\log P_0\in(-0.73,1.04)$ \\
O24 & 1c/m & $-3.571\pm0.007$ & $-4.650\pm0.007$ & $-0.160\pm0.010$ & $0.556$ & $0.015$ & $\log P_0\in(-0.65,1.17)$ \\
O24 & 1c/r & $-3.431\pm0.015$ & $-4.490\pm0.011$ & $-0.205\pm0.023$ & $0.556$ & $0.063$ & $\log P_0\in(-0.59,1.31)$ \\
O24 & 2c/b & $-3.591\pm0.019$ & $-6.159\pm0.008$ & $-0.127\pm0.014$ & $0.997$ & $0.015$ & $\log P_0\in(0.35,1.21)$ \\
O24 & 2c/m & $-3.439\pm0.016$ & $-6.039\pm0.008$ & $-0.185\pm0.014$ & $0.997$ & $0.028$ & $\log P_0\in(0.29,1.39)$ \\
O24 & 2c/r & $-3.088\pm0.019$ & $-5.808\pm0.010$ & $-0.221\pm0.018$ & $0.997$ & $0.046$ & $\log P_0\in(0.39,1.65)$ \\
O24 & 3c/b & $-3.574\pm0.023$ & $-6.133\pm0.008$ & $-0.150\pm0.017$ & $0.997$ & $0.010$ & $\log P_0\in(0.32,1.23)$ \\
O24 & 3c/m & $-3.439\pm0.018$ & $-6.012\pm0.008$ & $-0.203\pm0.015$ & $0.997$ & $0.023$ & $\log P_0\in(0.36,1.38)$ \\
O24 & 3c/r & $-3.098\pm0.019$ & $-5.782\pm0.010$ & $-0.227\pm0.018$ & $0.997$ & $0.040$ & $\log P_0\in(0.48,1.65)$ \\
\hline
\multicolumn{8}{l}{$W_{VK}=K-0.13(V-K)$}\\
\hline
O00 & 1c/m & $-3.660\pm0.010$ & $-3.808\pm0.006$ & $-0.141\pm0.013$ & $0.308$ & $0.043$ & $\log P_0\in(-0.72,0.94)$ \\
O00 & 2c/m & $-3.501\pm0.018$ & $-5.325\pm0.009$ & $-0.052\pm0.015$ & $0.789$ & $0.034$ & $\log P_0\in(0.08,1.18)$ \\
O00 & 3c/m & $-3.529\pm0.030$ & $-5.296\pm0.017$ & $-0.131\pm0.026$ & $0.789$ & $0.031$ & $\log P_0\in(0.20,1.23)$ \\
O24 & 1c/m & $-3.568\pm0.010$ & $-4.534\pm0.008$ & $-0.120\pm0.015$ & $0.556$ & $0.033$ & $\log P_0\in(-0.65,1.17)$ \\
O24 & 2c/m & $-3.401\pm0.014$ & $-5.906\pm0.007$ & $-0.130\pm0.012$ & $0.997$ & $0.017$ & $\log P_0\in(0.29,1.39)$ \\
O24 & 3c/m & $-3.398\pm0.016$ & $-5.879\pm0.008$ & $-0.145\pm0.014$ & $0.997$ & $0.013$ & $\log P_0\in(0.36,1.38)$ \\
\hline
\multicolumn{8}{l}{$K$ band}\\
\hline
O00 & 1c/m & $-3.592\pm0.010$ & $-3.648\pm0.007$ & $-0.143\pm0.013$ & $0.308$ & $0.042$ & $\log P_0\in(-0.72,0.94)$ \\
O00 & 2c/m & $-3.434\pm0.017$ & $-5.137\pm0.009$ & $-0.055\pm0.015$ & $0.789$ & $0.034$ & $\log P_0\in(0.08,1.18)$ \\
O00 & 3c/m & $-3.462\pm0.029$ & $-5.108\pm0.016$ & $-0.133\pm0.026$ & $0.789$ & $0.031$ & $\log P_0\in(0.20,1.23)$ \\
O24 & 1c/m & $-3.501\pm0.010$ & $-4.360\pm0.008$ & $-0.122\pm0.015$ & $0.556$ & $0.033$ & $\log P_0\in(-0.65,1.17)$ \\
O24 & 2c/m & $-3.334\pm0.014$ & $-5.707\pm0.007$ & $-0.132\pm0.012$ & $0.997$ & $0.018$ & $\log P_0\in(0.29,1.39)$ \\
O24 & 3c/m & $-3.332\pm0.015$ & $-5.680\pm0.007$ & $-0.147\pm0.013$ & $0.997$ & $0.013$ & $\log P_0\in(0.36,1.38)$ \\
\hline
\end{tabular}
\end{table*}

\subsection{Metallicity dependence and analytical form for $\ML$, $\PR$ and $\PA$ relations\label{subsec:relfeh}}

Before deriving $\ML$, $\PR$ and $\PA$ relations in analytical form, we first investigate their dependence on \feh{} for models with and without overshooting (O00 and O24). We fit our data with the following relation 
\begin{equation}
\log y=a(\log x - \log x_0) +b\,, \label{eq:basiclinear}
\end{equation}
with corresponding variables and centering, $\log x_0$ (for which we use median value of independent variable across models considered in the fit). 

\begin{figure*}
\includegraphics[width=\linewidth]{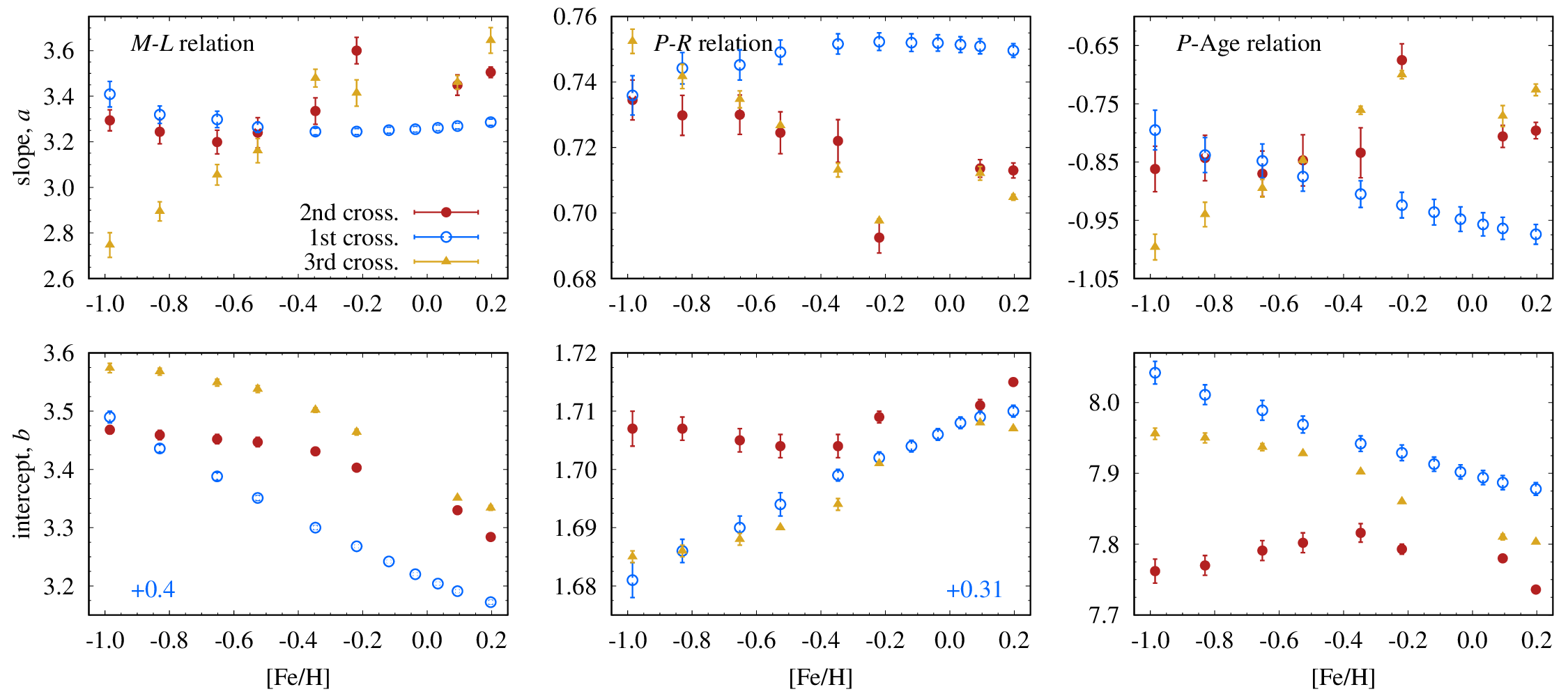}
\caption{The dependence of slope, $a$ (top) and intercept, $b$ (bottom) on metallicity for $\ML$, $\PR$ and $\PA$ relations (left, middle and right, respectively) for O00 (no overshooting) models on the 1st, 2nd and 3rd crossings. For the 1st crossing, some of the relations were shifted vertically, by the amount indicated in respective panels, for better visualisation.
\label{fig:relfehOV0}}
\end{figure*}

\begin{figure*}
\includegraphics[width=\linewidth]{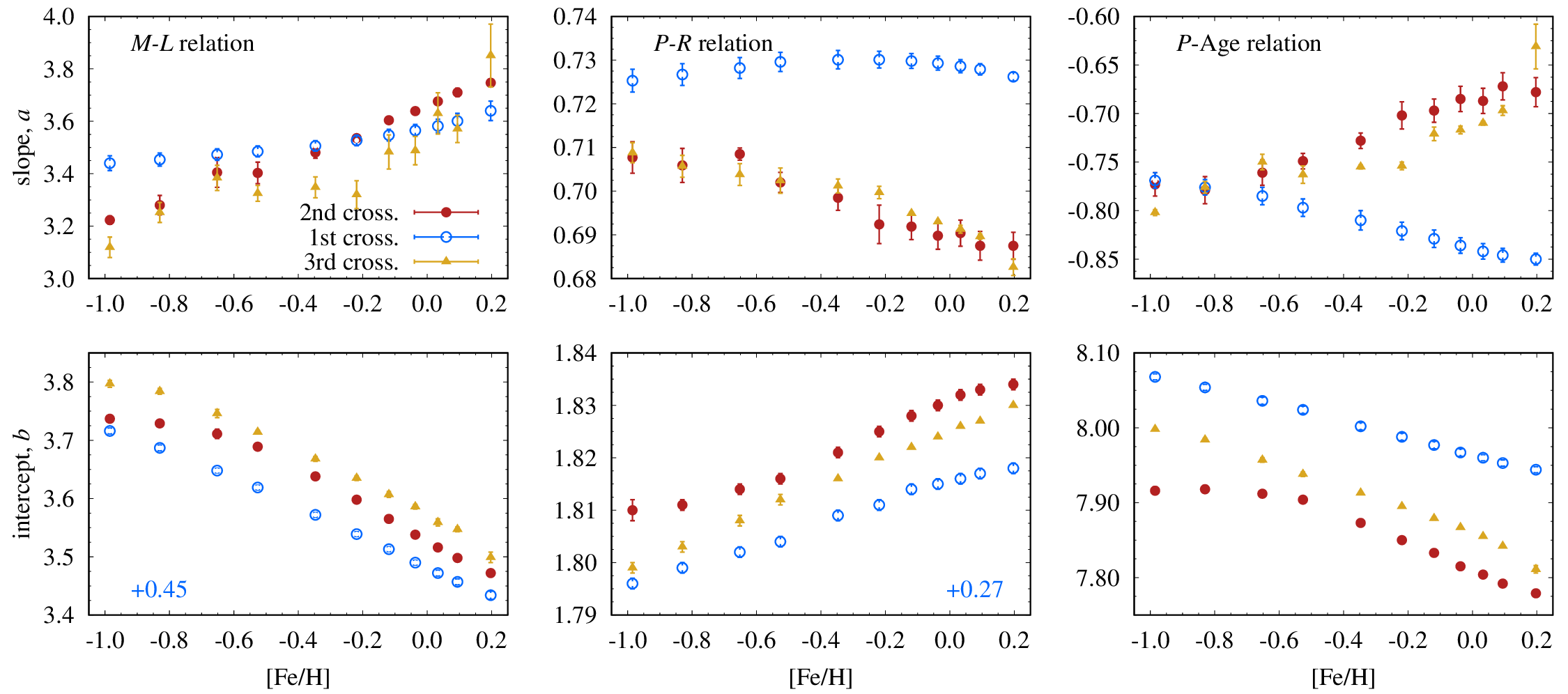}
\caption{Same as Fig.~\ref{fig:relfehOV0} but for models including overshooting (O24).
\label{fig:relfehOV4}}
\end{figure*}

Results for the three relations, slope, $a$, and intercept, $b$, vs.\ \feh{} are displayed in Figs~\ref{fig:relfehOV0} (O00) and ~\ref{fig:relfehOV4} (O24). Different symbols/colors correspond to different crossings. It is evident that both the slope and intercept depend on \feh{}, however the dependence for the intercept is significantly stronger. Considering 1st crossing, intercept is, to a very good approximation, a linear function of \feh{} both for O00 and O24 models. This is also the case for the 2nd/3rd crossings and O24 models. 

For models without overshooting (O00; Fig.~\ref{fig:relfehOV0}), relations are still mostly monotonous, but may be more nonlinear, eg., for \ML{} relation and 2nd crossing, $b$ slowly decreases with \feh, to \feh$\simeq-0.4$, after which the decrease with \feh{} is steeper. The exceptions are $\PR$ and $\PA$ relations for 2nd crossing, for which intercept first decreases/increases with \feh{} and then increases/decreases (for \PR/\PA, respectively). For both relations however, the dependence on metallicity is very weak. Also, for O00 models and 2nd/3rd crossings there are no data for $\feh=0.033$, $-0.037$ and $-0.119$, as the corresponding loops are too short to cross the IS (or there are too few crossings for a reliable fit).

For the slope (top panels in Figs~\ref{fig:relfehOV0} and \ref{fig:relfehOV4}), relations are more complex, in particular for O00 models, for which we often record non-monotonous progressions with \feh{}. For models with overshooting, all relations can be considered monotonous, but often nonlinear.

In the following, taking advantage of the fact that, compared to the intercept, the slope is only weakly sensitive to metallicity, we derive analytical expressions for the considered relations as:
\begin{equation}
\log y=a(\log x - \log x_0) +b +c\feh{}\,,\label{eq:linearfeh}
\end{equation}
ie., including linear dependence of intercept on \feh{}. For the two cases just discussed ($\PR$ and $\PA$ 2nd crossing relations for O00 models, with non-monotonous, but weak dependence of intercept on \feh{}), the relations will simply yield a close to zero metallicity term. We prefer such solution, rather than fitting separate relations in different metallicity ranges, since the metallicity dependence is weak, or cannot be established where data are missing due to too short loops.

For centerings in $\log P$, to allow more direct comparison between different relations, we use 4 values: for models with (O24) and without overshooting (O00) and for 1st crossing and loop (2nd/3rd) relations. They are explicitly given in tables to follow.

In addition to analytical relations presented in the following sections, data collected in Tab.~\ref{tab:data_all} can be used to plot any of the considered relations for specific metallicity, crossing, IS (hot/cool) and blue/red edge/midline.

\subsection{Mass-Luminosity relation\label{subsec:ml}}

\begin{figure*}
\includegraphics[width=\linewidth]{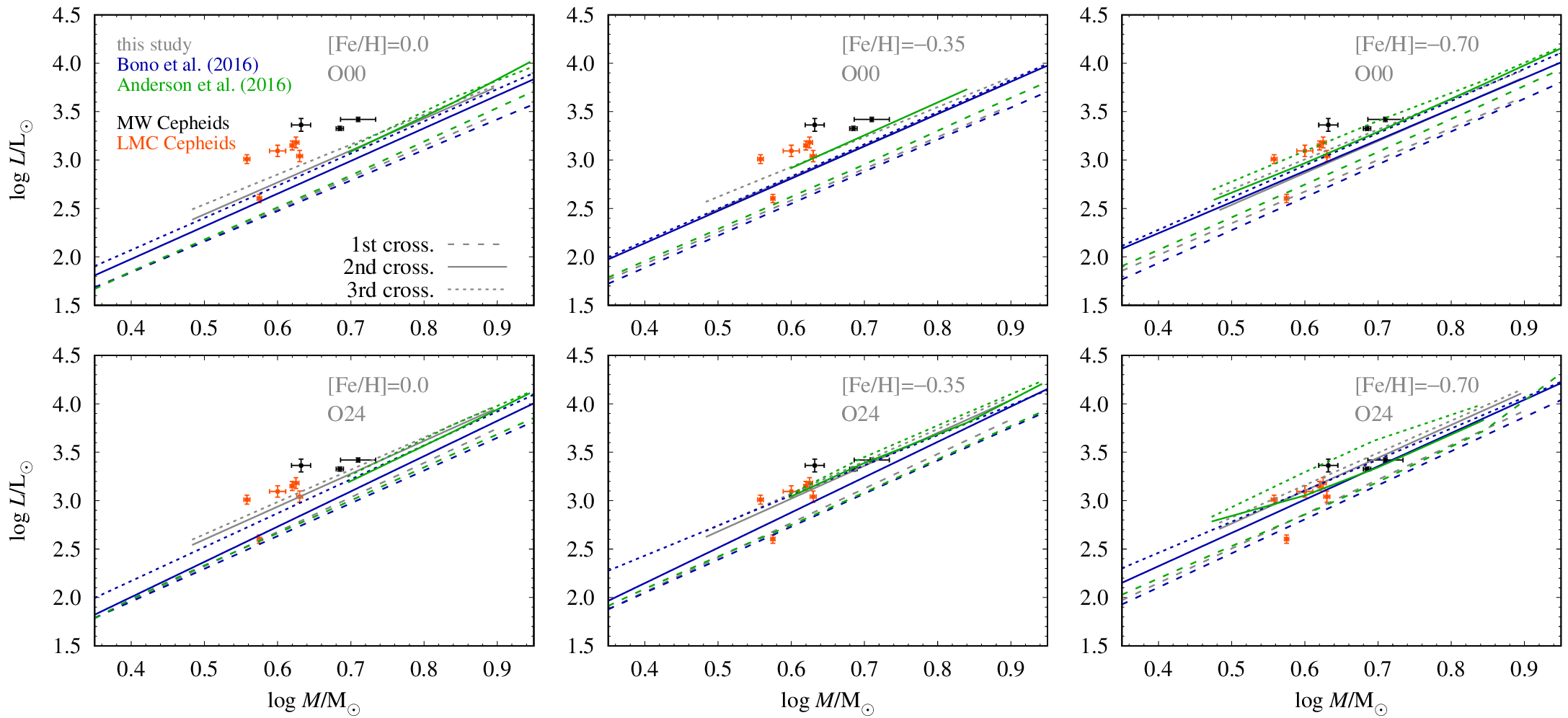}
\caption{$\ML$ relations for models without overshooting (O00, top) and including overshooting (O24, bottom), for MW, LMC and SMC metallicities ($\feh=0.0$, $-0.35$ and $-0.7$ in the left, middle, and right panels, respectively), confronted with determinations from \cite{Gallenne-2018-V1334Cyg, Gallenne-2025-SuCyg, Pilecki-2018, Evans-2024-Polaris} and other theoretical relations \citep[][different colors]{Bono-2016,Anderson-2016}.
\label{fig:ml}}
\end{figure*}

The derived analytical fits are collected in Tab.~\ref{tab:ml}. The metallicity term is always negative, meaning that the higher the metallicity, the lower the luminosity at a given mass. In Fig.~\ref{fig:ml} we present the relations for three metallicities, solar (left; typical for MW Cepheids), $\feh=-0.35$ (middle) and $\feh=-0.7$ (right panels). The latter two metallicities may be considered representative for the LMC and SMC, respectively -- for a summary of recent metallicity determinations see \cite{Hocde-2023}. Upper panels show relations without overshooting, bottom panels show models including overshooting (O24). Over-plotted are data for a few Cepheids with precise mass determinations, from the works of \cite{Gallenne-2018-V1334Cyg, Gallenne-2025-SuCyg, Pilecki-2018, Evans-2024-Polaris}. 

In Fig.~\ref{fig:ml} we also over-plot theoretical relations from two other studies. The first is \cite{Bono-2016} (B16 in the following) from which we use their canonical (no overshooting; top panels) and non-canonical relations \citep[including MS core overshooting; bottom panels, $0.2H_p$, see][]{Pietrinferni-2004}. Metallicities used are close to ours, ie., $\feh=0.053$, $-0.353$ and $-0.659$, in the left, middle and right panels, respectively. 

The second study is \cite{Anderson-2016} (A16 in the following) who studied the effects of rotation. In the top panel we use their data for non-rotating models. In the bottom panel, we use data for models with average rotation ($\omega_{\rm ini}=0.5\omega_{\rm crit}$). These models also include MS core overshooting ($0.1H_p$). In terms of metallicity, we display their models of $Z=0.014$, $0.006$ and $0.002$ in the left, middle and right panels, respectively. 

We stress that comparison with other theoretical work is not one-to-one. While in terms of physics included, B16 models are similar to ours, the parameters are slightly different and models were computed with different evolutionary codes. A16 models are very different as they include rotation. Still, it is interesting to note that the models qualitatively agree reasonably well. Our models without overshooting well agree with canonical models from B16 and non-rotating models from A16. The latter systematically predict larger luminosities, which is  understandable, as these models include overshooting ($0.1H_p$). We also note that for solar metallicity our relations are slightly brighter than B16. 

For our models including overshooting (O24), we compare them with non-canonical models from B16 and average rotation models from A16. Relations for the 1st crossing agree reasonably well. Differences are noticeable for the blue loop relations. Again, our relations are brighter than B16 relations. One factor contributing to the discrepancy is slightly larger overshooting extent in our models ($\sim0.23H_p$ vs.\ $0.2H_p$). Then, we note that the difference for 3rd crossing is lower than for the 2nd crossing, indicating a larger luminosity extent of blue loops in B16. Compared with A16, the relations are similar for solar and LMC metallicity but significantly different for the SMC metallicity, for which their 2nd crossing relation (for $\log M/\MS>0.6$) is dimmer, while 3rd crossing relation is brighter, indicating significantly larger luminosity span of the loops in rotating, low-metallicity models.

Comparing to observations, overshooting models (and non-canonical B16 models, and average rotation A16 models) represent a better match with observations. Still, considering solar metallicity and Galactic Cepheids (bottom left panel in Fig.~\ref{fig:ml}) observed Cepheids are too bright as compared to model predictions for given mass. The discrepancy decreases with decreasing metallicity. For the LMC metallicity, for its 4 Cepheids (OGLE-LMC-CEP-0227, -4506, -1718B, -2532) our 2nd/3rd crossing relations still predict luminosities that are on average 0.09\,dex/0.04\,dex dimmer. This discrepancy is only lifted as we decrease the metallicity even further, to one corresponding to the SMC. It is however difficult to assume that all eclipsing binary systems in the LMC have such atypical low metallicity. Another way to lift the discrepancy would be to increase MS core overshooting even further. To increase the luminosity of the loop by $\sim0.09$ dex, so that our 2nd crossing relation would match, on average, the four discussed Cepheids, one needs overshooting extent of $\fcor\approx0.025$ (or $0.29H_p$). This is a manifestation of the Cepheid mass discrepancy problem. We note however, that the comparison presented above, should be treated with caution, as it is based on only few Cepheids for which metallicity determinations are not available.

The mass discrepancy issue is present also for B16 non-canonical models, while rotating A16 models match the observations best. Relation for their 2nd crossing roughly overlaps our 2nd crossing relations, while relation for the 3rd crossing is significantly brighter, in agreement with \cite{Anderson-2014} conclusion, that inclusion of rotation (and increasing rotation rate) increases the luminosity separation between the 2nd and 3rd crossing.

By including Reimers mass loss with $\eta=0.4$ (O24\_ML4 models) mass is slightly decreased at nearly constant luminosity (see Sect.~\ref{subsec:massloss}). At a given mass, $\ML$ relation shifts towards slightly higher luminosities, reducing the just described discrepancy, however the shift is very small. As compared to relations without mass loss, at $\feh=-0.35$, the shift is only $0.009$\,dex, a factor of 10 smaller than the discrepancy with observations. Significantly stronger, pulsation-induced mass loss would be needed to reduce the discrepancy, see \cite{Neilson-2011}.  The $\ML$ relations for O24\_ML2 models, 2nd and 3rd crossings (1st crossing relation is the same as for O24 models, as mass loss becomes effective only on RGB) are also given in Tab.~\ref{tab:ml}. We intentionally fixed the slope of the relations to that of O24 models, to accommodate the difference in intercept (and in metallicity dependence).

$\ML$ relations for the cool IS are collected in Tab.~\ref{tabapp:ml} in Sect.~\ref{secapp:coolIS} of the Appendix. They are barely different from the hot IS relation, which is expected, since the relation does not involve pulsation period. Values of slope and metallicity term, $a$ and $c$, do agree within errors for nearly all listed relations. The intercepts, $b$ are systematically, but slightly ($\sim 0.01$) smaller for the cool IS relation, which is also expected as within the IS, for most of the considered masses and metallicities, luminosity slightly increases with increasing $\Teff$.  

\begin{deluxetable*}{lcrrrrcl}
\tablecaption{Coefficients of the $\ML$ relations as given by eq.~\eqref{eq:linearfeh}. First two columns identify data set, crossing number and edge along which relation is computed (`m' for the midline). In the last column, we provide the recommended parameter range in which the relation is applicable. All relations adopt hot IS. \label{tab:ml}}
\tablehead{
\colhead{data} & \colhead{cross./edge} & \colhead{$a$}  &  \colhead{$b$} & \colhead{$c$} & \colhead{$\log x_0$} & \colhead{rms}  & \colhead{remarks}
}
\startdata
\multicolumn{8}{l}{{\it Mass-Luminosity relations (hot IS)}}\\
O00 & 1c/m & $3.272\pm0.010$ & $2.815\pm0.003$ & $-0.262\pm0.005$ & $0.699$ & $0.018$ & $\log M/\MS\in(0.30,0.90)$ \\
O00 & 2c/m & $3.297\pm0.037$ & $3.356\pm0.007$ & $-0.140\pm0.012$ & $0.778$ & $0.030$ & $\log M/\MS\in(0.48,0.90)$ \\
O00 & 3c/m & $3.088\pm0.065$ & $3.401\pm0.012$ & $-0.222\pm0.021$ & $0.778$ & $0.039$ & $\log M/\MS\in(0.48,0.90)$ \\
O24 & 1c/m & $3.558\pm0.019$ & $3.030\pm0.005$ & $-0.262\pm0.010$ & $0.699$ & $0.021$ & $\log M/\MS\in(0.30,0.90)$ \\
O24 & 2c/m & $3.393\pm0.034$ & $3.542\pm0.005$ & $-0.236\pm0.010$ & $0.778$ & $0.023$ & $\log M/\MS\in(0.48,0.90)$ \\
O24 & 3c/m & $3.332\pm0.035$ & $3.579\pm0.005$ & $-0.248\pm0.011$ & $0.778$ & $0.021$ & $\log M/\MS\in(0.48,0.90)$ \\
 O24\_ML4 & 2c/m & $3.393$(fixed) & $3.553\pm0.004$ & $-0.230\pm0.007$ & $0.778$ & $0.024$ & $\log M/\MS\in(0.47,0.90)$ \\
 O24\_ML4 & 3c/m & $3.332$(fixed) & $3.589\pm0.003$ & $-0.244\pm0.006$ & $0.778$ & $0.020$ & $\log M/\MS\in(0.47,0.90)$ \\
\enddata
\end{deluxetable*}

\subsection{Period-Radius relation\label{subsec:pr}}

The derived analytical fits for $\PR$ relations for O00 and O24 models are collected in Tab.~\ref{tab:pr}. The metallicity term for nearly all relations is positive, meaning that the higher the metallicity the larger the radius at a given period. The effect is small, of order of $\sim$0.02 per dex in $\log R/\RS$. Only for O00 models on the 2nd crossing and blue edge relation, the metallicity term is negative (but very close to zero, indicating no metallicity effect; see also Fig.~\ref{fig:relfehOV0} and discussion in Sect.~\ref{subsec:relfeh}). In Tab.~\ref{tab:pr} we also provide average blue loop relations, based on data on 2nd and 3rd crossing (midline).

\begin{figure*}
\includegraphics[width=\linewidth]{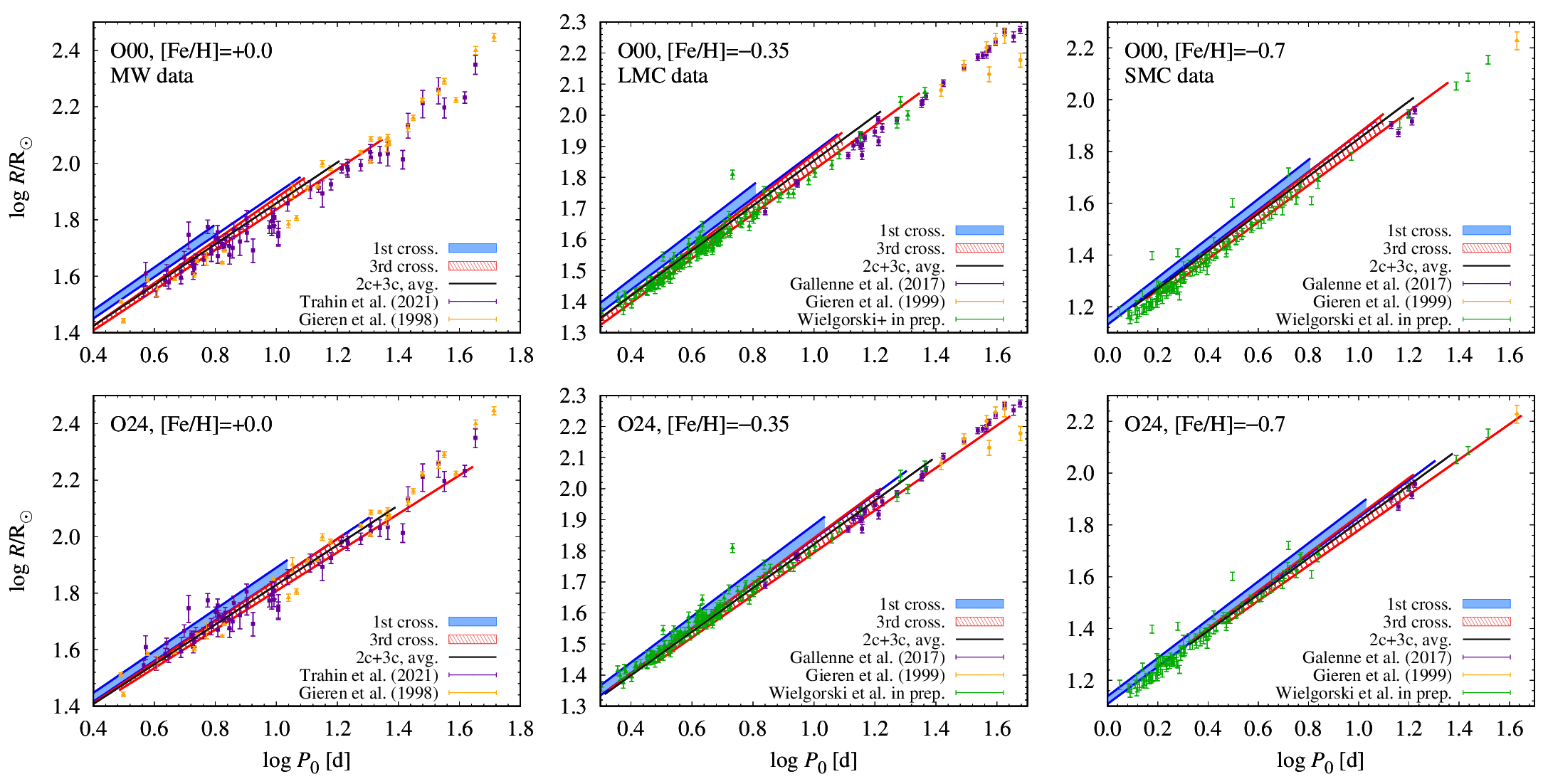}
\caption{$\PR$ relations for models without overshooting (O00, top) and including overshooting (O24, bottom), for MW, LMC and SMC metallicities ($\feh=0.0$, $-0.35$ and $-0.7$ in the left, middle, and right panels, respectively), confronted with determinations from \cite{Trahin-2021, Gieren-1998, Gieren-1999, Gallenne-2017} and Wielg\'orski et al. (in prep.).
\label{fig:pr}}
\end{figure*}

In Fig.~\ref{fig:pr} we present the relations in the same layout as in Fig.~\ref{fig:ml} for three metallicities and O00 and O24 models. Since relations involve pulsation period, they are represented with bands; upper and lower envelopes correspond to relations derived for the blue and red edge, respectively. Relations for the 2nd and 3rd crossing largely overlap; for clarity we plot the relations for 3rd crossing only (those for 2nd crossing are slightly shifted towards 1st crossing as compared to 3rd crossing relations). In addition, average blue loop relations are plotted with solid black line. Relations for the blue and red edge extend over different period ranges, recommended range in which the relations apply (see Sect.~\ref{subsec:pl}) is given in the last column of Tab.~\ref{tab:pr}.  Over-plotted are data for classical Cepheids with radii determinations through BW technique or its variants; for the MW Cepheids in the left panel \citep{Gieren-1998, Trahin-2021}, LMC and SMC Cepheids in the middle and right panels \citep{Gieren-1999, Gallenne-2017}. For the LMC and the SMC we also plot data from Wielg\'orski et al. (in prep.), which well sample the short period end. For the SMC, for clarity of the plot, only every 4th star is plotted.

\begin{deluxetable*}{lcrrrrcl}
\tablecaption{Coefficients of the $\PR$ relations as given by eq.~\eqref{eq:linearfeh}. First two columns identify data set, crossing number and edge along which relation is computed (`b/m/r' for the blue, midline and red edge). In the last column, we provide the recommended parameter range in which the relation is applicable. All relations adopt hot IS. \label{tab:pr}}
\tablehead{
\colhead{data} & \colhead{cross./edge} & \colhead{$a$}  &  \colhead{$b$} & \colhead{$c$} & \colhead{$\log x_0$} & \colhead{rms}  & \colhead{remarks}
}
\startdata
\multicolumn{8}{l}{{\it Period-Radius relations (hot IS)}}\\
O00 & 1c/b & $0.7579\pm0.0009$ & $1.4099\pm0.0006$ & $0.0217\pm0.0012$ & $0.308$ & $0.0038$ & $\log P_0\in(-0.79,0.81)$ \\
O00 & 1c/m & $0.7496\pm0.0013$ & $1.3964\pm0.0008$ & $0.0246\pm0.0017$ & $0.308$ & $0.0060$ & $\log P_0\in(-0.72,0.94)$ \\
O00 & 1c/r & $0.7375\pm0.0020$ & $1.3825\pm0.0013$ & $0.0342\pm0.0027$ & $0.308$ & $0.0094$ & $\log P_0\in(-0.66,1.08)$ \\
O00 & 2c/b & $0.7497\pm0.0032$ & $1.7266\pm0.0016$ & $-0.0027\pm0.0027$ & $0.789$ & $0.0055$ & $\log P_0\in(0.02,1.04)$ \\
O00 & 2c/m & $0.7263\pm0.0029$ & $1.7092\pm0.0015$ & $0.0062\pm0.0025$ & $0.789$ & $0.0063$ & $\log P_0\in(0.08,1.18)$ \\
O00 & 2c/r & $0.7059\pm0.0026$ & $1.6919\pm0.0016$ & $0.0205\pm0.0026$ & $0.789$ & $0.0065$ & $\log P_0\in(0.10,1.34)$ \\
O00 & 3c/b & $0.7556\pm0.0058$ & $1.7185\pm0.0031$ & $0.0107\pm0.0048$ & $0.789$ & $0.0050$ & $\log P_0\in(0.14,1.10)$ \\
O00 & 3c/m & $0.7323\pm0.0052$ & $1.7024\pm0.0028$ & $0.0210\pm0.0044$ & $0.789$ & $0.0058$ & $\log P_0\in(0.20,1.23)$ \\
O00 & 3c/r & $0.7133\pm0.0045$ & $1.6855\pm0.0026$ & $0.0341\pm0.0042$ & $0.789$ & $0.0056$ & $\log P_0\in(0.20,1.36)$ \\
O00 & 2c+3c/m & $0.7235\pm0.0028$ & $1.7057\pm0.0014$ & $0.0133\pm0.0023$ & $0.789$ & $0.0095$ & $\log P_0\in(0.10,1.22)$ \\
O24 & 1c/b & $0.7387\pm0.0013$ & $1.5624\pm0.0013$ & $0.0192\pm0.0020$ & $0.556$ & $0.0028$ & $\log P_0\in(-0.73,1.04)$ \\
O24 & 1c/m & $0.7308\pm0.0014$ & $1.5460\pm0.0012$ & $0.0208\pm0.0022$ & $0.556$ & $0.0037$ & $\log P_0\in(-0.65,1.17)$ \\
O24 & 1c/r & $0.7194\pm0.0017$ & $1.5290\pm0.0014$ & $0.0286\pm0.0027$ & $0.556$ & $0.0064$ & $\log P_0\in(-0.59,1.31)$ \\
O24 & 2c/b & $0.7295\pm0.0033$ & $1.8506\pm0.0013$ & $0.0125\pm0.0023$ & $0.997$ & $0.0033$ & $\log P_0\in(0.35,1.21)$ \\
O24 & 2c/m & $0.7041\pm0.0024$ & $1.8302\pm0.0011$ & $0.0233\pm0.0021$ & $0.997$ & $0.0037$ & $\log P_0\in(0.29,1.39)$ \\
O24 & 2c/r & $0.6788\pm0.0021$ & $1.8115\pm0.0011$ & $0.0379\pm0.0020$ & $0.997$ & $0.0039$ & $\log P_0\in(0.39,1.65)$ \\
O24 & 3c/b & $0.7262\pm0.0039$ & $1.8452\pm0.0013$ & $0.0165\pm0.0028$ & $0.997$ & $0.0027$ & $\log P_0\in(0.32,1.23)$ \\
O24 & 3c/m & $0.7042\pm0.0027$ & $1.8247\pm0.0012$ & $0.0265\pm0.0022$ & $0.997$ & $0.0027$ & $\log P_0\in(0.36,1.38)$ \\
O24 & 3c/r & $0.6824\pm0.0023$ & $1.8059\pm0.0013$ & $0.0404\pm0.0022$ & $0.997$ & $0.0028$ & $\log P_0\in(0.48,1.65)$ \\
O24 & 2c+3c/m & $0.7030\pm0.0019$ & $1.8275\pm0.0009$ & $0.0248\pm0.0016$ & $0.997$ & $0.0047$ & $\log P_0\in(0.31,1.39)$ \\
\enddata
\end{deluxetable*}

Relations without overshooting (top panels in Fig.~\ref{fig:pr}) do not match well with observations for any of the considered metallicities. The predicted radii are on average larger at a given pulsation period. Once overshooting is included (bottom panels) the relations for the 3rd (and overlapping 2nd) crossings match the observations reasonably well. Inclusion of MS core overshooting is thus necessary to reproduce the observed $\PR$ progressions. For the SMC, we again observe a discrepancy at the short-period end, where the blue-loop relations do not extend far enough; the relation for the 2nd crossing reaches periods shorter by about 0.1\,dex compared to the 3rd crossing relation, which is still not enough. Models without overshooting reproduce pulsation period distribution in the SMC significantly better, but fail at reproducing radii. At the long-period end our relations suffer from lack of models above 8\MS. 

Detailed and quantitative comparison of the theoretical $\PR$ relations with observations will be presented in Wielg\'orski et al. (in. prep.). 

Relations for the cool IS are collected in Tab.~\ref{tabapp:pr} in Sect.~\ref{secapp:coolIS} of the Appendix. Comparing the corresponding coefficients, we note that the slopes are slightly, but systematically less steep (on average $\sim1\sigma$ difference) the intercepts are systematically smaller ($4\sigma$ difference) and the metallicity terms are slightly, but systematically larger ($\sim1\sigma$ difference). The net effect is slightly smaller radii at a given pulsation period for cool IS relations. For O24, 2nd crossing relations (and $\feh=-0.35$), the cool relation predicts radii smaller by 0.003\,dex in $\log R/\RS$ at short period end ($\log P=0.4$; less than 1\% difference in radii) and 0.014\,dex smaller radii at long period ($\log P=1.4$; $\approx$3\% difference in radii) end -- both values are smaller than a typical error associated with observations displayed in Fig.~\ref{fig:pr}.

\subsection{Period-Age relation\label{subsec:pa}}

\begin{figure*}
\includegraphics[width=\linewidth]{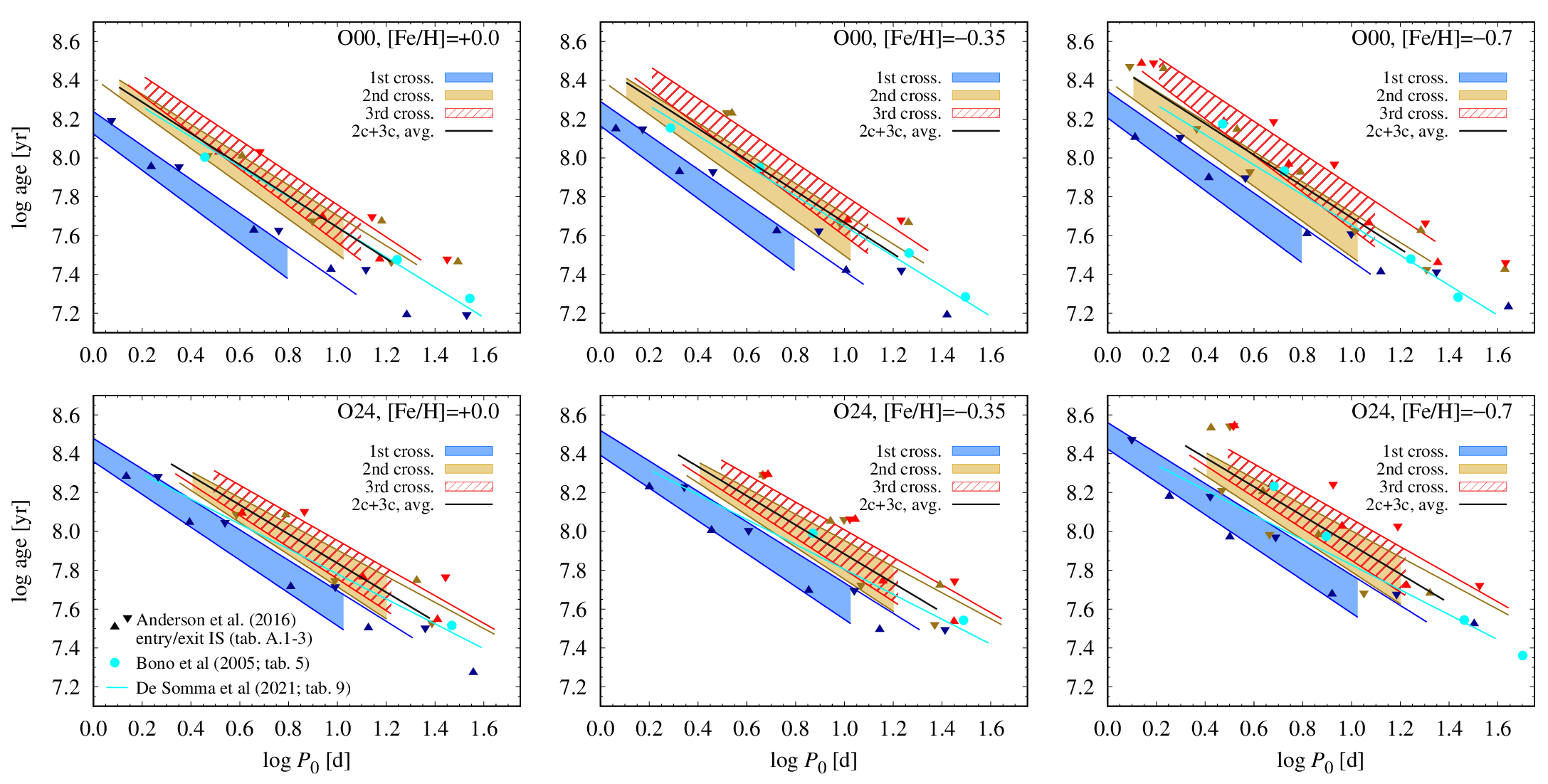}
\caption{$\PA$ relations for models without overshooting (O00, top) and including overshooting (O24, bottom), for MW, LMC and SMC metallicities ($\feh=0.0$, $-0.35$ and $-0.7$ in the left, middle, and right panels, respectively). Relations are plotted for each crossing, the upper and lower envelopes of the bands correspond to red and blue edge of the IS. Relations are confronted with other theoretical work. For \cite{Anderson-2016} we use data from their tabs.~A.1-3; up and down triangles correspond to entry/exit of the IS. For \cite{Bono-2005PA} we use data from their tab.~5 (canonical/non-canonical models in the top and bottom panels, respectively, $Z=0.0198$, $0.008$ and $0.004$ in the left, middle and right columns, respectively). For \cite{DeSomma-2021PA} we use their analytical metal dependent relations for canonical (case `A') and non-canonical (case `B') relations (their tab.~9) in the top and bottom panels, respectively.
\label{fig:pa}}
\end{figure*}

The analytical fits for $\PA$ relations are collected in Tab.~\ref{tab:pa}. As for $\PL$ and $\PR$ relations, recommended range in which the relations apply is given in the last column of Tab.~\ref{tab:pa}. In general, the metallicity term is negative, meaning that the higher the metallicity, the lower the age at a given pulsation period.  Only for O00 models on the 2nd crossing and blue edge relation, the metallicity term is positive (but very small, indicating weak metallicity effect; see also Fig.~\ref{fig:relfehOV0} and discussion in Sect.~\ref{subsec:relfeh}). In Fig.~\ref{fig:pa} we present the relations in the same layout as in Figs~\ref{fig:ml} and \ref{fig:pr} for three metallicities and O00 and O24 models. Relations are represented with bands; upper and lower envelope correspond to red and blue edges of the IS, respectively. Relations for all crossings are presented.  Relations for 1st crossing in general well separate from the blue loop relations. Relations for the 2nd and 3rd crossings largely overlap, those for 3rd crossing giving on average slightly larger ages at given pulsation period. In Tab.~\ref{tab:pa} we also provide joint relations for the 2nd and 3rd crossing that can be used when crossing number is unknown. To derive these relations we used the data for corresponding midlines. These relations are plotted with solid black lines in Fig.~\ref{fig:pa}.

\begin{deluxetable*}{lcrrrrcl}
\tablecaption{Coefficients of the $\PA$ relations as given by eq.~\eqref{eq:linearfeh}. First two columns identify data set, crossing number and edge along which relation is computed (`b/m/r' for the blue, midline and red edge). In the last column, we provide the recommended parameter range in which the relation is applicable. All relations adopt hot IS. \label{tab:pa}}
\tablehead{
\colhead{data} & \colhead{cross./edge} & \colhead{$a$}  &  \colhead{$b$} & \colhead{$c$} & \colhead{$\log x_0$} & \colhead{rms}  & \colhead{remarks}
}
\startdata
\multicolumn{8}{l}{{\it Period-Age relations (hot IS)}}\\
O00 & 1c/b & $-0.9351\pm0.0070$ & $7.8353\pm0.0045$ & $-0.1182\pm0.0088$ & $0.308$ & $0.036$ & $\log P_0\in(-0.79,0.81)$ \\
O00 & 1c/m & $-0.9063\pm0.0085$ & $7.9053\pm0.0053$ & $-0.1310\pm0.0109$ & $0.308$ & $0.046$ & $\log P_0\in(-0.72,0.94)$ \\
O00 & 1c/r & $-0.8733\pm0.0101$ & $7.9694\pm0.0064$ & $-0.1494\pm0.0134$ & $0.308$ & $0.055$ & $\log P_0\in(-0.66,1.08)$ \\
O00 & 2c/b & $-0.9076\pm0.0197$ & $7.6952\pm0.0098$ & $0.0180\pm0.0163$ & $0.789$ & $0.041$ & $\log P_0\in(0.02,1.04)$ \\
O00 & 2c/m & $-0.8277\pm0.0168$ & $7.7840\pm0.0084$ & $-0.0047\pm0.0141$ & $0.789$ & $0.043$ & $\log P_0\in(0.08,1.18)$ \\
O00 & 2c/r & $-0.7797\pm0.0139$ & $7.8687\pm0.0082$ & $-0.0231\pm0.0138$ & $0.789$ & $0.041$ & $\log P_0\in(0.10,1.34)$ \\
O00 & 3c/b & $-0.9467\pm0.0138$ & $7.7634\pm0.0065$ & $-0.1002\pm0.0103$ & $0.789$ & $0.028$ & $\log P_0\in(0.14,1.10)$ \\
O00 & 3c/m & $-0.8720\pm0.0142$ & $7.8484\pm0.0071$ & $-0.1262\pm0.0112$ & $0.789$ & $0.034$ & $\log P_0\in(0.20,1.23)$ \\
O00 & 3c/r & $-0.8317\pm0.0138$ & $7.9346\pm0.0078$ & $-0.1386\pm0.0124$ & $0.789$ & $0.039$ & $\log P_0\in(0.20,1.36)$ \\
O00 & 2c+3c/m & $-0.8062\pm0.0207$ & $7.8126\pm0.0105$ & $-0.0719\pm0.0171$ & $0.789$ & $0.076$ & $\log P_0\in(0.10,1.22)$ \\
O24 & 1c/b & $-0.8449\pm0.0037$ & $7.8896\pm0.0027$ & $-0.0929\pm0.0049$ & $0.556$ & $0.013$ & $\log P_0\in(-0.73,1.04)$ \\
O24 & 1c/m & $-0.8142\pm0.0043$ & $7.9683\pm0.0030$ & $-0.1033\pm0.0060$ & $0.556$ & $0.021$ & $\log P_0\in(-0.65,1.17)$ \\
O24 & 1c/r & $-0.7845\pm0.0050$ & $8.0425\pm0.0035$ & $-0.1164\pm0.0072$ & $0.556$ & $0.027$ & $\log P_0\in(-0.59,1.31)$ \\
O24 & 2c/b & $-0.8262\pm0.0102$ & $7.7176\pm0.0036$ & $-0.1100\pm0.0069$ & $0.997$ & $0.019$ & $\log P_0\in(0.35,1.21)$ \\
O24 & 2c/m & $-0.7483\pm0.0075$ & $7.8172\pm0.0033$ & $-0.1242\pm0.0061$ & $0.997$ & $0.020$ & $\log P_0\in(0.29,1.39)$ \\
O24 & 2c/r & $-0.6732\pm0.0065$ & $7.9060\pm0.0034$ & $-0.1413\pm0.0060$ & $0.997$ & $0.021$ & $\log P_0\in(0.39,1.65)$ \\
O24 & 3c/b & $-0.8154\pm0.0071$ & $7.7580\pm0.0022$ & $-0.1359\pm0.0049$ & $0.997$ & $0.010$ & $\log P_0\in(0.32,1.23)$ \\
O24 & 3c/m & $-0.7615\pm0.0057$ & $7.8612\pm0.0023$ & $-0.1445\pm0.0044$ & $0.997$ & $0.013$ & $\log P_0\in(0.36,1.38)$ \\
O24 & 3c/r & $-0.7110\pm0.0055$ & $7.9556\pm0.0028$ & $-0.1577\pm0.0048$ & $0.997$ & $0.016$ & $\log P_0\in(0.48,1.65)$ \\
O24 & 2c+3c/m & $-0.7490\pm0.0082$ & $7.8387\pm0.0035$ & $-0.1344\pm0.0065$ & $0.997$ & $0.031$ & $\log P_0\in(0.31,1.39)$ \\
\enddata
\end{deluxetable*}

Our results are compared with three other theoretical work. For \cite{Bono-2005PA} we use their data for canonical models (no rotation, no overshooting, no mass loss) in the top panels and non-canonical models (MS overshooting included, $0.2H_p$) in the bottom panels. In consecutive columns, data for $Z=0.0198$, $0.008$ and $0.004$ are plotted. These models adopt BaSTI tracks from \cite{Pietrinferni-2004}, which use \cite{GN-93} reference solar composition, hence higher $Z$ values. We also compare our results with metallicity dependent $\PA$ relations from \cite{DeSomma-2021PA} which are based on updated BaSTI tracks from \cite{Hidalgo-2018}. In the top panel we show their canonical (case `A') relation (no overshooting, no rotation, no mass loss) while in the bottom panel we plot their non-canonical (case `B') relation, which is based on $\ML$ relation 0.2\,dex brighter than the canonical one. These relations represent average for the blue loop. For A16, just as in Fig.~\ref{fig:ml}, we use data for their non-rotating (top) and average rotation models (bottom; data along blue and red edges). Again we stress these relations were calculated with different codes assuming different physical input.

For 1st crossing, we compare with A16 and the agreement is very good. In the top panel, as expected, A16 relations are slightly shifted towards larger ages at a given period, as these models include mild overshooting (0.1$H_p$). The same applies to the 2nd and 3rd crossing A16 relations displayed in the top panel of Fig.~\ref{fig:pa}. For models including rotation, which are confronted with our O24 models, the relations for 2nd and 3rd crossing are similar for the MW and LMC metallicity; A16 relations predicting slightly larger ages. For the SMC metallicity, A16 relations for the 2nd and 3rd crossing are well separated, those for 2nd crossing giving smaller and those for 3rd crossing giving larger ages as compared to our relations. For this particular case we also noted the largest differences for $\ML$ relation, see Fig.~\ref{fig:ml}. 

Comparing to relations based on BaSTI tracks, from \cite{Bono-2005PA} and \cite{DeSomma-2021PA}, we first note that these relations are quite similar. Significant differences are observed for their non-canonical models. For the LMC and SMC metallicity, the relations based on updated BaSTI tracks at shorter periods predict shorter ages.  Comparing \cite{DeSomma-2021PA} relation with our joint relation for the loop (black lines) we observe consistent slopes. For their canonical and our O00 models we observe a nearly perfect match at solar metallicity. For lower metallicities, our relations predict larger ages at given period, the difference increasing with decreasing \feh{}. Similar for their non-canonical and our O24 models, although here the difference is noticeable already at solar metallicity.

Relations for the cool IS are collected in Tab.~\ref{tabapp:pa} in Sect.~\ref{secapp:coolIS} of the Appendix. Comparing the corresponding coefficients, we note that the slopes are very slightly (on average 1$\sigma$ difference), but systematically less steep, the intercepts are systematically and significantly larger and the metallicity terms have comparable values. The net effect is higher age at a given period for cool IS relations. For O24, 2nd crossing relations, the cool relation predicts ages about 5\% longer at short period end ($\log P=0.4$) and about 12\% longer at long period end ($\log P=1.4$), assuming LMC metallicity.

\section{Summary and conclusions\label{sec:discussion}}

We have presented evolutionary and pulsation scenarios for classical, F-mode Cepheids computed consistently with \MESA{} and, for linear periods and IS, with \MESA-\RSP. Our attention was focused on the effects of MS core and RGB envelope overshooting as well as mass loss (Reimers formula). The effects of \npg{} nuclear reaction rate (JINA vs. NACRE) and reference solar mixture (A09 vs.\ GS98) were also briefly discussed (Sect.~\ref{secapp:net} and \ref{secapp:gs98}) and are, in general, very small, except for no overshooting, solar metallicity case. Dedicated studies of the impact of the adopted reference solar composition, as well as nuclear reaction rates -- particularly when updated determinations become available -- are clearly a valuable direction for future work.

All calculated evolutionary tracks are available online. In total, there are several thousand models with masses ranging from $2-8\MS$, calculated in steps of $0.5\MS$, and for some models in steps of $0.1\MS$, for 11 metallicity values in the range from $\feh=-1.0$ to $\feh=+0.2$ (Tab.~\ref{tab:fehz}). The calculated model grids (Tab.~\ref{tab:grids}) extensively explore the range of overshooting parameters, both from the MS core and from the convective  envelope on the RGB. Some of the grids also take mass loss into account. These models (in particular GF grids, with $0.1$\MS{} step in mass) are well suited for preliminary modeling of specific Cepheids, eg., Cepheids in eclipsing binaries \citep[GF models are used exactly for that purpose in our modeling of Cepheids in eclipsing binary systems from][to be presented in Ziolkowska et al., in prep.]{Pilecki-2018}, or for general and broader comparisons with observations than those presented here. Although this work focuses on Cepheids, the published tracks start at ZAMS and can be used to model stars and study evolutionary properties at earlier stages of evolution: on the MS and on the RGB.

Concerning overshooting, we have mapped in detail the blue loop morphology as a function of mass, metallicity and overshooting parameters (Sect.~\ref{subsec:loops}). Without, or with weak envelope overshooting, or large MS core overshooting, the loop extent is a non-monotonous function of mass and metallicity, in particular at higher metallicities. Consequently, the study of metal rich systems require dense sampling of metallicity and overshooting parameter space.

For IS edges, midlines, $\PL$, $\ML$, $\PR$ and $\PA$ relations, convenient, metallicity dependent analytical relations are provided for a limited set of models (mostly O00 and O24) -- see Tabs.~\ref{tab:ais}, \ref{tab:plrelation}, \ref{tab:ml}, \ref{tab:pr} and \ref{tab:pa}. The relations are also available in a tabular form, separately for each considered metallicity and for all computed model sets, including for two versions of the IS (hot and cool) -- Tab.~\ref{tab:data_all}. Crossing times and period change rates are collected in Tab.~\ref{tab:agextime}.

For all relations, we find that both slope and zero point depend on metallicity.  The dependence is much stronger for the zero point and is almost always monotonic with metallicity. Consequently, our analytical fits adopt metallicity independent slopes and zero points depending on metallicity in linear manner. 

Since metallicity dependence of the $\PL$ relation, quantified with $\gamma$ parameter, is of wide interest, we investigated it in detail. We find that the effect depends on the underlying mass-luminosity relation. Considering the Wesenheit, $W_{VI}$ index, the effect is the weakest fore models that do not include overshooting ($\gamma=-0.11$\,mag\,dex$^{-1}$ at 2nd crossing). When overshooting is included we get significantly higher effect, $\gamma=-0.19$\,mag\,dex$^{-1}$ for the 2nd crossing. For even brighter (at the same mass) 3rd crossing models we get $\gamma=-0.19$\,mag\,dex$^{-1}$ and $\gamma=-0.20$\,mag\,dex$^{-1}$ for models without and including overshooting, respectively.  We have also investigated the effect in different pass bands, following the work of \cite{Breuval-2022}. In agreement with observations, we find that the effect is nearly independent of pass band. However the average value (across considered pass bands) we obtain ($\gamma=-0.16$\,mag\,dex$^{-1}$) is smaller than in \cite{Breuval-2022} ($\gamma=-0.28$\,mag\,dex$^{-1}$).

While in general we note a satisfactory agreement with observations and other theoretical work, some discrepancies were also recorded.

A well known and old puzzle is the Cepheid mass discrepancy, which also manifests in our models (Sect.~\ref{subsec:ml}, Fig.~\ref{fig:ml}). For a given luminosity, the predicted Cepheid masses are too large as compared with observations. Inclusion of MS core overshooting of reasonable extent and of Reimers mass loss is not enough to lift the discrepancy and other scenarios, such as rotation or pulsation induced mass loss must be considered.

We also encountered problems with reproducing observed Cepheid period distributions, both at the long and short period ends. The former is a technical one -- we refrain from computing models with masses larger than 8\MS{} as we lack convergence due to the development of thin convective shells at the end of MS. A better treatment of the convective boundaries seems necessary to address this problem; for details see \citetalias{Ziolkowska-2024}. At the short period end, our blue loop models do not reach periods as short as observed in the SMC, so in the low metallicity regime. While shorter periods are possible when overshooting extent is decreased, the models do not match observed radii then (Sect.~\ref{subsec:pr}).

The comparison with observations and with other theoretical studies presented here is necessarily limited and largely qualitative. More detailed and quantitative comparisons are planned in forthcoming work. In the first of these (Wielg\'orski et al., in prep.), we will focus on the $\PR$ relation, taking advantage of the largest available sample of classical Cepheids in the Magellanic Clouds with radii determined using the BW technique.

The presented analysis is not exhaustive, and further research is planned. Based on the models already calculated, it is possible to take color terms into account in the analyzed relations and to analyze surface abundances. Calculating IS and evolutionary and pulsation relations for 1O Cepheids is also planned. The next step is to take model rotation into account and analyze how the implementation of rotation in \mesa{} affects the properties of evolutionary tracks, in particular compared to other codes that take rotation into account (Smolec et al., in prep). It is important to consider rotation in models with and without convective overshooting in order to investigate the expected parameter degeneracies. Comprehensive comparison with observations, based on population synthesis is also needed. However, the latter would require extending the calculations to include more massive models in order to account for Cepheids with longer periods. This, however, requires the development of a better methods for determining the boundaries of convective zones in \MESA.

Another necessary direction for the development of evolutionary codes, such as \MESA{}, is to take into account the postulated pulsation-enhanced mass loss \citep[][]{Neilson-2012}. For all relations linking evolutionary and pulsation calculations, self-consistency is extremely important. \MESA, in conjunction with \RSP, guarantees a high level of consistency by using the same microphysical properties (opacity, equation of state). Although in the presented pulsation calculations the envelope parameters are taken directly from evolutionary models, the envelope model itself is chemically homogeneous and constructed independently. This is generally a very good approximation for classical pulsating stars. Nevertheless, fully consistent pulsation calculations directly on the evolutionary model are now also possible \citep[][]{Farag-2026arXiv}.

\begin{acknowledgments}
This research is supported by the National Science Center, Poland, Sonata BIS project 2018/30/E/ST9/00598. We thank the referee for a careful reading of the manuscript and for helpful suggestions.
\end{acknowledgments}




\software{\MESA{} \citep{Paxton-2011,Paxton-2013,Paxton-2015,Paxton-2018,Paxton-2019,Jermyn-2023} \texttt{MESA} SDK  \citep{Townsend-2022}, Numpy \citep{Harris-2020}.}

\appendix

\section{Exponential vs.\ step convective core overshooting\label{secapp:stepvsexp}}

We use exponential formalism for overshooting, in which overshoot extent is characterized with $\fcor$ parameter (Sect.~\ref{subsec:tools}). Many studies in the literature use step overshooting, in which extent of the overshooting is expressed as a fraction of the local pressure scale height, $\beta H_p$. To allow a more direct comparison of such work with our results, we took advantage of the fact that both formalisms are implemented in \MESA{} and determined the relationship between $\fcor$ and $\beta$, for MS convective core overshooting. We investigated whether this relation depends on mass and metallicity.

\begin{figure}[ht!]
\includegraphics[width=\linewidth]{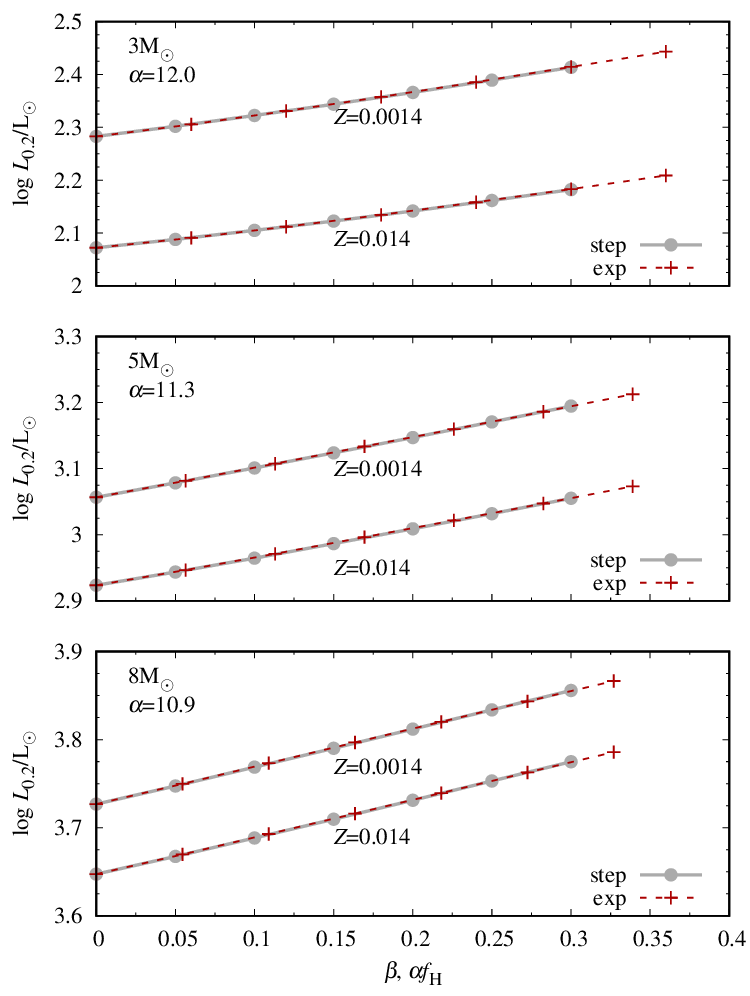}
\caption{Calibration of the relation between exponential and step MS convective core overshooting. In each panel luminosity at $0.2$ center H mass fraction, $L_{0.2}$, is plotted as a function of step and scaled exponential overshoot parameters. Three panels show results for 3, 5 and 8\MS. In each panel two families of curves are present for $Z=0.014$ and $Z=0.0014$.
\label{fig:calov}}
\end{figure}

To this aim we computed MS evolution for 3, 5 and 8\MS{} and two metallicities, $Z=0.014$ ($\feh\simeq0.0$) and $Z=0.0014$ ($\feh\simeq-1.0$). For each $M$/$Z$, a sequence of models adopting exponential overshooting with $\fcor$ varying from $0$ to $0.03$ with $0.005$ step was computed. Similar for step overshooting, $\beta$ was changed from $0$ to $0.3$ with $0.05$ step. Then, for each computed track, we determined luminosity when the central hydrogen content dropped to a mass fraction of $0.2$, $L_{0.2}$ (results do not depend on this particular choice). In Fig.~\ref{fig:calov}, $\log L_{0.2}/\LS$ is plotted vs.\ $\beta$ with filled dots and gray shaded lines. Top, middle and bottom panels correspond to 3, 5 and 8\MS{} models, respectively, and the two lines correspond to two metallicities (labeled). We observe a nearly linear increase of luminosity with increasing $\beta$. For exponential overshooting, these relations may be perfectly reproduced, given the exponential overshoot parameter, $\fcor$, is multiplied by a factor $\alpha$, determined through least square approach requesting $\log L_{0.2}/\LS (\beta)=\log L_{0.2}/\LS(\alpha\fcor)$. For a given $M$, $\alpha$ is little sensitive to $Z$: rounding to a single decimal place results are the same for both $Z$. On the other hand, $\alpha$ decreases slightly with increasing $M$ from $12.0$ at 3\MS{} to $10.9$ at 8\MS. The scaled relations for exponential overshooting are plotted with plus symbols and dashed red lines in Fig.~\ref{fig:calov} and indeed they overlap with those for step overshooting. For rough estimates, a factor of $11.5$ may be used to translate exponential to step overshooting parameter.

\section{Compatibility with newer version of \MESA{}}\label{secapp:oldvsnew}

To demonstrate that our results fully hold with the recent release version of \MESA, 24.08.1\footnote{When the work was nearing completion, another release appeared, 25.12.1}, we used it to compute a limited number of models assuming two cases: without convective overshooting and with overshooting ($\fcor=0.02$, $\fenv=0.04$). We note that it is not possible to make a one-to-one comparison that would use exactly the same physical setup. This is due to the redefinition of nuclear reaction nets used by default in the newer version of \MESA{} -- the ability to set the rates preferences (REACLIB vs.\ NACRE) has been removed. Still, both old and new calculations are based primarily on nuclear reaction rates from REACLIB and adopt the same rates for \npg{} \citep{jina} and \cag{} \citep{Kunz-2002}.

\begin{figure*}
\includegraphics[width=\linewidth]{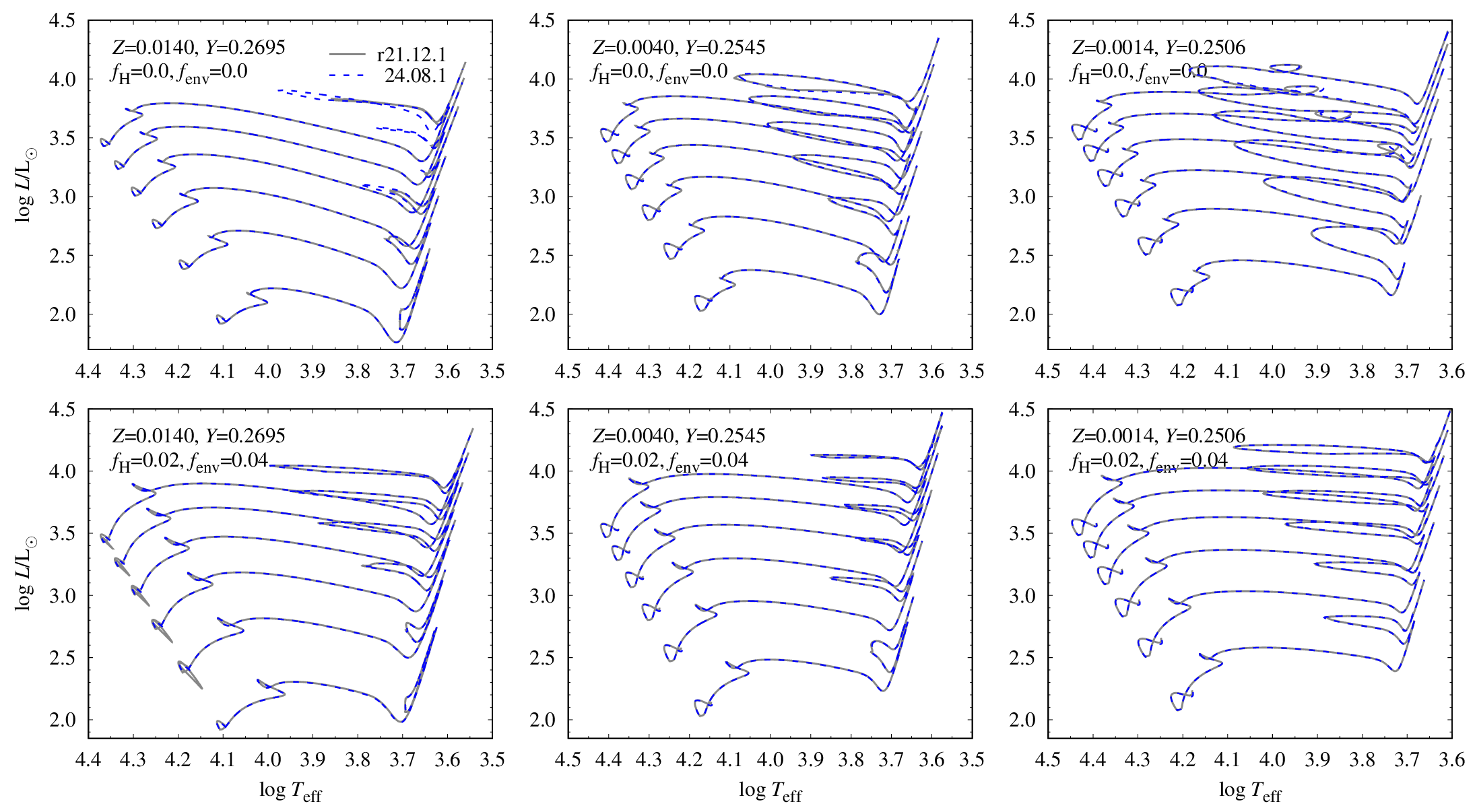}
\caption{Evolutionary tracks computed with \MESA{}-r21.12.1 (solid gray lines) and \MESA-24.08.1 (dashed blue lines) from ZAMS till AGB for models of 3, 4, 5, 6, 7 and 8\MS{} and different metallicities, $Z=0.014$ (left), $Z=0.004$ (middle) and $Z=0.0014$ (right). Models in the top row do not include overshooting, while in the bottom row overshooting is included.
\label{fig:oldnew}}
\end{figure*}

The comparison is illustrated in the HRDs in Fig.~\ref{fig:oldnew}. We observe a very satisfactory agreement of tracks computed with both \MESA{} versions. In particular, for models with overshooting (bottom panels of Fig.~\ref{fig:oldnew}), for all masses and metallicities the tracks nearly overlap. The shape and extent of the loops are qualitatively the same. Luminosity differences at expected 2nd and 3rd crossings of the IS, are always less than $0.005$\,dex. Above statements also hold for models without overshooting and with intermediate and low metallicity, $Z=0.004$ and $Z=0.0014$. Only for solar metallicity and $M\geq5\MS$ (top left panel in Fig.~\ref{fig:oldnew}) we observe qualitative differences between tracks during core helium burning. Blue loops computed with the newer version of \MESA{} are significantly longer. For 7\MS{}, the loop did not develop with the older version, while it is prominent with the new version. Still the luminosity levels at core helium burning are comparable. We note that the differences we observe for the discussed models are consistent with differences we recorded in \citetalias{Ziolkowska-2024} (see figs.~14 and 22 in the Appendix) while considering different nuclear reaction rates. Significant sensitivity was also recorded for models without convective overshooting at solar metallicity, which allows us to conclude that in this regime core helium burning is strongly sensitive to nuclear reaction rates. Since these are not one-to-one consistent in the two \MESA{} versions, such differences are not surprising.

\section{The effects of \npg{} nuclear reaction rate\label{secapp:net}}

\begin{figure*}[ht!]
\includegraphics[width=\linewidth]{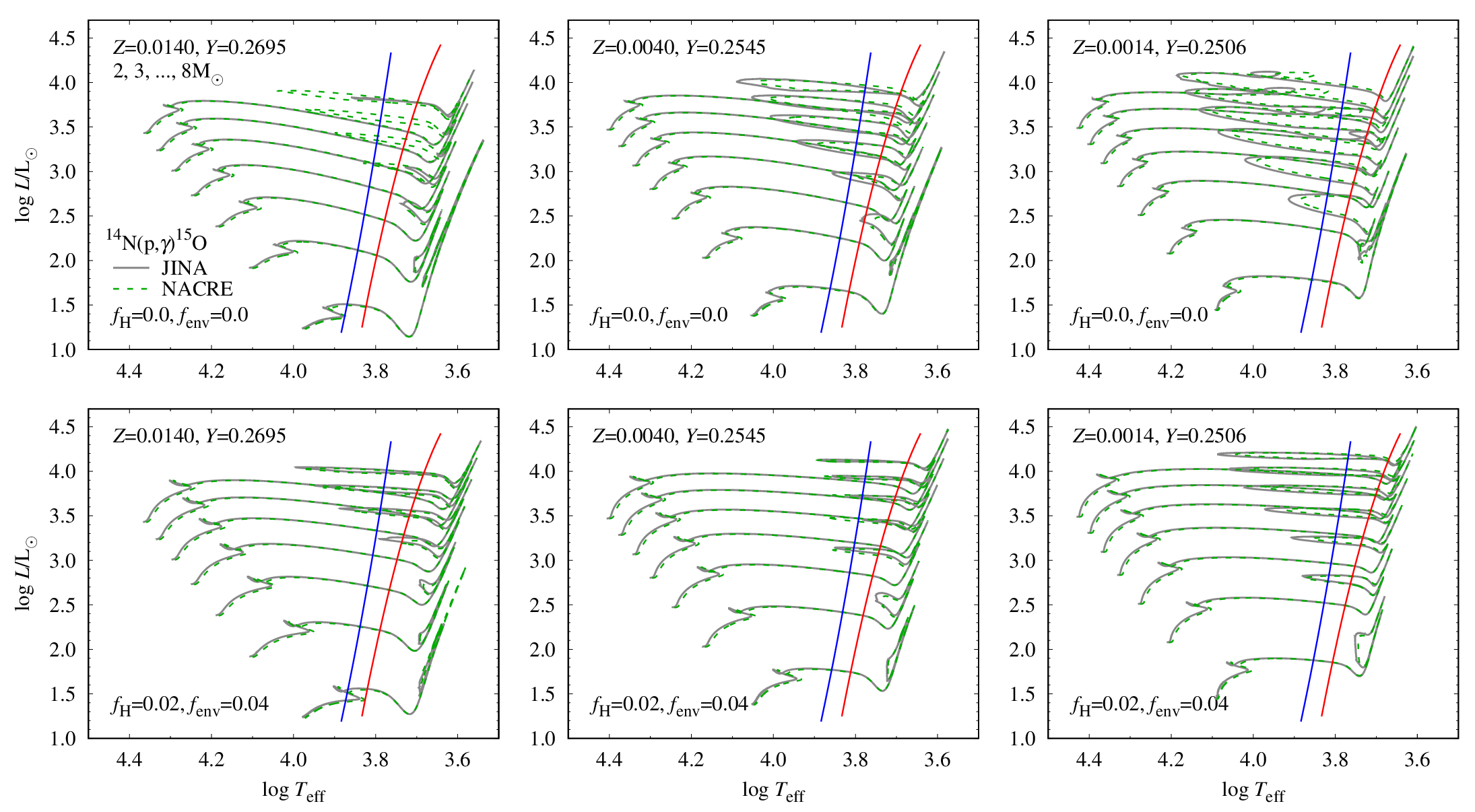}
\caption{The effect of \npg{} reaction rate (the default JINA REACLIB, solid gray lines vs.\ the NACRE rate, dashed green lines) across metallicity ($Z=0.014$, $Z=0.004$ and $Z=0.0014$, in the left, middle and right columns, respectively) and overshooting (no overshoot and core and envelope overshoot enabled, in the top and bottom rows, respectively) scenarios. Fiducial hot IS strip is over-plotted for a reference.
\label{figapp:HRnpg}}
\end{figure*}

\npg{} is the slowest reaction in the CNO chain, setting the rate of the whole cycle. It is known to have a significant impact on evolutionary tracks of low and intermediate mass stars \citep[see eg.,][]{Weiss-2005, Pietrinferni-2010, Cassisi-2014}. The JINA REACLIB rate \citep{jina} we are using as default, at lower temperatures ($T_9<0.2$) is lower than the NACRE \citep{nacre}, leading to slightly brighter MS phase. The differences in evolutionary tracks, when using these two rates are presented in Fig.~\ref{figapp:HRnpg} for high ($Z=0.014$), intermediate ($Z=0.004$) and low ($Z=0.0014$) metallicity (left, middle and right columns, respectively) and for models without overshooting (top row), and models including core and envelope overshooting ($\fcor=0.02$, $\fenv=0.04$; bottom row). The differences are most significant for models without overshooting at solar metallicity. For $M\geq5\MS$ the NACRE models have extended loops that enter the IS, while the JINA REACLIB models do not enter the IS. In contrast, for intermediate and low metallicity, the NACRE models give slightly shorter and slightly more luminous blue loops. When overshooting is turned on, the differences between JINA and REACLIB models are much smaller, however at some parameter regimes (eg., $Z=0.004$ and 5, 6\MS{} models) significant (longer loops for NACRE models).

Overall the differences between JINA and NACRE tracks are noticeable, but small, with the exception of high metallicity models without convective core overshooting.

\section{Evolutionary tracks with GS98 reference solar composition}\label{secapp:gs98}

\begin{figure}[ht!]
\includegraphics[width=\linewidth]{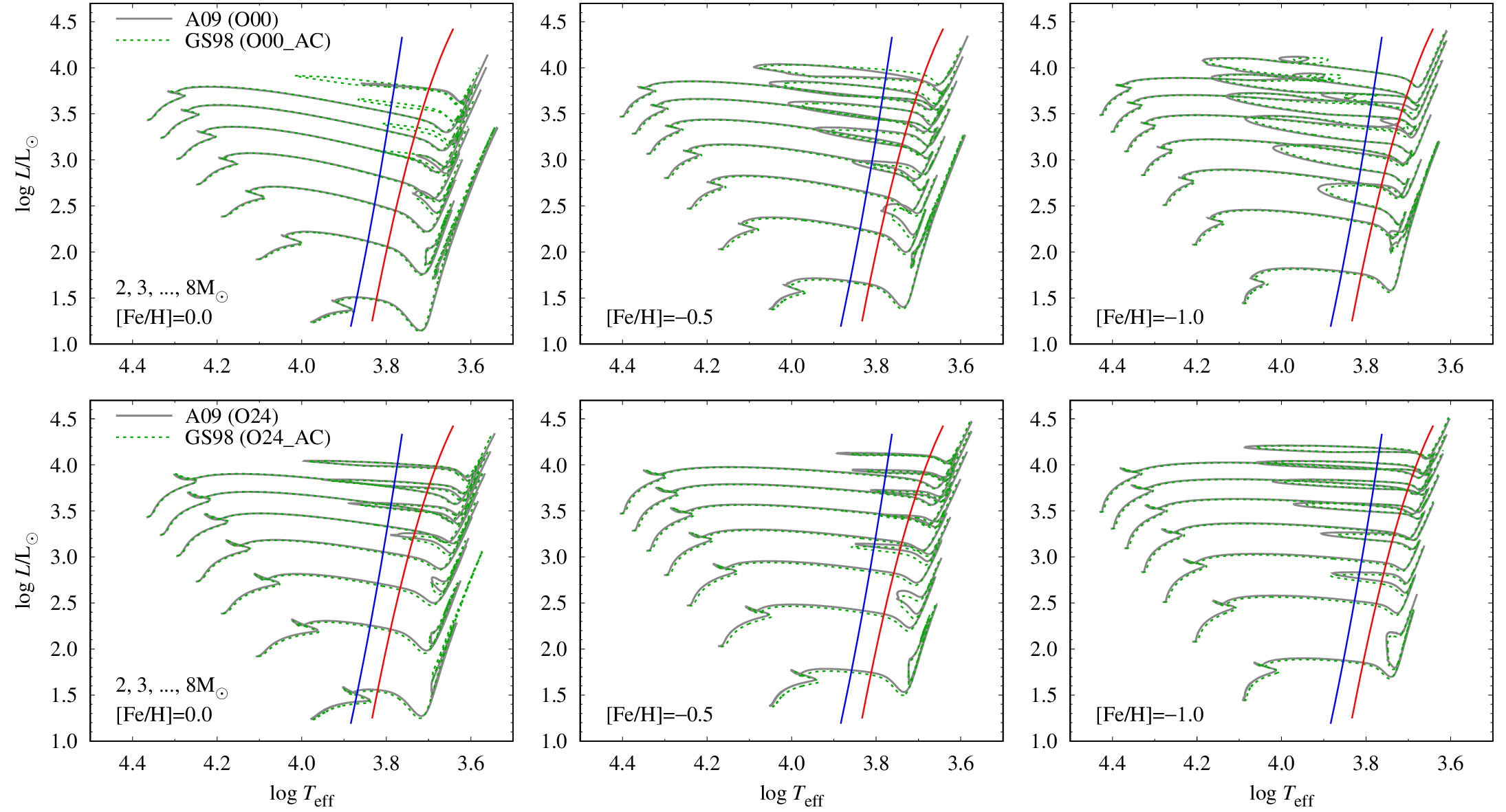}
\caption{The effect of different reference solar compositions, A09 (solid gray lines) vs.\ GS98 (dashed green lines) across metallicity (\feh{}=0.0, $-0.5$ and $-1.0$, in the left, middle and right columns, respectively)
and overshooting (no overshoot and core and envelope overshooting included, in the top and bottom rows, respectively) scenarios. Fiducial hot IS strip is over-plotted for a reference.
\label{figapp:gs98}}
\end{figure}

Two sets of tracks, O00\_AC and O24\_AC, were computed assuming  \cite{GS-98} reference solar composition. In these models, mixing length parameter was calibrated separately in \citetalias{Ziolkowska-2024} and $X$, $Y$ and $Z$ values were adjusted to match the same \feh{} values as adopted for other tracks in this work, see last column in Tab.~\ref{tab:fehz} and tabs. 1 and 3 in \citetalias{Ziolkowska-2024}. In Fig.~\ref{figapp:gs98}, we compare these tracks with those assuming our standard reference solar composition, ie.\ A09.

Tracks assuming GS98 composition are in general slightly dimmer. The track morphology, including the extent of the loops, is nearly the same in models of intermediate and low metallicity and in all models including core and envelope overshooting (O24 models). The only noticeable differences are recorded at solar metallicity and models without overshooting. Here, at higher masses, the loops are significantly longer when GS98 composition is used, while they are short, not even entering the IS (except for 8\MS) for A09 composition.

\section{Relations for the cool IS}\label{secapp:coolIS}

In Tabs~\ref{tabapp:plrelation}, \ref{tabapp:ml}, \ref{tabapp:pr} and \ref{tabapp:pa} we present the $\PL$, $\ML$, $\PR$ and $\PA$ relations, respectively, computed assuming cool IS.

\begin{table*}
\caption{Coefficients of the $\PL$ relations as given by eq.~\eqref{eq:plfeh}. First two columns identify data set, crossing number and edge along which relation is computed (`b/m/r' for the blue, midline and red edge). In the last column, we provide the recommended parameter range in which the relation is applicable. All relations adopt cool IS. \label{tabapp:plrelation}}
\begin{tabular}{lcrrrrcl}
\tablecolumns{8}
data & crossing & $\alpha$  &  $\delta$ & $\gamma$ & $\log P_0$ & rms  & remarks\\
\hline
\multicolumn{8}{l}{$W_{VI}=I-1.55(V-I)$}\\
\hline
O00 & 1c/m & $-3.639\pm0.008$ & $-3.875\pm0.005$ & $-0.200\pm0.011$ & $0.308$ & $0.034$ & $\log P_0\in(-0.65,1.00)$ \\
O00 & 2c/m & $-3.477\pm0.020$ & $-5.388\pm0.010$ & $-0.121\pm0.017$ & $0.789$ & $0.044$ & $\log P_0\in(0.11,1.24)$ \\
O00 & 3c/m & $-3.506\pm0.031$ & $-5.360\pm0.017$ & $-0.205\pm0.027$ & $0.789$ & $0.041$ & $\log P_0\in(0.25,1.28)$ \\
O24 & 1c/b & $-3.611\pm0.007$ & $-4.712\pm0.007$ & $-0.154\pm0.010$ & $0.556$ & $0.009$ & $\log P_0\in(-0.69,1.08)$ \\
O24 & 1c/m & $-3.541\pm0.010$ & $-4.594\pm0.008$ & $-0.176\pm0.014$ & $0.556$ & $0.033$ & $\log P_0\in(-0.54,1.23)$ \\
O24 & 1c/r & $-3.356\pm0.019$ & $-4.396\pm0.014$ & $-0.226\pm0.028$ & $0.556$ & $0.073$ & $\log P_0\in(-0.51,1.35)$ \\
O24 & 2c/b & $-3.567\pm0.014$ & $-6.125\pm0.007$ & $-0.141\pm0.012$ & $0.997$ & $0.015$ & $\log P_0\in(0.30,1.33)$ \\
O24 & 2c/m & $-3.302\pm0.019$ & $-5.959\pm0.009$ & $-0.214\pm0.017$ & $0.997$ & $0.041$ & $\log P_0\in(0.33,1.54)$ \\
O24 & 2c/r & $-3.087\pm0.017$ & $-5.694\pm0.010$ & $-0.212\pm0.017$ & $0.997$ & $0.044$ & $\log P_0\in(0.38,1.68)$ \\
O24 & 3c/b & $-3.568\pm0.016$ & $-6.103\pm0.008$ & $-0.163\pm0.014$ & $0.997$ & $0.012$ & $\log P_0\in(0.35,1.35)$ \\
O24 & 3c/m & $-3.296\pm0.019$ & $-5.933\pm0.009$ & $-0.233\pm0.017$ & $0.997$ & $0.036$ & $\log P_0\in(0.42,1.54)$ \\
O24 & 3c/r & $-3.100\pm0.019$ & $-5.665\pm0.011$ & $-0.212\pm0.019$ & $0.997$ & $0.043$ & $\log P_0\in(0.46,1.67)$ \\
\hline
\multicolumn{8}{l}{$W_{VK}=K-0.13(V-K)$}\\
\hline
O00 & 1c/m & $-3.624\pm0.013$ & $-3.760\pm0.008$ & $-0.154\pm0.017$ & $0.308$ & $0.046$ & $\log P_0\in(-0.65,1.00)$ \\
O00 & 2c/m & $-3.461\pm0.014$ & $-5.269\pm0.008$ & $-0.071\pm0.013$ & $0.789$ & $0.028$ & $\log P_0\in(0.11,1.24)$ \\
O00 & 3c/m & $-3.486\pm0.029$ & $-5.243\pm0.015$ & $-0.156\pm0.025$ & $0.789$ & $0.030$ & $\log P_0\in(0.25,1.28)$ \\
O24 & 1c/m & $-3.532\pm0.012$ & $-4.479\pm0.009$ & $-0.127\pm0.017$ & $0.556$ & $0.034$ & $\log P_0\in(-0.54,1.23)$ \\
O24 & 2c/m & $-3.336\pm0.014$ & $-5.844\pm0.007$ & $-0.161\pm0.013$ & $0.997$ & $0.024$ & $\log P_0\in(0.33,1.54)$ \\
O24 & 3c/m & $-3.337\pm0.015$ & $-5.820\pm0.007$ & $-0.182\pm0.013$ & $0.997$ & $0.018$ & $\log P_0\in(0.42,1.54)$ \\
\hline
\multicolumn{8}{l}{$K$ band}\\
\hline
O00 & 1c/m & $-3.557\pm0.014$ & $-3.586\pm0.008$ & $-0.156\pm0.017$ & $0.308$ & $0.046$ & $\log P_0\in(-0.65,1.00)$ \\
O00 & 2c/m & $-3.391\pm0.015$ & $-5.066\pm0.008$ & $-0.074\pm0.013$ & $0.789$ & $0.029$ & $\log P_0\in(0.11,1.24)$ \\
O00 & 3c/m & $-3.416\pm0.028$ & $-5.042\pm0.015$ & $-0.158\pm0.024$ & $0.789$ & $0.030$ & $\log P_0\in(0.25,1.28)$ \\
O24 & 1c/m & $-3.464\pm0.012$ & $-4.292\pm0.009$ & $-0.129\pm0.017$ & $0.556$ & $0.034$ & $\log P_0\in(-0.54,1.23)$ \\
O24 & 2c/m & $-3.265\pm0.014$ & $-5.629\pm0.007$ & $-0.162\pm0.013$ & $0.997$ & $0.024$ & $\log P_0\in(0.33,1.54)$ \\
O24 & 3c/m & $-3.265\pm0.014$ & $-5.606\pm0.007$ & $-0.182\pm0.013$ & $0.997$ & $0.018$ & $\log P_0\in(0.42,1.54)$\\
\hline
\end{tabular}
\end{table*}


\begin{deluxetable*}{lcrrrrcl}
\tablecaption{Coefficients of the $\ML$ relations as given by eq.~\eqref{eq:linearfeh}. First two columns identify data set, crossing number and edge along which relation is computed (`m' for the midline). In the last column, we provide the recommended parameter range in which the relation is applicable. All relations adopt cool IS. \label{tabapp:ml}}
\tablehead{
\colhead{data} & \colhead{cross./edge} & \colhead{$a$}  &  \colhead{$b$} & \colhead{$c$} & \colhead{$\log x_0$} & \colhead{rms}  & \colhead{remarks}
}
\startdata
\multicolumn{8}{l}{{\it Mass-Luminosity relations (cool IS)}}\\
O00 & 1c/m & $3.272\pm0.011$ & $2.800\pm0.003$ & $-0.268\pm0.005$ & $0.699$ & $0.019$ & $\log M/\MS\in(0.30,0.90)$ \\
O00 & 2c/m & $3.329\pm0.036$ & $3.346\pm0.006$ & $-0.147\pm0.011$ & $0.778$ & $0.030$ & $\log M/\MS\in(0.48,0.90)$ \\
O00 & 3c/m & $3.104\pm0.065$ & $3.385\pm0.012$ & $-0.230\pm0.021$ & $0.778$ & $0.043$ & $\log M/\MS\in(0.48,0.90)$ \\
O24 & 1c/m & $3.552\pm0.018$ & $3.020\pm0.005$ & $-0.264\pm0.010$ & $0.699$ & $0.019$ & $\log M/\MS\in(0.30,0.90)$ \\
O24 & 2c/m & $3.423\pm0.033$ & $3.534\pm0.005$ & $-0.243\pm0.010$ & $0.778$ & $0.024$ & $\log M/\MS\in(0.48,0.90)$ \\
O24 & 3c/m & $3.335\pm0.036$ & $3.571\pm0.006$ & $-0.254\pm0.011$ & $0.778$ & $0.024$ & $\log M/\MS\in(0.48,0.90)$ \\
O24\_ML2 & 2c/m & $3.423$(fixed) & $3.546\pm0.004$ & $-0.238\pm0.007$ & $0.778$ & $0.026$ & $\log M/\MS\in(0.47,0.90)$ \\
 O24\_ML2 & 3c/m & $3.335$(fixed) & $3.582\pm0.003$ & $-0.248\pm0.006$ & $0.778$ & $0.023$ & $\log M/\MS\in(0.47,0.90)$ \\
\enddata
\end{deluxetable*}

\begin{deluxetable*}{lcrrrrcl}
\tablecaption{Coefficients of the $\PR$ relations as given by eq.~\eqref{eq:linearfeh}. First two columns identify data set, crossing number and edge along which relation is computed (`b/m/r' for the blue, midline and red edge). In the last column, we provide the recommended parameter range in which the relation is applicable. All relations adopt cool IS. \label{tabapp:pr}}
\tablehead{
\colhead{data} & \colhead{cross./edge} & \colhead{$a$}  &  \colhead{$b$} & \colhead{$c$} & \colhead{$\log x_0$} & \colhead{rms}  & \colhead{remarks}
}
\startdata
\multicolumn{8}{l}{{\it Period-Radius relations (cool IS)}}\\
O00 & 1c/b & $0.7564\pm0.0010$ & $1.4041\pm0.0007$ & $0.0222\pm0.0013$ & $0.308$ & $0.0044$ & $\log P_0\in(-0.75,0.86)$ \\
O00 & 1c/m & $0.7462\pm0.0016$ & $1.3904\pm0.0010$ & $0.0274\pm0.0020$ & $0.308$ & $0.0071$ & $\log P_0\in(-0.65,1.00)$ \\
O00 & 1c/r & $0.7341\pm0.0022$ & $1.3737\pm0.0014$ & $0.0429\pm0.0030$ & $0.308$ & $0.0103$ & $\log P_0\in(-0.63,1.13)$ \\
O00 & 2c/b & $0.7437\pm0.0031$ & $1.7202\pm0.0019$ & $-0.0001\pm0.0030$ & $0.789$ & $0.0055$ & $\log P_0\in(0.04,1.08)$ \\
O00 & 2c/m & $0.7181\pm0.0028$ & $1.7023\pm0.0015$ & $0.0109\pm0.0024$ & $0.789$ & $0.0064$ & $\log P_0\in(0.11,1.24)$ \\
O00 & 2c/r & $0.6994\pm0.0023$ & $1.6835\pm0.0014$ & $0.0296\pm0.0024$ & $0.789$ & $0.0058$ & $\log P_0\in(0.14,1.39)$ \\
O00 & 3c/b & $0.7507\pm0.0059$ & $1.7129\pm0.0033$ & $0.0148\pm0.0049$ & $0.789$ & $0.0057$ & $\log P_0\in(0.18,1.14)$ \\
O00 & 3c/m & $0.7241\pm0.0050$ & $1.6963\pm0.0026$ & $0.0268\pm0.0042$ & $0.789$ & $0.0061$ & $\log P_0\in(0.25,1.28)$ \\
O00 & 3c/r & $0.7076\pm0.0046$ & $1.6773\pm0.0027$ & $0.0424\pm0.0044$ & $0.789$ & $0.0067$ & $\log P_0\in(0.24,1.43)$ \\
O00 & 2c+3c/m & $0.7160\pm0.0026$ & $1.6991\pm0.0014$ & $0.0183\pm0.0022$ & $0.789$ & $0.0094$ & $\log P_0\in(0.16,1.28)$ \\
O24 & 1c/b & $0.7379\pm0.0014$ & $1.5567\pm0.0013$ & $0.0203\pm0.0020$ & $0.556$ & $0.0024$ & $\log P_0\in(-0.69,1.08)$ \\
O24 & 1c/m & $0.7274\pm0.0016$ & $1.5390\pm0.0013$ & $0.0227\pm0.0024$ & $0.556$ & $0.0046$ & $\log P_0\in(-0.54,1.23)$ \\
O24 & 1c/r & $0.7160\pm0.0019$ & $1.5205\pm0.0014$ & $0.0357\pm0.0029$ & $0.556$ & $0.0069$ & $\log P_0\in(-0.51,1.35)$ \\
O24 & 2c/b & $0.7231\pm0.0025$ & $1.8429\pm0.0012$ & $0.0149\pm0.0021$ & $0.997$ & $0.0035$ & $\log P_0\in(0.30,1.33)$ \\
O24 & 2c/m & $0.6934\pm0.0023$ & $1.8225\pm0.0012$ & $0.0292\pm0.0021$ & $0.997$ & $0.0041$ & $\log P_0\in(0.33,1.54)$ \\
O24 & 2c/r & $0.6780\pm0.0021$ & $1.8027\pm0.0012$ & $0.0436\pm0.0022$ & $0.997$ & $0.0046$ & $\log P_0\in(0.38,1.68)$ \\
O24 & 3c/b & $0.7225\pm0.0028$ & $1.8381\pm0.0013$ & $0.0189\pm0.0023$ & $0.997$ & $0.0030$ & $\log P_0\in(0.35,1.35)$ \\
O24 & 3c/m & $0.6941\pm0.0024$ & $1.8175\pm0.0012$ & $0.0332\pm0.0021$ & $0.997$ & $0.0029$ & $\log P_0\in(0.42,1.54)$ \\
O24 & 3c/r & $0.6829\pm0.0024$ & $1.7964\pm0.0015$ & $0.0447\pm0.0025$ & $0.997$ & $0.0042$ & $\log P_0\in(0.46,1.67)$ \\
O24 & 2c+3c/m & $0.6925\pm0.0017$ & $1.8199\pm0.0009$ & $0.0310\pm0.0015$ & $0.997$ & $0.0048$ & $\log P_0\in(0.37,1.54)$ \\
\enddata
\end{deluxetable*}

\begin{deluxetable*}{lcrrrrcl}
\tablecaption{Coefficients of the $\PA$ relations as given by eq.~\eqref{eq:linearfeh}. First two columns identify data set, crossing number and edge along which relation is computed (`b/m/r' for the blue, midline and red edge). In the last column, we provide the recommended parameter range in which the relation is applicable. All relations adopt cool IS. \label{tabapp:pa}}
\tablehead{
\colhead{data} & \colhead{cross./edge} & \colhead{$a$}  &  \colhead{$b$} & \colhead{$c$} & \colhead{$\log x_0$} & \colhead{rms}  & \colhead{remarks}
}
\startdata
\multicolumn{8}{l}{{\it Period-Age relations (cool IS)}}\\
 O00 & 1c/b & $-0.9338\pm0.0075$ & $7.8657\pm0.0046$ & $-0.1220\pm0.0094$ & $0.308$ & $0.039$ & $\log P_0\in(-0.75,0.86)$ \\
 O00 & 1c/m & $-0.8959\pm0.0093$ & $7.9338\pm0.0056$ & $-0.1404\pm0.0118$ & $0.308$ & $0.049$ & $\log P_0\in(-0.65,1.00)$ \\
 O00 & 1c/r & $-0.8718\pm0.0103$ & $8.0028\pm0.0066$ & $-0.1545\pm0.0141$ & $0.308$ & $0.057$ & $\log P_0\in(-0.63,1.13)$ \\
 O00 & 2c/b & $-0.8863\pm0.0182$ & $7.7356\pm0.0110$ & $0.0208\pm0.0169$ & $0.789$ & $0.039$ & $\log P_0\in(0.04,1.08)$ \\
 O00 & 2c/m & $-0.8058\pm0.0157$ & $7.8187\pm0.0081$ & $-0.0125\pm0.0136$ & $0.789$ & $0.042$ & $\log P_0\in(0.11,1.24)$ \\
 O00 & 2c/r & $-0.7592\pm0.0133$ & $7.9037\pm0.0083$ & $-0.0250\pm0.0137$ & $0.789$ & $0.042$ & $\log P_0\in(0.14,1.39)$ \\
 O00 & 3c/b & $-0.9351\pm0.0155$ & $7.7959\pm0.0079$ & $-0.1112\pm0.0120$ & $0.789$ & $0.032$ & $\log P_0\in(0.18,1.14)$ \\
 O00 & 3c/m & $-0.8502\pm0.0143$ & $7.8788\pm0.0072$ & $-0.1400\pm0.0116$ & $0.789$ & $0.037$ & $\log P_0\in(0.25,1.28)$ \\
 O00 & 3c/r & $-0.8178\pm0.0167$ & $7.9698\pm0.0097$ & $-0.1414\pm0.0157$ & $0.789$ & $0.050$ & $\log P_0\in(0.24,1.43)$ \\
 O00 & 2c+3c/m & $-0.7911\pm0.0193$ & $7.8461\pm0.0100$ & $-0.0809\pm0.0164$ & $0.789$ & $0.076$ & $\log P_0\in(0.16,1.28)$ \\
 O24 & 1c/b & $-0.8425\pm0.0040$ & $7.9214\pm0.0029$ & $-0.0965\pm0.0054$ & $0.556$ & $0.015$ & $\log P_0\in(-0.69,1.08)$ \\
 O24 & 1c/m & $-0.8063\pm0.0049$ & $8.0001\pm0.0033$ & $-0.1090\pm0.0067$ & $0.556$ & $0.024$ & $\log P_0\in(-0.54,1.23)$ \\
 O24 & 1c/r & $-0.7801\pm0.0051$ & $8.0748\pm0.0036$ & $-0.1187\pm0.0073$ & $0.556$ & $0.027$ & $\log P_0\in(-0.51,1.35)$ \\
 O24 & 2c/b & $-0.8071\pm0.0082$ & $7.7572\pm0.0036$ & $-0.1087\pm0.0063$ & $0.997$ & $0.020$ & $\log P_0\in(0.30,1.33)$ \\
 O24 & 2c/m & $-0.7191\pm0.0070$ & $7.8535\pm0.0033$ & $-0.1342\pm0.0060$ & $0.997$ & $0.020$ & $\log P_0\in(0.33,1.54)$ \\
 O24 & 2c/r & $-0.6705\pm0.0070$ & $7.9427\pm0.0039$ & $-0.1411\pm0.0068$ & $0.997$ & $0.024$ & $\log P_0\in(0.38,1.68)$ \\
 O24 & 3c/b & $-0.8129\pm0.0060$ & $7.7954\pm0.0026$ & $-0.1345\pm0.0045$ & $0.997$ & $0.013$ & $\log P_0\in(0.35,1.35)$ \\
 O24 & 3c/m & $-0.7369\pm0.0056$ & $7.8969\pm0.0026$ & $-0.1569\pm0.0047$ & $0.997$ & $0.015$ & $\log P_0\in(0.42,1.54)$ \\
 O24 & 3c/r & $-0.7134\pm0.0065$ & $7.9987\pm0.0038$ & $-0.1522\pm0.0063$ & $0.997$ & $0.020$ & $\log P_0\in(0.46,1.67)$ \\
 O24 & 2c+3c/m & $-0.7213\pm0.0074$ & $7.8754\pm0.0035$ & $-0.1449\pm0.0063$ & $0.997$ & $0.032$ & $\log P_0\in(0.37,1.54)$ \\
\enddata
\end{deluxetable*}

\clearpage

\section{Data for evolutionary and pulsation relations in tabular form}

\movetabledown=2.5in
\begin{rotatetable*}
\begin{deluxetable*}{lllrrrrrrrrrrrrrrr}
\digitalasset
\tablewidth{0pt}
\tablecaption{Data for evolutionary and pulsation relations, see Tab.~\ref{tab:datatab_content} for the description of table content. \label{tab:data_all}}
\tablehead{\colhead{edge} & \colhead{set} & \colhead{cross.} & \colhead{$M$} & \colhead{$Z$} & \colhead{$X$} & \colhead{$\log {\rm age}$} & \colhead{$\log\Teff$} & \colhead{$\log L/\LS$} & \colhead{$\log R/\RS$} & \colhead{$Y_c$} & \colhead{$P_0$} & \colhead{$P_1$} & \colhead{$V$} & \colhead{$I$} & \colhead{$J$} & \colhead{$H$} & \colhead{$K$}}
\colnumbers
\startdata
b & O00 &  1c & 2.00000 & 0.0200 & 0.70150 & 9.0015 & 3.8723 & 1.4429 & 0.4996 & 0.98074 &  0.124209 &  0.091755 &  1.0712 &  0.8172 &  0.6346 &  0.5054 &  0.5340 \\ 
b & O00 &  1c & 2.50000 & 0.0200 & 0.70150 & 8.7381 & 3.8561 & 1.8146 & 0.7178 & 0.98072 &  0.245918 &  0.182314 &  0.1425 & -0.1483 & -0.3530 & -0.4996 & -0.4660 \\ 
b & O00 &  1c & 3.00000 & 0.0200 & 0.70150 & 8.5221 & 3.8448 & 2.0839 & 0.8750 & 0.98073 &  0.396588 &  0.293212 & -0.5213 & -0.8522 & -1.0595 & -1.2168 & -1.1793 \\ 
b & O00 &  1c & 3.50000 & 0.0200 & 0.70150 & 8.3430 & 3.8360 & 2.3025 & 1.0020 & 0.98074 &  0.581778 &  0.427686 & -1.0645 & -1.4319 & -1.6530 & -1.8272 & -1.7863 \\ 
b & O00 &  1c & 4.00000 & 0.0200 & 0.70150 & 8.1915 & 3.8286 & 2.4894 & 1.1102 & 0.98074 &  0.806172 &  0.588297 & -1.5288 & -1.9244 & -2.1596 & -2.3482 & -2.3053 \\ 
b & O00 &  1c & 4.50000 & 0.0200 & 0.70150 & 8.0606 & 3.8223 & 2.6526 & 1.2045 & 0.98075 &  1.071743 &  0.775714 & -1.9318 & -2.3498 & -2.6088 & -2.8146 & -2.7692 \\ 
b & O00 &  1c & 5.00000 & 0.0200 & 0.70150 & 7.9469 & 3.8166 & 2.8003 & 1.2896 & 0.98076 &  1.388323 &  0.996019 & -2.2953 & -2.7365 & -3.0158 & -3.2372 & -3.1891 \\ 
b & O00 &  1c & 5.50000 & 0.0200 & 0.70150 & 7.8470 & 3.8116 & 2.9353 & 1.3672 & 0.98076 &  1.760830 &  1.251841 & -2.6269 & -3.0885 & -3.3861 & -3.6217 & -3.5714 \\ 
b & O00 &  1c & 6.00000 & 0.0200 & 0.70150 & 7.7586 & 3.8069 & 3.0606 & 1.4391 & 0.98077 &  2.198798 &  1.549039 & -2.9332 & -3.4110 & -3.7260 & -3.9745 & -3.9224 \\ 
b & O00 &  1c & 6.50000 & 0.0200 & 0.70150 & 7.6797 & 3.8027 & 3.1766 & 1.5056 & 0.98077 &  2.703313 &  1.888000 & -3.2161 & -3.7109 & -4.0409 & -4.3015 & -4.2474 \\ 
b & O00 &  1c & 7.00000 & 0.0200 & 0.70150 & 7.6094 & 3.7987 & 3.2857 & 1.5681 & 0.98078 &  3.287183 &  2.277713 & -3.4815 & -3.9942 & -4.3375 & -4.6096 & -4.5530 \\ 
b & O00 &  1c & 7.50000 & 0.0200 & 0.70150 & 7.5459 & 3.7951 & 3.3869 & 1.6259 & 0.98078 &         x &         x & -3.7274 & -4.2575 & -4.6125 & -4.8948 & -4.8358 \\ 
b & O00 &  1c & 8.00000 & 0.0200 & 0.70150 & 7.4890 & 3.7917 & 3.4833 & 1.6810 & 0.98079 &         x &         x & -3.9613 & -4.5049 & -4.8720 & -5.1628 & -5.1026 \\ 
m & O00 &  1c & 2.00000 & 0.0200 & 0.70150 & 9.0081 & 3.8472 & 1.4459 & 0.5512 & 0.98074 &  0.151345 &  0.111948 &  1.0865 &  0.7433 &  0.5395 &  0.3794 &  0.4142 \\ 
m & O00 &  1c & 2.50000 & 0.0200 & 0.70150 & 8.7388 & 3.8330 & 1.7934 & 0.7534 & 0.98072 &  0.281870 &  0.208649 &  0.2220 & -0.1737 & -0.3987 & -0.5833 & -0.5433 \\ 
\multicolumn{18}{l}{\ldots}\\
b & O00 &  1c & 2.00000 & 0.0160 & 0.71150 & 8.9823 & 3.8703 & 1.4865 & 0.5253 & 0.98459 &  0.136614 &  0.101124 &  0.9688 &  0.7109 &  0.5224 &  0.3889 &  0.4182 \\ 
b & O00 &  1c & 2.50000 & 0.0160 & 0.71150 & 8.7175 & 3.8550 & 1.8399 & 0.7326 & 0.98458 &  0.259331 &  0.192344 &  0.0863 & -0.2095 & -0.4168 & -0.5667 & -0.5327 \\ 
b & O00 &  1c & 3.00000 & 0.0160 & 0.71150 & 8.5041 & 3.8440 & 2.1042 & 0.8868 & 0.98458 &  0.413414 &  0.305650 & -0.5668 & -0.9056 & -1.1158 & -1.2783 & -1.2397 \\ 
b & O00 &  1c & 3.50000 & 0.0160 & 0.71150 & 8.3281 & 3.8352 & 2.3212 & 1.0128 & 0.98459 &  0.603961 &  0.443964 & -1.1062 & -1.4805 & -1.7045 & -1.8838 & -1.8421 \\ 
b & O00 &  1c & 4.00000 & 0.0160 & 0.71150 & 8.1792 & 3.8279 & 2.5068 & 1.1202 & 0.98459 &  0.834346 &  0.608972 & -1.5670 & -1.9683 & -2.2079 & -2.4017 & -2.3579 \\ 
\multicolumn{18}{l}{\ldots}\\
\enddata
\tablecomments{This Table is published in its entirety in the electronic edition of the {\it Astrophysical Journal Supplement Series} and online on Zenodo. Results for hot and cool IS are available in separate tables online. A portion is shown here for guidance regarding its form and content.}
\end{deluxetable*}
\end{rotatetable*}

\end{document}